# High Performance Computing with FPGAs and OpenCL

**Hamid Reza Zohouri**

Supervisor: Prof. Satoshi Matsuoka

# Abstract

With the impending death of Moore's law, the High Performance Computing (HPC) community is actively exploring new options to satisfy the never-ending need for faster and more power efficient means of computing. Even though GPUs have been widely employed in world-class supercomputers in the past few years to accelerate different types of computation, the high power usage and slow power efficiency improvements of these devices remains a limiting factor in deploying larger supercomputers, especially on the path to Exascale computing. Field-Programmable Gate Arrays (FPGAs) are an emerging alternative to GPUs for this purpose. These devices, despite being older than GPUs, have rarely been used in the HPC industry and have mainly been employed in embedded and low-power markets. Since the traditional applications of FPGAs have vastly different requirements compared to typical HPC application, the usability, productivity and performance of FPGAs for HPC applications is unknown.

In this study, our aim is to determine whether FPGAs can be considered as a viable solution for accelerating HPC applications, and if so, how they fare against existing processors in terms of performance and power efficiency in different HPC workloads. We take advantage of the recent improvements in High Level Synthesis (HLS) that, unlike traditional Hardware Description Languages (HDL) that are known to be notoriously hard to use and debug, allow FPGAs to be more easily programmed by software programmers using familiar software programming languages. Specifically, we use Intel FPGA SDK for OpenCL that allows modern Intel FPGAs to be programmed as an accelerator, similar to GPUs.

In the first step, we evaluate the performance and power efficiency of FPGAs in different benchmarks, each being a representative of a specific HPC workload. For this purpose, we port a subset of the Rodinia benchmark suite for two generations of Intel FPGAs, and then optimize each benchmark based on the specific architectural characteristics of these FPGAs. Then, we compare the performance and power efficiency of these devices against same-generation CPUs and GPUs. We show that even though a direct port of CPU and GPU kernels for FPGAs usually performs poorly on these devices, with FPGA-specific optimizations, up to two orders of magnitude performance improvement can be achieved, resulting in better performance to that of CPUs in all cases, and competitive performance to that of GPUs in most. Furthermore, we show that FPGAs have a clear power efficiency edge in every case, reaching up to 16.7 and 5.6 times higher power efficiency compared to their same-generation CPUs and GPUs, respectively.

Based on our experience from the initial evaluation, we determine that for stencil computation, which is one of the most important computation patterns in HPC, FPGAs can not only compete with GPUs in terms of power efficiency, but also in terms of pure performance. Taking advantage of the unique architectural advantages of FPGAs for stencil computation, we design and implement a parameterized OpenCL-based template kernel that can be used to accelerate 2D and 3D star-shaped stencils on FPGAs regardless of stencil order. Our design, apart from using both spatial and temporal blocking, also employs multiple HLS-specific optimizations for FPGAs to maximize performance. Moreover, we devise a performance model that allows us to quickly tune the performance parameters of our design, significantly reducing the time and resources necessary for placement and routing. We show that our design allows FPGAs to achieve superior performance to that of CPUs, GPUs and Xeon Phi devices in 2D stencil computation, and competitive performance for 3D. Specifically, we show that using an Intel Arria 10 GX 1150 device, for 2D and 3D star-shaped stencils of first to fourth-order, we can achieve over 700 and 270 GFLOP/s of compute performance, respectively. Furthermore, we show that our implementation outperforms all existing implementations of stencil computation on FPGAs.

This thesis makes multiple contributions to the emerging field of using FPGAs in HPC, and the optimization techniques discussed in this work can be used as guidelines for optimizing most types of applications on FPGAs using HLS, even for non-HPC applications.

# Acknowledgements

I would like to thank my supervisor, Prof. Satoshi Matsuoka, for giving me the chance to study at Tokyo Tech and be a part of Matsuoka Lab., and providing me with valuable guidance throughout my PhD. I would also like to thank my mentors, Dr. Naoya Maruyama and Dr. Artur Podobas, for their valuable mentorship and helping me walk the path towards a successful PhD.

I would also like to express my gratitude to Matsuoka Lab. research administrators, especially Chisato Saito, for their continuous help and support regarding administration work and life in Japan. Furthermore, I would like to thank my friends and colleagues at Matsuoka Lab. for their help with the many different research and life-related problems I encountered during my stay in Japan.

I would like to thank MEXT for providing me with the scholarship that allowed me to come to Japan and study at Tokyo Tech, one of the best universities in the world. Furthermore, I would like to thank JST, JSPS and OIL for their generous fundings that allowed me to perform my research using state-of-the-art hardware and software. I would also like to thank Intel for donating software licenses through their university program that proved crucial in my research.

Finally, I would like to thank my parents for their continuous support of my life, and their patience and understanding during the years I had to be away from home so that I could focus on my studies; I certainly would not have been able to reach this point without them.

**Hamid Reza Zohouri**

**August 2018**

# Table of Contents

# List of Figures

# List of Tables

# 1 Introduction

## 1.1 Motivation

For many years, non-stop improvements in computer technology in terms of both performance and power efficiency have been driven by the Moore's Law [1] and Dennard Scaling [2]. However, with Moore's Law losing steam, and Dennard Scaling coming to an end, the age of Dark Silicon [3] is closer than ever. High Performance Computing (HPC), which relies on latest cutting-age hardware to satisfy the never-ending need for higher performance and power efficiency, is going to be most impacted by this new age. This has forced the HPC community to employ specialized accelerators in the past few years. So far, GPUs have been the most popular accelerator to be used for HPC applications. However, these devices are also impacted by the impending death of Moore's law just like CPUs. Apart from that, GPUs are power-hungry devices that can consume up to 300 Watts and power efficiency improvements in GPUs is reaching its limit. With power consumption and efficiency being the main bottleneck in designing and employing large HPC machines, the usability of GPUs in large supercomputers is subject to many power and cooling limitations.

FPGAs are one of the accelerators that are recently emerging as more power-efficient alternatives to GPUs. Even though these devices are older than GPUs, they have been traditionally designed for low-power and embedded markets and have had limited computational capabilities. Furthermore, these devices were traditionally programmed using Hardware Description Languages (HDL), mainly Verilog and VHDL, that are based on a vastly different programming model compared to standard software programming languages like C and Fortran. This issue has always been a major roadblock in adoption of FPGAs among software programmers.

For many years, High Level Synthesis (HLS) tools have been developed to make FPGAs usable by software programmers. Such tools allow software programmers to describe their FPGA design in a standard software programming language, and then convert this *high-level* description to a *low-level* description based on Verilog or VHDL. Many such tools have been developed since the inception of HLS; however, rarely any of them have been endorsed or supported by the major FPGA manufacturers, namely Intel PSG (formerly Altera) and Xilinx. Recently, Xilinx acquired AutoESL [4] and based on that, developed Vivado HLS [5] that allows conversion of C and C++ code to low-level FPGA descriptions. Later, Altera (now Intel PSG) introduced their OpenCL SDK [6] to provide a similar possibility for software programmers based on the open-source and royalty-free OpenCL programming language. Eventually, Xilinx followed suit and introduced their OpenCL SDK named SDAccel [7]. With official HLS tools being directly developed and supported by FPGA manufacturers, a sudden shift in the HLS ecosystem happened that enabled more widespread adoption of FPGAs among software programmers.

In 2014, the very first large-scale adoption of FPGAs in a cloud system was kick-started by Microsoft under the Catapult project [8]. Microsoft specifically chose to employ FPGAs instead of GPUs due to lower power and space requirement of FPGAs, which allowed them to achieve a notable improvement in the performance of the Bing search engine, with minimal changes in the design of their data center. Later, Intel introduced their new Arria 10 FPGA family which, for the first time in the history of FPGAs, included DSPs with native support for floating-point operations [9]. This radical change in FPGA architecture paved the way for adoption of FPGAs in the HPC market that largely relies on floating-point computation. Since the past year, FPGAs have also become available in commercial cloud platforms like Amazon AWS [10] and Nimbix [11].

## 1.2 Problem Statement

FPGAs are relatively new in the HPC ecosystem and it is not clear how suitable they are for accelerating HPC applications. On top of that, existing HLS tools are much less mature compared to widely-used software compilers and hence, it is not known how well they perform on a given FPGA for different application types. Moreover, optimization techniques for different types of applications have been widely studied on CPUs and GPUs, while there is little existing work on optimizing HPC applications on FPGAs using HLS.

Despite the recent advancements in FPGA technology, these devices are still behind GPUs in terms of both compute performance and external memory bandwidth. For example, an Arria 10 GX 1150 FPGA with full DSP utilization operating at the peak DSP operating frequency of 480 MHz [9] provides a peak single-precision floating-point compute performance of 1.45 GFLOP/s. Furthermore, typical Arria 10 FPGA boards [12] are coupled with two banks of DDR4 memory running at 2133 MHz (1066 double data-rate), and a 64-bit bus to each bank, which provides only 34.1 GB/s of external memory bandwidth. Compared to the same-generation NVIDIA GTX 980 Ti GPU, with a compute performance of 6900 GFLOP/s and external memory bandwidth of 336.6 GB/s, the Arria 10 FPGA is at a 4.75x disadvantage in terms of compute performance, and a ~10x disadvantage in terms of external memory bandwidth. However, the TDP of the Arria 10 FPGA is 3.9x lower (70 Watts vs. 275 Watts), potentially allowing this FPGA to achieve better power efficiency than the GTX 980 Ti GPU if high computational efficiency can be achieved on the FPGA.

Considering the major architectural differences between FPGAs and CPUs/GPUs, it is not clear how well existing CPU and GPU code perform on FPGAs and how well typical CPU or GPU-based optimizations affect performance on FPGAs, if at all. Unlike CPUs and GPUs, FPGAs do not have a cache hierarchy; however, modern FPGAs provide a large amount of on-chip memory (6.6 MB on Arria 10 GX 1150) which can be used as scratchpad memory. Furthermore, high-performance CPU and GPU applications largely rely on the multi-threading capabilities of these devices while being forced to align with SIMD and vectorization limitations of such hardware that is the result of their fixed architecture. However, no such restrictions exist on FPGAs due to their reconfigurable nature, giving the programmer much more design flexibility at the cost of larger design exploration space and long placement and routing time.

## 1.3 Proposal and Contributions

In the first part of our study, to study the usability and performance of FPGAs in HPC, we evaluate FPGAs in a set of benchmarks that are representative of typical HPC workloads. We port a subset of the well-known Rodinia benchmark suite [13] for Intel FPGAs and compare the performance and power efficiency of two FPGA generations to that of their same-generation CPUs and GPUs. In this part of our study, we make the following contributions:

- We devise a general performance model for computation on FPGAs and use this model as a guide for optimizing FPGA kernels. Based on this model, we show that the traditional NDRange OpenCL programming that is used on GPUs and takes advantage of thread-level parallelism is usually not suitable for FPGAs. Instead, the Single Work-item model that takes advantage of pipelined parallelism matches better with the underlying FPGA architecture and achieves better performance in most cases.
- We present a comprehensive list of HLS-based optimization techniques for FPGAs, ranging from basic compiler-assisted optimizations to advanced manual optimizations, and describe how each of them is expected to affect performance on an FPGA based on our model.
- We show that a direct port of kernels that are optimized for CPUs and GPUs perform poorly on FPGAs. However, by using advanced optimizations techniques that take advantage of the unique architectural features of FPGAs, we can achieve over an order of magnitude performance improvement compared to direct ports.
- We show that in some applications, FPGAs can achieve competitive performance to that of their same-generation GPUs, and in all of our studied applications, they achieve better power efficiency up to 5.6 times higher. Furthermore, FPGAs can achieve better performance and power efficiency compared to their same-generation CPUs in every case.

Based on our experience from the first part of our study, we conclude that one of the computation patterns in HPC that FPGAs can excel at is stencil computation. Hence, we further focus on this computation pattern to maximize the performance of applications based on this type of computation on FPGAs. In this part of our study, we make the following contributions:

- We create an FPGA-based accelerator for stencil computation that uses two parameterized OpenCL template kernels, one for 2D stencils and one for 3D, to quickly implement different stencils. Apart from performance parameters, stencil radius is also parameterized in our kernel so that high-order stencils, which are widely used in HPC applications, can also be accelerated on FPGAs using our design.
- Unlike many previous work on accelerating stencil computation on FPGAs that take advantage of temporal blocking but avoid spatial blocking to achieve maximize performance at the cost of restricting the size of the input in multiple dimensions, we combine spatial and temporal blocking and show that it is possible to achieve high performance without such restrictions.

- We tackle the issues arisen from the added design complexity due to multiple levels of blocking and multiply-nested loops in our design by taking advantage of multiple HLS-based FPGA-specific optimizations.
- We devise a performance model for our FPGA-based stencil computation accelerator that allows us to quickly tune the performance parameters in our design and minimize the number of configurations that need to be placed and routed. This significantly reduces the amount of time and computational resources that is necessary for parameter tuning on FPGAs.
- We show that for first to fourth-order star-shaped 2D and 3D stencils, we can achieve over 700 GFLOP/s and 270 GFLOP/s of compute performance, respectively, on an Arria 10 GX 1150 device. This level of performance is superior to that of CPUs, Xeon Phi and GPUs for 2D stencil computation, and competitive or better in 3D. Furthermore, the FPGA remains the most power efficient device in nearly all cases.
- Using our performance model, we project the performance of our evaluated stencils for the upcoming Intel Stratix 10 FPGAs and show that these devices can achieve up to 4.2 TFLOP/s and 1.8 TFLOP/s of compute performance, for 2D and 3D stencil computation, respectively. This level of performance is expected to be superior to that of modern GPUs for 2D stencils, and competitive for 3D, with superior power efficiency in every case.

## 1.4 Thesis Outline

The remaining chapters of this thesis are outlined as follows:

- **Background:** In this chapter, we briefly discuss the architecture of FPGAs, the OpenCL programming model and Intel FPGA SDK for OpenCL.
- **General Performance Model and Optimizations for FPGAs:** In this chapter, we first discuss our general performance model for computation on FPGAs and based on that, demonstrate the differences between NDRange and Single Work-item programming models and give guidelines as to which is preferred depending on the target application. Then, we present a set of code optimization techniques, ranging from basic compile-assisted optimizations to advanced manual optimizations and describe how each maps to our model.
- **Evaluating FPGAs for HPC Applications Using OpenCL:** In this chapter, we discuss the details of porting and optimizing a subset of the Rodinia benchmark suite based on the optimization techniques from the previous chapter and show the effect of different levels of optimization on performance. Then, we compare each benchmark on each FPGA with its same-generation CPU and GPU in performance and power efficiency.
- **High-Performance Stencil Computation on FPGAs Using OpenCL:** In this chapter, we first discuss our implementation of first-order stencil computation on FPGAs using combined spatial and temporal blocking. Then we extend this implementation to high-order stencils. In the next step, we present our performance model for our stencil accelerator, which is used to prune our parameter search space. Finally, we project the performance of our evaluated stencils for the upcoming Stratix 10 FPGAs and compare the performance and power efficiency of our design on Stratix V, Arria 10 and Stratix 10 FPGAs with multiple CPU, GPU and Xeon Phi devices.

- **Summary and Insights:** In the final chapter, we summarize our work and present insights we obtained from the study.

# 2 Background

## 2.1 Field-Programmable Gate Arrays (FPGAs)

### 2.1.1 FPGA Architecture

FPGAs are generally regarded as a middle-ground between ASICs and general-purpose processors. This notion comes from the reconfigurable nature of these devices, making them more flexible than ASICs (at the cost of lower area and power efficiency) and more power efficient than general-purpose processors (at the cost of lower flexibility and more complex programming). Even though, deep down, FPGAs have a fixed architecture, they are largely composed of SRAM cells arranged in form of Loop-Up Tables (LUT), a plethora of registers, and programmable routing. Because of this, these devices can be rapidly reconfigured to implement different logic, just by changing the content of the LUTs and the routing configuration. Apart from the *soft-logic* LUTs, modern FPGAs also include *hard-logic* components such as Digital Signal Processors (DSP), large memory blocks (Block RAMs) and different I/O controllers (DDR, PCI-E, network, etc.). These components implement specialized logic that would otherwise take up too much space if implemented using LUTs.

Fig. 2-1 shows the architecture of the Intel Arria 10 FPGA [14]. In this FPGA, the soft-logic consists of Adaptive Logic Modules (ALM), and the hard-logic consists of DSPs, Block RAMs, multiple controllers, Transceivers and Phase-Locked Loops (PLL).

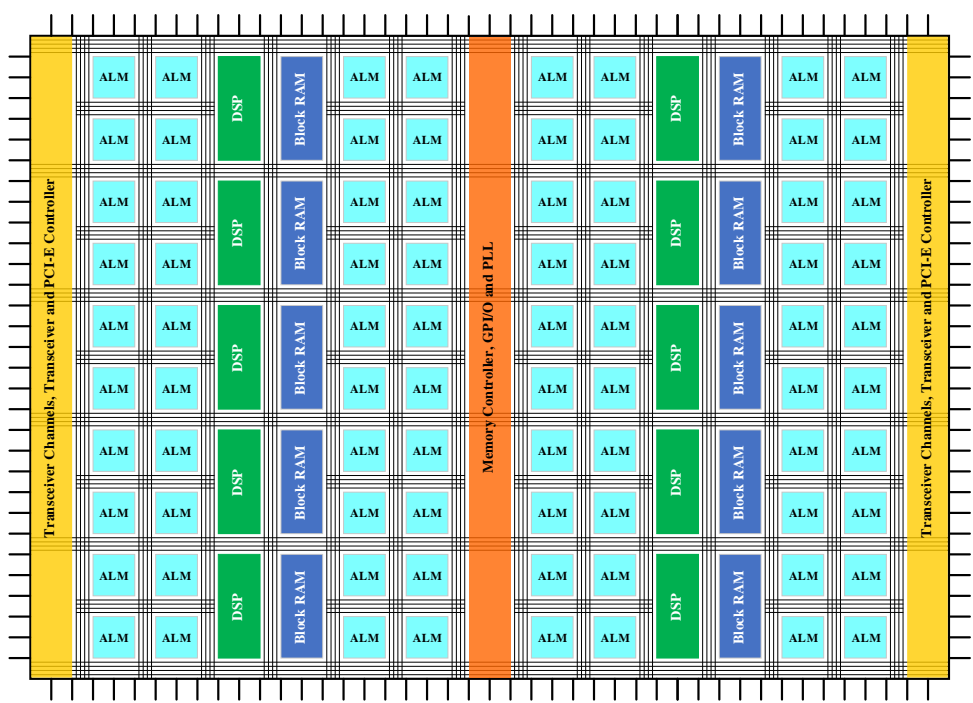

**Figure 2-1 Intel Arria 10 FPGA architecture**

In the Arria 10 FPGA, each ALM consists of multiple-input LUTs, adders and carry logic, and registers (Flip-Flops). The internal architecture of the ALMs in the Arria 10 FPGA is depicted in Fig. 2-2. Each Adaptive LUT is capable of implementing multiple combinations of different functions including one 6-input function, two 5-input functions with two shared inputs, two 4-input functions with independent inputs, etc.

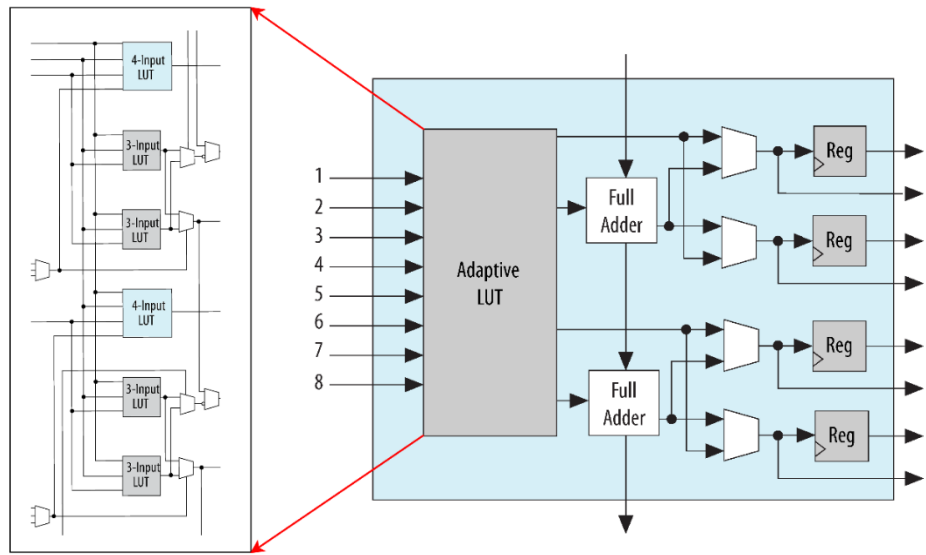

**Figure 2-2 Intel Arria 10 ALM architecture; combined from Fig. 5 in [14] and Fig. 7 in [15]**

Fig. 2-3 shows the block diagram of the DSPs in the Intel Arria 10 FPGA. Each of these DSPs is capable of implementing an IEEE-754-compliant single-precision floating-point addition (FADD), multiplication (FMUL), or Fused Multiply and Add (FMA) operation, or one 27-bit-by-27-bit integer or fixed-point multiplication. Furthermore, multiple DSPs can be chained to implement dot products or other complex operations.

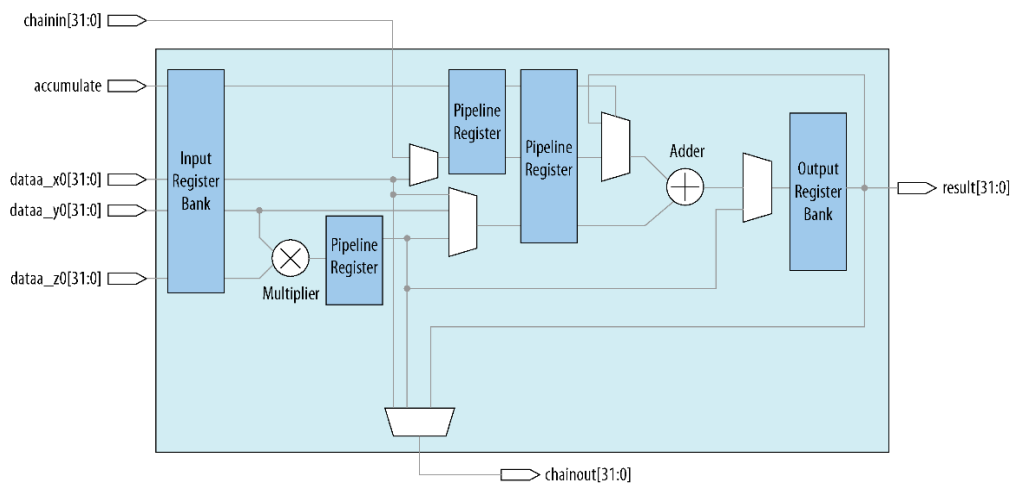

**Figure 2-3 Intel Arria 10 DSP architecture; taken from Fig. 27 in [15]**

Finally, each Block RAM in the Intel Arria 10 device, called an M20K block, is capable of storing a maximum of 20 Kbits of data. Each block has two ports that operate independently, and can satisfy one read and one write operation simultaneously. Data can be stored in each block with a maximum width of 40 bits, in which case the address size will be 9 bits (512 addresses). Apart from implementing multiple-ported RAM or ROMs, each M20K can also be used to implement First-In First-Out buffers (FIFO) or shift registers. Multiple M20K blocks can also be chained to implemented larger buffers.

### 2.1.2 FPGA Synthesis

Traditionally, to create an FPGA design, the application is described using a Hardware Description Language (HDL) like Verilog or VHDL, and then multiple steps are carried out until an FPGA bitstream is created. First, the hardware description is *synthesized* into a netlist. All coding errors are determined in this step. In the next step, the *mapping* process maps all functions in the netlist to functions that are available as hard-logic on the FPGA; any other function will be implemented using soft-logic (LUTs). After that, the *placement* process determines which of the multiple instances of each function on the FPGA should be used for implementing the functions that are required by the design. If a design requires more instances of a specific function than are available on the FPGA, placement will fail. In the next step, the *routing* process will determine which routing resources are used and how they are connected so that all functions are correctly connected to each other and all timing constraints are met. Since routing resources are limited, routing could also fail in case of routing congestion. Finally, the FPGA bitstream is generated. The bitstream is generally transferred to the FPGA using JTAG to implement the synthesized design on the hardware. This chain of operations is the equivalent of *compilation* for software programs, and all this functionality is provided by the FPGA manufacturers' tools. In case of Intel, these functions are provided by Intel Quartus Prime Software. For the Intel Stratix V FPGA, total synthesis time is typically 3 to 5 hours but can reach up to 8 hours, while for the larger Arria 10 device it is typically 8 to 12 hours but can take over a day for very large designs that suffer from severe routing congestion.

## 2.2 OpenCL Programming Language

OpenCL [16] is an open-source and royalty-free standard for programming heterogeneous systems in a host/device fashion. An OpenCL-based application can be split into two separate parts: one is the *host* code that executes on the host CPU and can be written in any programing language as long as a compatible compiler exists, and the other is a C-based *device* code that is more commonly called the *kernel* code. OpenCL provides the necessary APIs for controlling the accelerator and communicating between the host processor and the accelerator. This programming language can be considered as a device-agnostic alternative to the NVIDIA CUDA programming language.

The typical flow of operation in OpenCL is that first, all necessary data is allocated in host memory. Then, this data is transferred to the device memory using the respective OpenCL functions. In the next step, the kernel is loaded and executed on the device, where inputs are read from and outputs are written to the device memory. Finally, output data is transferred from the device memory to the host.

### 2.2.1 OpenCL Threading Model

In OpenCL, each thread is called a *work-item* and multiple work-items are grouped to form a *work-group*. To execute an application, the thread space is distributed over multiple work-groups. Within each work-group, work-items are synchronized using *barriers* and data can be shared between the work-items using the fast on-chip *local memory*. However, the only way to share data between different work-groups is through the slow off-chip memory. The number of work-items in a work-group is called the *local work size*, and the total number of work-items necessary to fully execute an application is called the *global work size*. Work-items and work-groups can be arranged in multiple dimensions, up to three, in an index space called an *NDRange*.

### 2.2.2 OpenCL Memory Model

In OpenCL, multiple memory types are defined:

**Global:** This memory space resides on the device off-chip (external) memory and is generally the largest (up to a couple Gigabytes) but slowest memory that exists on an OpenCL-capable accelerator. The content of this memory space is visible to all work-items of all work-groups. Global memory consistency is only guaranteed after a kernel is executed completely.

**Local:** This memory space resides on the on-chip memory of the OpenCL device and can be used to share data between the work-items within a work-group. Each work-group has its own local memory space, and the local memory space of a work-group is not visible to other work-groups. The total amount of local memory available on OpenCL accelerators is typically a few to a couple Megabytes. Local memory consistency is only guaranteed at *barriers*.

**Constant:** This memory space resides on device external memory; however, this is a read-only memory space and is generally cached in device on-chip memory for faster access.

**Private:** Any work-item-specific buffer or array is of this memory type. Private data generally resides on the fast device registers; however, due to very limited size (a few hundred Kilobytes per device), data stored in this memory space can leak to global memory, incurring a large performance penalty.

## 2.3 Intel FPGA SDK for OpenCL

Intel FPGA SDK for OpenCL provides the necessary APIs and run-time to program and use PCI-E-attached or System-on-Chip (SoC) FPGAs similar to a GPU or other accelerators. The necessary IP Cores to communicate between the FPGA, external DDR memory, and PCI-E, alongside with necessary PCI-E and DMA drivers for communication between the host and the FPGA are also provided by the board manufacturers in form of a Board Support Package (BSP). This relieves the programmer from the burden of having to manually set up the IP Cores and create the drives, as is done with traditional HDL-based FPGA designs. Some BSPs also provide the possibility to send and receive data using FPGA on-board network ports.

## 2.3.1 Intel FPGA SDK for OpenCL Flow

Fig. 2-4 shows the flow of Intel FPGA SDK for OpenCL (formerly Altera SDK for OpenCL) to compile the host code and convert the kernel code to an FPGA-compatible bitstream.

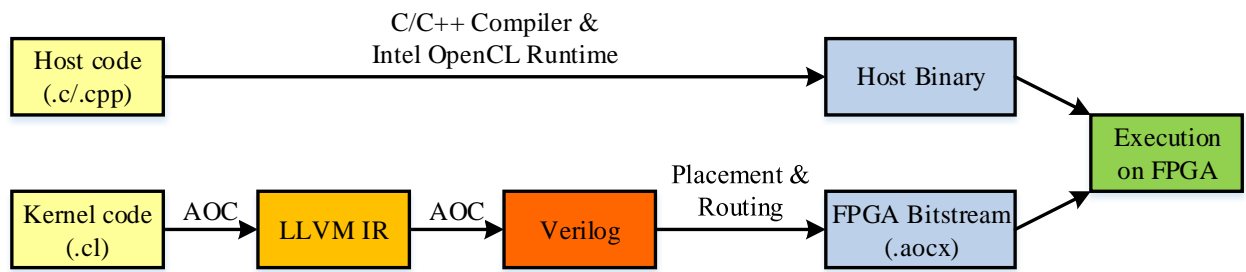

**Figure 2-4 Intel FPGA SDK for OpenCL flow; AOC is Intel FPGA SDK for OpenCL Offline Compiler**

Unlike CPUs and GPUs, run-time compilation of OpenCL kernels is not possible for FPGAs due to very long placement and routing time. Hence, the OpenCL kernel needs to be compiled offline into an FPGA bitstream, and then loaded at run-time by the host code to reprogram the FPGA and execute the application.

## 2.3.2 NDRange Programming Model on FPGAs

GPUs typically consist of a set of coarse-grained processing units (called Compute Units in AMD GPUs and Streaming Multiprocessors in NVIDIA GPUs), with each such unit containing a fixed number of fine-grained shader processors and a fixed amount of scratchpad memory and L1 cache. In the NDRange programming model, a work-group is generally mapped to one of the coarse-grained units, and each work-item is mapped to one of the fine-grained shared processors. However, the aforementioned decomposition to coarse and fine-grained units does not exists on FPGAs. By default, using the NDRange programming model on FPGAs will *not* result in thread-level parallelism and instead, the compiler will generate one compute unit implemented as a deep pipeline, with all work-items from all work-groups executing on that pipeline. Each region between two barriers in an NDRange kernel will be mapped to a separate pipeline, with each pipeline being flushed at the barrier. The compiler also automatically performs *work-group pipelining*, allowing multiple work-groups to be in-flight in the same compute unit simultaneously to maximize the efficiency of all the pipelines in the compute unit, at the cost of higher Block RAM usage. Fig. 2-5 a) shows how two consecutive threads/work-items are pipelined with a distance from each other, called the Initiation Interval (II). The initiation interval is adjusted at run-time by the run-time scheduler that the compiler implements on the FPGA to minimize pipeline stalls and maximize pipeline efficiency.

Intel FPGA SDK for OpenCL provides a SIMD attribute for the NDRange programming model that allows achieving work-item-level parallelism on the FPGA. Using this attributes, the pipeline is widened and pipeline stages are replicated so that multiple work-items can be issued in parallel by the scheduler in a compute unit. This programming model also provide the possibly to replicate the compute unit so that work-group-level parallelism can be achieved. Mixing SIMD and compute unit replication allows the programmer to achieve a GPU-like architecture on an FPGA, with the compute units acting as the coarse-grained processing units, and each set of the vectorized pipeline stages acting as a fine-grained unit.

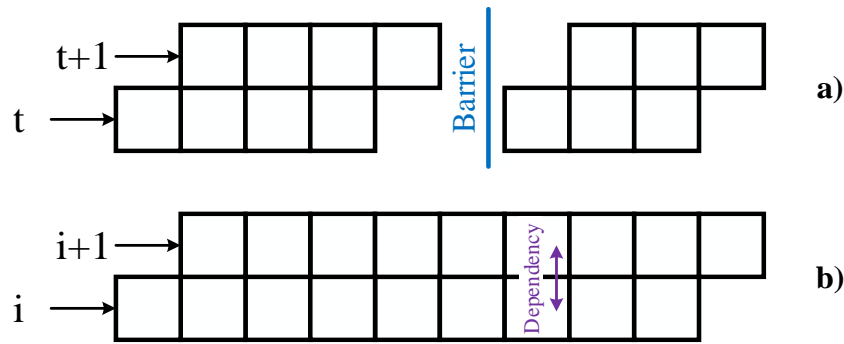

**Figure 2-5 Pipeline generation for (a) NDRange and (b) Single Work-item kernels**

### 2.3.3 Single Work-item Programming Model on FPGAs

Apart from the NDRange model, Intel FPGA SDK for OpenCL also provides another programming model called the *Single Work-item* model. In the model, the entire kernel is executed by one work-item and instead, loop iterations are pipelined to achieve high performance. When this programming model is used, each set of nested loops in the kernel is mapped to a separate pipeline by the compiler. Fig. 2-5 b) depicts how two consecutive iterations of a loop are pipelined one after another. Using this model, no run-time scheduler will be created in the hardware anymore and instead, iteration scheduling is static and initiation interval is determined at compile-time depending on loop-carried and memory load/store dependencies. Iteration-level parallelization in form of vectorization can be achieved in this programming model by *loop unrolling*.

Fig. 2-6 a) shows a basic example of an NDRange kernel and b) shows its equivalent Single Work-item kernel. Converting an NDRange kernel to Single Work-item can be done by wrapping the NDRange kernel in a *for* loop from zero to global work size in every dimension.

### 2.3.4 OpenCL Memory Types on FPGAs

Using Intel FPGA SDK for OpenCL, OpenCL **global** memory resides on the FPGA external memory, which is usually a few banks of DDR3 or DDR4 memory. OpenCL **local** and **private** memory, depending on the size and access pattern of the buffer, will be implemented using registers or Block RAMs. Finally, **constant** memory is also implemented using Block RAMs with a fixed size that can be controlled using a compilation argument.

```
__kernel void ndrange(__global float* a, __global float* b)
{
   int i = get_global_id(0);
   a[i] = b[i];
}
```
a)

```
__kernel void single_wi(__global float* a, __global float* b, int global_size)
{
   for (int i = 0; i < global_size; i++)
   {
      a[i] = b[i];
   }
}
```
b)

Figure 2-6 NDRange (a) vs. Single Work-item (b) code example

# 3 General Performance Model and Optimizations for FPGAs

In this chapter, we will discuss our general performance model for FPGAs starting from a single-pipeline model and then extending it for data parallelism. Then, we outline the difference between the two programing models available in Intel FPGA SDK for OpenCL based on this model. In the next step, we discuss multiple HLS-based optimization techniques for FPGAs ranging from basic compiler-assisted optimizations to advanced manual optimizations, and explain how each relates to our model. The contents of this chapter have been partially published in [17].

## 3.1 General Performance Model

### 3.1.1 Single-pipeline Model

For a given pipeline with a depth of *P*, a loop trip count of *L* (i.e. number of inputs) and an initiation interval of *II*, the total number of clock cycles to finish computation is:

$$T_{cycle} = P + II \times (L - 1) \tag{3-1}$$

Here, *P* cycles are required until the pipeline is filled and the first output is generated, and after that, a new output is generated every *II* cycles. To convert this value to time, we have:

$$T_{seconds} = \frac{T_{cycle}}{f_{max}} = \frac{P + II \times (L - 1)}{f_{max}} \tag{3-2}$$

In Eq. (3-2), $f_{max}$ is the operating frequency of the FPGA that is determined after placement and routing and is typically between 150 to 350 MHz on Intel Stratix V and Arria 10 devices. Among the parameters in Eq. (3-2), *P* is controlled by the compiler; however, as a general rule of thumb, simpler code will result in a shorter pipeline and lower *P*. *L* is also application-dependent and cannot be directly controlled by the user. $f_{max}$ is also generally a function of circuit complexity and size. Loop-carried dependencies and feedbacks in the design will adversely affect $f_{max}$. Moreover, the bigger the design is and the closer utilization of each resource is to 100%, the more $f_{max}$ will be lowered due to placement and routing complications. In Section 3.2.4.4, we will show an advanced optimization technique that can significantly improve $f_{max}$ in Single Work-item kernel. The only remaining parameter is *II*. This parameter is the one that can be most directly influenced by the programmer and hence, most of the performance optimization effort will be spent on improving this parameter.

*II* is influenced by multiple factors: loop-carried dependencies, shared on-chip resources like shared ports to local memory buffers implemented as multi-ported RAM/ROM, and accesses to/from external memory and on-chip channels since they can be stalled. These sources can be split into two groups: sources that affect compile-time initiation interval ($II_c$), and sources that affect run-time initiation interval ($II_r$). For Single Work-item kernels, the effect of loop-carried dependencies and shared on-chip resources is determined at compile-time and $II_c$ is adjusted accordingly. In NDRange kernels, loops are *not* pipelined and hence, no such analysis is done and we can assume $II_c = 1$. However, in both kernel types, accesses to external memory and channels will still influence $II_r$. By default, the compiler inserts enough stages in the pipeline to hide the *minimum* latency of these operations, and accesses that take longer at run-time result in a pipeline *stall*. To estimate compile-time initiation interval ($II_c$), we consider each kernel type separately (Fig. 3-1):

- **Single Work-item kernels:** $II_c$ in this case depends on the number of stall cycles per iteration ($N_d$) determined by the compiler and will be equal to $N_d + 1$. Hence, Eq. (3-1) transforms into:

$$T_{cycle} = P + (N_d + 1) \times (L - 1) \tag{3-3}$$

- **NDRange kernels:** In these kernels, even though we can assume $II_c = 1$, we need to take the overhead of barriers into account. Total run time for an NDRange kernel with $N_b$ barriers is:

$$\begin{aligned}T_{cycle} &= \sum_{i=0}^{N_b}(P_i + L_i - 1) = \left(\sum_{i=0}^{N_b} P_i\right) + (N_b + 1) \times (L - 1) \\ &= P + (N_b + 1) \times (L - 1)\end{aligned} \tag{3-4}$$

In Eq. (3-4), $P_i$ and $L_i$ show the pipeline length and number of inputs (work-items) for each pipeline in an NDRange kernel. Since the number of work-items is fixed per kernel, $L_i$ for every pipeline is the same and equal to $L$. Furthermore, we will call the accumulated length of all the pipelines, $P$. After simplifying the equation, we reach a statement that is very similar to Eq. (3-3). **In practice, the number of barriers in an NDRange kernel plays a similar role to that of stalls inserted in the pipeline due to dependencies in a Single Work-item kernel, and we can assume $II_c$ is equal to $(N_b + 1)$ instead of one.**

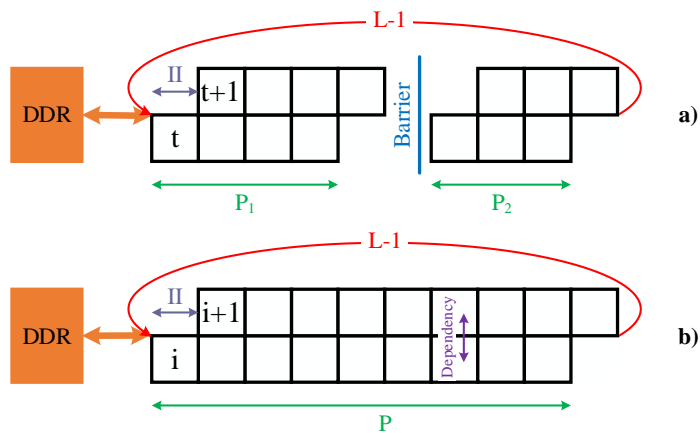

Figure 3-1 NDRange (a) vs. Single Work-item (b) pipeline model

To take the effect of external memory accesses into account and estimate run-time initiation interval ($II_r$), we use a simple model for external memory. For $N_m$ bytes read from and written to external memory per cycle and an external memory bandwidth per clock cycle of $BW$, we have:

$$II_r > \frac{N_m}{BW} \tag{3-5}$$

Here, $BW$ is determined by the specifications of the external memory on the FPGA board. Furthermore, since our model is simplified and does not take coalescing, alignment and contention from different external memory accesses into account, the right-hand size of (3-5) only shows the *minimum $II_r$*.

Putting everything together, we have:

$$II > \max(II_c, II_r) \Rightarrow II > \max\left(\begin{cases} N_d + 1 \\ N_b + 1 \end{cases}, \frac{N_m}{BW}\right) \tag{3-6}$$

We ignore the role of stalls caused by on-chip channels here since if the channels are deep enough and the rate of channel reads and writes is similar, channel stalls will be very rare.

### 3.1.2 Extension for Data Parallelism

To extend our model for cases where data parallelism in form of loop unrolling, SIMD or compute unit replication is employed with a degree of parallelism of $N_p$ (Fig. 3-2), run time can be calculated as:

$$T_{cycle} = P' + II \times \frac{(L - N_p)}{N_p} \tag{3-7}$$

In this case, the pipeline depth generally increases compared to the case where data parallelism is not present. However, for an $L \gg P'$, Eq. (3-7) points to a performance improvement of nearly $N_p$ times since the loop trip count is effectively reduced by a factor of $N_p$. On the other hand, data parallelism also increases memory pressure by a factor of $N_p$ and hence, $II$ will be affected as follow:

$$II > \max\left(\begin{cases} N_d + 1 \\ N_b + 1 \end{cases}, \frac{N_m \times N_p}{BW}\right) \tag{3-8}$$

Based on Eq. (3-7) and (3-8), when data parallelism is present, assuming that sufficient external memory bandwidth is available, performance will improve by a factor close to $N_p$.

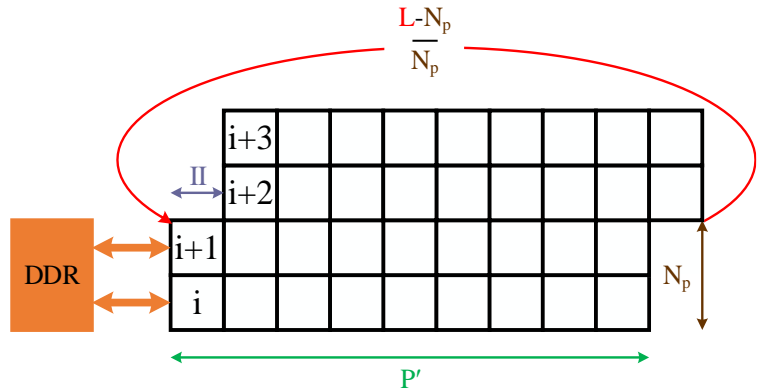

Figure 3-2 Pipeline model for data parallelism

### 3.1.3 General Optimization Guidelines

Based on our model, we conclude that to improve performance for an HLS-based design on FPGAs, our optimization effort should be focused on:

- Reducing stalls ($N_d$) in Single Work-item kernels
- Reducing number of barriers ($N_b$) in NDRange kernels
- Reducing external memory accesses ($N_m$)
- Increasing data parallelism ($N_p$)

### 3.1.4 Single Work-item vs. NDRange Kernels

One important factor in OpenCL-based designs for Intel FPGAs is to decide whether to use Single Work-item or NDRange kernels. Fig. 3-3 shows the most important difference between these two kernel types. Even though we explained earlier that in NDRange kernels, threads are pipelined and no thread-level parallelism exists by default, the programming model itself assumes that threads are running in parallel and hence, the issue distance between the threads neither appears nor can be influenced in the kernel code. Due to this reason, local memory-based optimizations in NDRange kernels require barriers since there is no direct way of transferring data between threads in an NDRange kernel. In contrast, in Single Work-item kernels there is a minimum distance of one clock cycle between loop iterations. It is possible to take advantage of this issue distance to directly transfer data from one loop iteration to another, especially to resolve loop-carried dependencies. This type of communication can be realized using single-cycle reads and writes from and to the plethora of registers that are available in every FPGA. Based on this analysis, we conclude that **in Single Work-item kernels, it might be possible to fully resolve iteration dependencies and reduce $N_d$ to zero; however, in NDRange kernels where local memory-based optimizations are employed, barriers will always be required and $N_b$ will never become zero. Hence, based on Eq. (3-6), a Single Work-item kernel can potentially have a lower effective $II_c$ compared to its NDRange equivalent.** This shows the clear advantage of the Single Work-item programming model compared to NDRange for FPGAs. Moreover, shift registers, which are an efficient storage type on FPGAs, can only be inferred in Single Work-item kernels (Section 3.2.4.1).

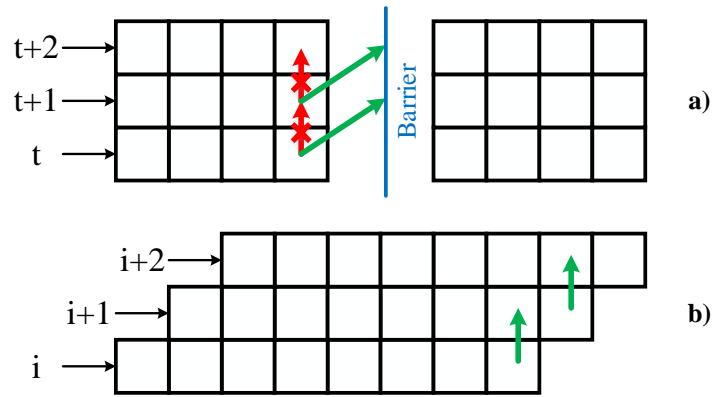

**Figure 3-3 Data sharing in (a) NDRange and (b) Single Work-item kernels**

On the other hand, NDRange kernels also have their own advantages in certain applications. In cases where an $II_c = 1$ can be achieved, a Single Work-item kernel is preferred. However, for cases where this cannot be achieved, NDRange kernels could potentially achieve better performance. This difference stems from the fact that $II_c$ in Single Work-item kernels is static and is determined based on the worst case loop-carried or load/store dependency at compile-time, while in NDRange kernels, initiation interval is determined at run-time by the thread scheduler. Because of this, in cases where $II_c = 1$ cannot be achieved in the Single Work-item implementation of an application, the thread scheduler in the NDRange equivalent might be able to achieve a lower *average initiation interval* by reordering the threads at run-time, compared to the Single Work-item equivalent with a fixed worst-case initiation interval.

In summary, the following points should be taken into account to decide whether to choose NDRange kernels for a design, or Single Work-item kernels:

- Applications with non-fully-pipelineable loops (e.g. loops with variable exit conditions or complex loop-carried/load/store dependencies) or random external memory accesses can potentially perform better using the NDRange programming model
- Every other application will potentially perform better using the Single Work-item programming model, especially if registers or shift registers can be used to efficiently resolve loop-carried dependencies.

## 3.2 HLS-based Optimization Techniques for FPGAs

In this section, we will discuss multiple different optimization techniques for HLS-based designs on FPGAs, ranging from basic compiler-assisted optimizations to advanced manual optimizations that require significant code refactoring. The basic optimizations discussed here are techniques that are introduced in Intel's OpenCL documents [18, 19], while most of the advanced ones are not directly discussed in these documents. We only discuss kernel optimization here. Some of the discussed optimizations are only applicable to one kernel type;

"NDR" and "SWI" are used to mark optimizations specific to NDRange and Single Work-item kernels, respectively. Optimizations applicable to both have been marked as "BOTH".

### 3.2.1 Basic Compiler-assisted Optimizations

**3.2.1.1 *restrict* keyword (BOTH):**

The *restrict* keyword can be added to global pointers in the kernel to prevent the compiler from assuming false pointer aliasing. This optimization usually has little to no effect in NDRange kernels; however, it is a crucial optimization in Single Work-item kernels, which, if not used, can result in very high initiation interval or even full sequential execution. This optimization can improve performance by reducing $N_d$, and consequently, reducing $II_c$.

**3.2.1.2 *ivdep* pragma (SWI):**

The *ivdep* pragma is used to prevent the compiler from assuming false load/store dependencies on global buffers. This pragma in only applicable to Single Work-item kernels and should be used with extreme care since incorrect usage can result in incorrect output. Even though this pragma can also be used for local buffers, it is very rare for the compiler to detect a false dependency on such buffers. This optimization can improve performance by reducing $N_d$, and consequently, reducing $II_c$.

**3.2.1.3 Removing thread scheduler (SWI):**

The run-time thread scheduler is only needed for NDRange kernels and is not required for Single Work-item kernels. However, in cases where a Single Work-item kernel is launched from the host using the *clEnqueueNDRangeKernel()* function, the scheduler is required for correct kernel launch. On the other hand, for cases where the *clEnqueueTask()* function is used to launch such kernel, a compiler-provided attribute can be used to remove the scheduler to save some FPGA resources. The only downside of this optimization is that *clEnqueueTask()* has been deprecated in OpenCL 2.0. This optimization does not directly affect performance and only reduces area utilization by a small amount.

**3.2.1.4 Setting work-group size (NDR):**

Intel FPGA SDK for OpenCL Offline Compiler provides attributes for NDRange kernels to set the exact or maximum kernel work-group size. This allows the compiler to minimize the size of local memory buffers based on the user-supplied information. Furthermore, the SIMD attribute can only be used if the exact work-group size is set by the programmer. If this information is not supplied, the compiler will assume a default work-group size of 256, potentially wasting valuable Block RAM resources if a smaller work-group size is used at run-time. Using this attribute comes at the cost of kernel execution failure if the run-time-provided work-group size supplied by the host does not align with the information supplied to the compiler in the kernel. This optimization does not have a direct effect on performance; however, the area reduction from this optimization can potentially allow performance improvement by using more data parallelism or larger block size.

**3.2.1.5 Data parallelism (BOTH):**

For NDRange kernels, data parallelism can be achieved by using the SIMD or *num_compute_units()* attributes. SIMD will provide work-item-level parallelism, while *num_compute_units()* provides work-group-level parallelism. Using SIMD has a lower area overhead since it does not require full compute unit replication, and only the pipeline stages are replicated so that multiple work-items can be issued and processed in parallel. Furthermore, internal and external memory accesses can be coalesced in this case, allowing higher internal and external memory bandwidth with minimum amount of contention and area overhead. However, using this attribute is subject to multiple limitations. First, the SIMD length must be a power of two up to maximum of 16 (which is an artificial compiler limitation since any arbitrary SIMD length should be implementable on an FPGA). Second, no thread-id-dependent branches should exist in the code. Finally, work-group size should be set by the programmer (Section 3.2.1.4) and be divisible by SIMD length. Using *num_compute_units()* has none of these limitations; however, it comes at the cost of higher area overhead due to complete compute unit replication, and lower memory throughput compared to SIMD due to multiple narrow accesses competing for the external memory bandwidth instead of one wide coalesced access. For Single Work-item kernels, a SIMD-like effect can be achieved using loop unrolling, without any of the limitations that exist for using SIMD. However, area overhead of loop unrolling will be minimized when a loop with a trip count known at compile-time is either fully unrolled, or partially unrolled with a factor that the trip count is divisible by. As explained in Section 3.1.2, these techniques can improve performance by a factor close to the degree of parallelism if sufficient external memory bandwidth is available.

A direct effect of data parallelism in form of SIMD for NDRange kernels and unrolling for Single Work-item kernels is that external memory accesses which are consecutive in the dimension that SIMD or unrolling is applied on will be coalesced by the compiler into wider accesses at compile-time, allowing better utilization of the external memory bandwidth. Compared to having multiple narrow accesses per iteration to external memory, a few wide accesses result in much less contention on the memory bus and much more efficient utilization of the external memory bandwidth. However, using SIMD and unrolling over non-consecutive external memory accesses could instead lead to many narrow access ports and lower performance due to large amount of contention on the memory bus. Using either of these techniques also has a similar effect on local memory buffers. Using SIMD and unrolling over consecutive local memory accesses leads to access coalescing and data interleaving with minimal area overhead, while applying these over non-consecutive accesses will result in high replication factors for local buffers and waste of FPGA area.

### 3.2.2 Basic Manual Optimizations

**3.2.2.1 Shift register for floating-point reduction (SWI):**

On most FPGAs, floating-point operations cannot be performed in one clock cycle (unless at the cost of extremely low operating frequency). Because of this, for floating-point reduction operations where the same variable appears on both sides of the assignment (i.e. the reduction variable), data dependency on this variable prevents pipelining with an initiation interval of one and instead, the initiation interval equals the latency of the floating-point operation (e.g. 8 clocks for floating-point addition on Intel Stratix V). As suggested in Intel's documents [18], this dependency can be eliminated by inferring a shift register with a size equal to the latency of the floating-point operation. In this case, in every iteration data is read from the head of the shift register and written to its tail, with the shift register being shifted afterwards. The use of an array of reduction variables instead of just one such variable effectively eliminates the dependency, reducing $N_d$ to zero and consequently, reducing $II_c$ to one for the reduction loop. To obtain the final output, another reduction is needed on the content of the shift register. It is worth noting that unlike what is suggested in [18], the size of the shift register does not need to be one index bigger than the latency of the reduction operation and in our experience, even if the size is exactly equal to the latency of the operation, the dependency can be eliminated without lowering operating frequency. The transformation from an unoptimized floating-point reduction to optimized version with shift register is shown in Fig. 3-4.

```
float final_sum = 0.0f;

for (int i = 0; i < size; i++)
{
   final_sum += in[i];
}
```
<div align="center">**a)**</div>

```
#define FADD_LATENCY    8   // latency of floating-point operation

// shift register definition and initialization
float shift_reg[FADD_LATENCY] = {0.0f}, final_sum = 0.0f;

for (int i = 0; i < size; i++)
{
   // add and write to shift register
   shift_reg[FADD_LATENCY - 1] = shift_reg[0] + in[i];

   // shifting
   #pragma unroll
   for (int j = 0; j < FADD_LATENCY - 1; j++)
   {
      shift_reg[j] = shift_reg[j + 1];
   }
}

//final reduction
#pragma unroll
for (int i = 0; i < FADD_LATENCY; i++)
{
   final_sum += shift_reg[i];
}
```
<div align="center">**b)**</div>

**Figure 3-4 Shift register optimization for floating-point reduction**

This optimization is usually not enough to provide good performance on its own and it needs to be followed by data parallelism (Section 3.2.1.5). Even though the resulting optimized loop can be partially unrolled by using the unroll pragma to further improve the performance, doing so will break the shift register optimization and requires that the size of the shift register is increased further to accommodate for the unrolling. With large unroll factors, this method can result in large area overhead to implement the shift register. As a much better optimized alternative, it is possible to add data parallelism to the optimized code from 3-4 b) by performing manual unrolling as depicted in Fig. 3-5. In this case, the original loop is split into two loops, with the inner loop being fully unrolled and having a trip count equal to the unroll factor, and the exit condition of the outer loop being adjusted accordingly. In this case, the shift register optimization is not required for the inner loop since full unrolling effectively eliminates the dependency on the reduction variable, and it only needs to be applied to the outer loop. This method makes it possible to achieve efficient data parallelism alongside with the shift register optimization for floating-point reduction, without needing to increase the size of the shift register.

```c
#define FADD_LATENCY    8  // latency of floating-point operation
#define UNROLL          16 // unroll factor

// shift register definition and initialization
float shift_reg[FADD_LATENCY] = {0.0f}, final_sum = 0.0f;

// loop exit condition calculation
int exit = (size % UNROLL == 0) ? (size / UNROLL) : (size / UNROLL) + 1;
for (int i = 0; i < exit; i++)
{
   // unrolled addition
   float sum = 0.0f;
   #pragma unroll
   for (int j = 0; j < UNROLL; j++)
   {
      int index = i * UNROLL + j;
      sum += (index < size) ? in[index] : 0.0f;
   }

   // write to shift register
   shift_reg[FADD_LATENCY - 1] = shift_reg[0] + sum;

   // shifting
   #pragma unroll
   for (int j = 0; j < FADD_LATENCY - 1; j++)
   {
      shift_reg[j] = shift_reg[j + 1];
   }
}

//final reduction
#pragma unroll
for (int i = 0; i < FADD_LATENCY; i++)
{
   final_sum += shift_reg[i];
}
```

**Figure 3-5 Optimized floating-point reduction with unrolling**

As a final note, on the Intel Arria 10 FPGA, it is possible to use single-cycle floating-point accumulation and hence, the shift register optimization is not required on this FPGA. However, due to the requirements for correct inference of single-cycle accumulation by the compiler on Arria 10, it is required that data parallelism is implemented using the aforementioned method rather than applying partial unrolling directly to the reduction loop.

**3.2.2.2 Calculating constants on host instead of kernel (BOTH):**

For cases where a value is calculated on the kernel and remains constant throughout the kernel execution, calculation of this constant can be moved to the host code to save FPGA area. This optimization is specifically useful in cases where calculation of a constant involves complex mathematical functions (division, remainder, exponentiation, trigonometric functions, etc.) and could use a significant amount of FPGA area. This optimization does not directly lead to performance improvements; however, area savings from this optimization could allow more parallelism or higher $f_{max}$.

**3.2.2.3 Avoiding branches on global memory addresses and accesses (BOTH):**

For cases where in a kernel, the external memory address that is accessed could change based on run-time variables, instead of choosing the correct address using branches, it is best if both accesses are performed and results are stored in temporary variables and instead, the correct output is chosen from the temporary variables. This will prevent dynamic addressing and potentially allow the compiler to coalesce accesses when SIMD or unrolling is used. A similar problem exists when a branch involves choosing a value from two different global memory addresses. Also in this case, moving the accesses out of the branch and storing their value in two temporary variables, and instead using the temporary variables in the branch could allow correct coalescing when SIMD or loop unrolling is used. Apart from the area saving due to lower number of ports going to external memory, this optimization could also improve performance by decreasing $N_m$ and consequently, reducing $II_r$. However, for cases where SIMD or unrolling are not used and compile-time coalescing is not required, the best choice is to choose the correct address and only perform one access to global memory to minimize the number of access ports.

### 3.2.3 Advanced Compiler-assisted Optimizations

**3.2.3.1 Manual external memory banking (BOTH):**

By default, Intel FPGA SDK for OpenCL Offline Compiler interleaves all the global buffers between the two (or more) DDR memory banks available on an FPGA board so that the bandwidth of all the banks is efficiently shared between all the buffers. In cases where multiple global buffers exist with different access rates, or a few narrow accesses to global memory exist in the kernel, this automatic interleaving achieves best memory performance. However, in our experience, for cases where only two wide global memory accesses (with or without some accompanying narrow ones) exist in the kernel, each to a different global buffer, this automatic interleaving does not perform optimally and disabling it can improve performance if the buffers are each pinned to a different memory bank. To achieve this, the kernel should be compiled with a specific compiler switch to disable automatic interleaving [19], and the global buffers in the host code should be created with an additional flag that allows the user to manually determine which buffer should reside on which memory bank. In this case, performance is improved by increasing the effective *BW* from Eq. (3-5), and consequently, reducing $II_r$. Furthermore, for cases where multiple global memory types exist on the board (DDR, QDR, HBM, etc.), this technique can be used to manually allocate some of the global buffers on the non-default memory type(s).

**3.2.3.2 Disabling cache (BOTH):**

By default, Intel FPGA SDK for OpenCL Offline Compiler generates a private cache for every global memory access in a kernel if it cannot determine the exact access pattern. This cache is implemented using FPGA Block RAMs and is *not* shared between different accesses to the same global buffer. Despite its simplicity and small size (512 Kbits), this cache can be effective for designs that have good spatial locality that is not exploited by the programmer. However, in two cases this cache not only will not improve performance, but can potentially even reduce it:

- In cases where random accesses exist in the kernel with minimal spatial locality, the cache hit-rate will be very low and hence, disabling it can improve performance by avoiding the overhead of the cache mechanism. The hit-rate of the cache can be determined by using Intel FPGA Dynamic Profiler for OpenCL.
- In cases where data locality is manually exploited by the programmer by using on-chip memory, which will be the case for all well-optimized designs, the cache will not be required anymore and using it will only waste valuable Block RAM resources. In such cases, the cache can be disabled to save area.

To selectively disable the cache for a global buffer, it can be *falsely* marked as *volatile* in the OpenCL kernel. To completely disable the cache for all global buffers in a kernel, "--opt-arg -nocaching" can be added to the kernel compilation parameters. In our experience, this cache is usually not created in NDRange kernels but it is nearly always created in Single Work-item.

**3.2.3.3 *Autorun* kernels (SWI):**

Intel FPGA SDK for OpenCL provides a specific *autorun* attribute for Single Work-item kernels that do not have an interface to the host or the FPGA external memory but can communicate with other *autorun* or *non-autorun* kernels using on-chip channels [19]. This kernel type does not need to be invoked from the host and automatically launches as soon as the FPGA is programmed. Furthermore, the kernel is automatically restarted whenever it finishes execution. This type of kernel has two main use cases:

- For designs in which data is sent and received directly via the FPGA on-board peripherals, and no interaction from the host is required, this kernel type can be used so that the FPGA can act as a standalone processor. This type of design is specifically useful for network-based processing where data is streamed in and out through the FPGA on-board network ports.
- For streaming designs that require replication of a Single Work-item kernel, this attribute can be used alongside with the multi-dimensional version of the *num_compute_units()* attribute (different from the single-dimensional one used for NDRange kernels). In this case, a *get_compute_id()* function is supplied by the compiler that can be used to obtain the unique ID of each kernel copy at compile-time and then, each kernel copy can be customized using this ID. This attribute is specifically useful for streaming designs in form of multi-dimensional systolic array or single-dimensional ring architectures. Apart from the obvious area reduction duo to lack of interface to host and memory for this kernel type, in our experience, using this kernel type also results in efficient *floor-planning* and good scaling of operating frequency even with tens of kernel copies.

### 3.2.3.4 Flat compilation (BOTH):

Using Intel FPGA SDK for OpenCL, the FPGA is automatically reprogrammed at run-time with the pre-compiled FPGA bitstream before kernel execution. On the Intel Stratix V device, this reconfiguration is performed using Configuration via Protocol (CvP) [20]. However, CvP update is not supported on the Intel Arria 10 device [21] and hence, the FPGA is reprogrammed at run-time using Partial Reconfiguration (PR) via PCI-E. In this case, the logic related to the OpenCL BSP is the *static* part of the design which resides on a fixed section of the FPGA and is never reconfigured. The rest of the FPGA area can be used by the OpenCL kernel, which acts as the *dynamic* part of the design and is reconfigured at run-time. Using PR for OpenCL on Arria 10 imposes extra placement and routing constraints on the design to ensure correct run-time reconfiguration; this comes at the cost of more placement and routing complications on this device and consequently, worse timing or even an outright unfittable or unrouteable design, especially when FPGA logic or Block RAM utilization is high. Furthermore, in our experience, run-time partial reconfiguration through PCI-E has a high failure rate, resulting in the program or even the OS crashing in many cases. To avoid these issues, the possibility of *flat* compilation on Arria 10 is also provided by the compiler, which disables PR and place and routes the BSP and the OpenCL kernel as one flat design. This eliminates all extra constraints for PR and allows the programmer to use the FPGA area more efficiently and achieve best timing. This compiler optimization can improve performance by both providing better area utilization efficiency for large designs that would have failed to fit or route with the PR flow, and also improving $f_{max}$. However, using flat compilation comes at the cost of two shortcomings. One is the longer run-time reconfiguration time since the FPGA has to be reconfigured via JTAG instead of PCI-E (15-20 seconds vs. less than 5 seconds). The other is that since all clock constraints except for the kernel clock in the BSP are relaxed for flat compilation, other clocks like PCI-E and DDR might fail to meet timing and hence, the user has to try multiple seeds and manually check the timing report to make sure all timing constraints are met. In practice, we have seen that for large NDRange designs, it might not be possible to meet the timing constraints of the non-constrained clocks regardless of how many different seeds are tried. Hence, this optimization should be limited to highly-optimized Single Work-item designs and it is probably best to use the default PR flow for NDRange.

### 3.2.3.5 Target $f_{max}$ and seed sweep (BOTH):

By default, Intel FPGA SDK for OpenCL Offline Compiler balances pipeline stages in the design by inserting extra registers and FIFOs towards a target $f_{max}$ of 240 MHz. This target can be increased so that the compiler would insert more registers and FIFOs in the pipeline, potentially allowing a higher operating frequency, at the cost of higher logic and Block RAM utilization. This optimization is not viable for cases where logic and Block RAM utilization is already high, since the extra area usage could instead lead to more routing congestion and worse $f_{max}$. Furthermore, careful attention is required when changing the target $f_{max}$ for Single Work-item kernels since if the target is set too high, the compiler might have to increase the initiation interval of some loops to achieve the target and this could instead result in performance slow-down despite higher operating frequency. The compiler also provides the possibility to change the placement and routing seed, which can result in better (or worse) $f_{max}$, at the cost of no extra area utilization. These two optimizations can be used as the *last step* of optimization to maximize $f_{max}$ for a given design by compiling multiple instances of the design, each with a different seed and $f_{max}$ target, and choosing the one with highest operating frequency.

### 3.2.4 Advanced Manual Optimizations

**3.2.4.1 Shift register as local data storage (SWI):**

Each FPGA has a plethora of registers available that can be used for temporary data storage. The main advantage of registers over Block RAMs for data storage is that access latency to registers is one clock cycle, allowing efficient data sharing between loop iterations without increasing the loop initiation interval. However, since each ALM has a few registers, and chaining registers requires using FPGA routing resources, implementing large on-chip buffers using registers is not efficient. On the other hand, Block RAMs are suitable for implementing large on-chip buffers, but latency of dynamic access to Block RAMs is not one clock cycle and hence, reading from and writing to on-chip buffers implemented using Block RAMs in a loop can result in read-after-write dependencies and high initiation intervals. With all this, if certain requirements are met, large buffers can be implemented using Block RAMs with an access latency of one clock cycle:

- All accesses to the buffer use static addresses that are known at compile-time
- The content of the buffer is *shifted* once per loop iteration by a fixed amount

This will result in inference of an FPGA-specific on-chip storage called a *shift register* (also called *sliding window* or *line buffer* in literature). Shift registers are suitable for applications that involve the point of computation being shifted over a regular grid, including but not limited to stencil computation, image filtering, and sequence alignment. Due to the above requirements, this on-chip storage type can only be described in Single Work-item kernels.

Fig. 3-6 shows how a Block RAM is used to implement a shift register for 2D stencil computation. In this case, after an initial warm-up period to fill the shift register, all data points involved in the computation of the stencil reside in the shift register buffer at any given clock cycle. By incrementing the starting address of the buffer in the Block RAM, the center of the stencil is effectively shifted forward in the grid, while the relative distance of all neighbors from the starting address remains the same, allowing static addressing in a loop. Since static addressing does not require address decoding, and shifting the buffer forward only involves incrementing the starting address, accesses to shift registers can be done in one clock cycle. Shift registers are one of the most important architectural advantages of FPGAs compared to other hardware.

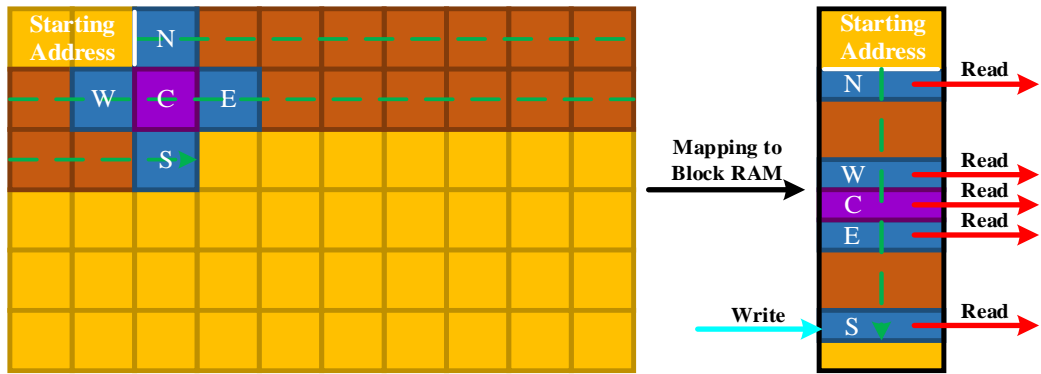

**Figure 3-6 Shift register inference**

The performance improvement of shift registers is two-fold: compared to using standard Block RAM-based buffers with dynamic access, using shift registers allows the programmer to avoid read-after-write dependencies, reducing $N_d$ to zero and $II_c$ to one. Furthermore, as shown in Fig. 3-6, a shift register can act as an efficient cache for neighboring cells in a grid, reducing redundant off-chip memory accesses ($N_m$) and improving $II_r$.

**3.2.4.2 Reducing local memory accesses (BOTH):**

Local memory-based optimizations are used on every hardware to improve performance by reducing accesses to the slower external memory and instead, storing and accessing frequently-used data in the faster local memory. Due to limited amount of local memory available on each given device, careful attention is required to ensure only widely-used buffers reside in this memory type, and are replaced as soon as they are not required anymore. On CPUs, the cache hierarchy acts as local memory and is mostly hardware-managed; however, programming techniques such as loop tiling can help improve cache hit rate and performance. On modern GPUs, a limited set of registers, some scratchpad memory, and two levels of cache are available. The user-managed resources (registers and scratchpad) make up the smaller chunk of the local memory resources, while the larger chunk consisted of caches is again hardware-managed. On FPGAs, things are different: all local memory resources on FPGAs are user-managed and no explicit cache hierarchy exists. This makes careful utilization and configuration of local memory resources on these devices even more crucial.

Apart from the size of local buffers, the number of accesses to such buffers also plays a crucial role in local memory usage on FPGAs. For small buffers implemented using registers, multiple read and write ports can be connected to the same buffer instance with small area overhead; however, as the number of accesses goes up, the *fan-in* and *fan-out* of the buffer also increase, resulting in routing compilations and lowered $f_{max}$, or an outright unrouteable design. For larger buffers implemented using Block RAMs, whether in form of shift registers or multi-ported RAM/ROMs, since each Block RAM only has two physical ports, in most cases the buffer needs to be physically replicated on the FPGA to provide the necessary number of access ports for parallel non-stallable accesses to the buffer. For cases with an inefficient access pattern to local memory, this replication can easily result in exhaustion of Block RAM resources on the FPGA. If this happens, the compiler will then restart the compilation and instead shares Block RAM ports between multiple accesses so that replication factor, and consequently, Block RAM utilization, is reduced and the design can fit. Needless to say, port sharing requires arbitration and results in stallable accesses, reducing performance by increasing $II_r$ in pretty much every case.

One compiler-assisted solution to reduce the Block RAM replication factor is to double-pump the Block RAMs. In this case, the number of ports to each Block RAM is effectively doubled since the Block RAM is driven with 2x the clock of the OpenCL kernel. However, correct communication between the OpenCL kernel and the Block RAMs in this case will require extra logic and since there is a physical limit for the operating frequency of Block RAMs (500-600 MHz on current devices), the maximum operating frequency of the OpenCL kernel will be limited to half of that limit (250-300 MHz). The compiler generally employs this technique by default; in fact, for cases where two or more writes to a specific local buffer exist in a kernel, since write ports need to be connected to all replicated instances of the buffer, there is no choice other than double-pumping the Block RAMs to achieve non-stallable accesses. However, the compiler also provides certain attributes for manual double-pumping and port sharing so that the programmer can also influence the compiler's decisions in this case.

A more important method to reduce Block RAM replication factor is to reduce number of accesses to the buffer. This can be done by using temporary registers, or even transposing the local buffer or re-ordering nested loops to allow efficient access coalescing in presence of loop unrolling or SIMD. This optimization, apart from the obvious area reduction, can also allow reducing or completely removing stallable accesses to local memory buffers and improving performance by reducing $II_r$.

Fig. 3-7 a) shows a code snippet in which a local buffer implemented using Block RAMs is first initialized, and then a reduction operation is performed, with the output being stored in the local buffer. As is evident from the code snippet, one read port and two write ports are required to the *temp* buffer in this part of the kernel. These accesses alone will force the compiler into double-pumping the Block RAMs used to implement this buffer. Furthermore, the buffer will need to be replicated one extra time for every two reads from the buffer (two ports of each double-pumped Block RAM are connected to the write accesses and the remaining two ports can be used for reads). However, by moving the local buffer outside of the reduction loop and replacing it with a temporary register as is shown in Fig. 3-7 b), one read and one write are removed, eliminating the need for buffer replication if up to three reads from the buffer exist in the rest of the kernel, and avoiding the need for double-pumping in case of only one read from the buffer in the rest of the kernel.

```
__global float* a, b;                    __global float* a, b;
__local float temp[M];                   __local float temp[M];

for(int i = 0; i < M; i++)               for(int i = 0; i < M; i++)
{                                        {
   temp[i] = 0;                             float reg = 0;
}
                                            for(int j = 0; j < N; j++)
for(int i = 0; i < M; i++)                  {
{                                              reg += a[i] * b[j];
   for(int j = 0; j < N; j++)               }
   {                                        temp[i] = reg;
      temp[i] += a[i] * b[j];            }
   }
}
                    a)                                       b)
```

Figure 3-7 Reducing Block RAM replication by using temporary registers

Another case of reducing Block RAM replication is depicted in Fig. 3-8. In code snippet a), since the inner loop is unrolled on the higher dimension of the local buffer, the accesses to that buffer cannot be coalesced and hence, eight write ports are required to the buffer which results in port sharing. However, by transposing the buffer, i.e. swapping its dimensions as shown in b), this issue can be avoided. In the latter case, only one large coalesced write to the buffer will be required and hence, instead of replicating the buffer, the large write will be made possible by interleaving the buffer across multiple Block RAMs. In this case, for buffers that are small enough to fit in less than eight Block RAMs, eight Block RAMs will still be required to implement the buffer so that enough ports are available; however, for larger buffers, the overhead of interleaving will be minimal. It is worth noting that re-ordering the *i* and *j* loops will also result in correct access coalescing to the local *temp* buffer; however, it will break access coalescing to the global buffer *a*, significantly reducing external memory performance.

```
__global float* a;                       __global float* a;
__local float temp[N][M];                __local float temp[M][N];

for(int i = 0; i < M; i++)               for(int i = 0; i < M; i++)
{                                        {
   #pragma unroll 8                         #pragma unroll 8
   for(int j = 0; j < N; j++)               for(int j = 0; j < N; j++)
   {                                        {
      temp[j][i] = a[i * COL + j];            temp[i][j] = a[i * COL + j];
   }                                        }
}                                        }
                    a)                                       b)
```

Figure 3-8 Reducing Block RAM replication by transposing the buffer

### 3.2.4.3 Loop collapse (SWI):

Multiply-nested loops are a recurring pattern in HPC applications. Having such loops incurs extra overhead on FPGAs since, relative to the depths of the loop nest, more registers and Block RAMs will be needed to store the state of the different variables in the loop nest. As an FPGA-specific optimization, loop nests can be *collapsed* into a single loop to avoid this extra area overhead. Even though the same optimization is also regularly performed on OpenMP code running on CPUs, the goal of the optimization on CPUs and FPGAs is completely different. This conversion is shown in Fig. 3-9.

```
for (int y = 0; y < M; y++)
{
    for (int x = 0; x < N; x++)
    {
        compute(x,y);
    }
}
```

```
int x = 0, y = 0;
while (y != M)
{
    compute(x,y);

    x++;
    if (x == N)
    {
        x = 0;
        y++;
    }
}
```

a)                                        b)

**Figure 3-9 Loop collapse optimization**

This optimization does not have a direct effect on performance; however, the area reduction resulting from this optimization can open up FPGA resources, providing room for extra parallelism or larger block size and consequently, higher performance. Furthermore, this optimization simplifies the pipeline and can result in lower pipeline latency (*P*). In the recent versions of Intel FPGA SDK for OpenCL, a new *loop_coalesce* pragma has been introduced that allows this optimization to be performed directly by the compiler. However, using this pragma is subject to certain limitations and does not work for all loop nests. Furthermore, the "exit condition optimization" which will be explained next is only possible after manual loop collapsing.

### 3.2.4.4 Exit condition optimization (SWI):

Apart from the extra area overhead of multiply-nested loops on FPGAs, having such loops also has another disadvantage: since we want all loops in a loop nest to have an initiation interval of one to achieve maximum performance, the exit condition of all the loops in the loop nest need to be evaluated in one clock cycle. Since the exit conditions depend on each other, a long chain of comparisons and updates on the loop variables is created that will adversely affect the design critical path and reduce $f_{max}$. In fact, in our experience, **the design critical path of Single Work-item kernels that do not have loop-carried dependencies is nearly always located in the chain of functions required to determine the exit condition of the deepest loop nest in the kernel.** Even though the "loop collapse" optimization introduced earlier reduces the loop nest to one loop, it does not change the loop exit condition. To improve this critical path that gets longer with the depth of the original loop nest, we can replace the exit condition of the collapsed loop with incrementation and comparison on a global index variable that does not depend on the loop variables of the original nested loop. Fig. 3-10 shows the resulting code after applying this optimization to the collapsed loop from Fig. 3-9 b).

```
int x = 0, y = 0, index = 0;
while (index != M * N)
{
   index++;
   compute(x,y);

   x++;
   if (x == N)
   {
      x = 0;
      y++;
   }
}
```

**Figure 3-10 Exit condition optimization**

    For the sake of clarity, we used a basic example in figures 3-9 and 3-10 to show how the loop collapse and exit condition optimizations are applied. However, in real-world scenarios, these optimizations will be applied to much more complex loop nests involving multiple layers of index and block variables. In such cases, the global index variable should be compared with the number of times the original loop nest would have iterated in total. Calculating this number will likely require a complex equation involving the iteration variable and exit condition of all the original loops in the loop nest. This operation can be done in the host code, and the exit condition for the collapsed loop can be passed to the kernel as argument to avoid extra area waste on the FPGA for mathematical computations that will only be performed once per kernel run. **It is worth noting that for deep loop nests, even after applying the exit condition optimization, the design critical path will still consist of the chain of updates and comparisons on the remaining index variables.**

# 4 Evaluating FPGAs for HPC Applications Using OpenCL

In this chapter, we will port a subset of the Rodinia benchmark suite [13] as a representative of typical HPC applications using the optimization techniques introduced in the previous chapter, and report speed-up compared to baseline, and performance and power efficiency compared to CPUs and GPUs. The contents of this chapter have been partially published in [17] and [22].

## 4.1 Background

Multiple benchmark suites have been proposed as representatives of HPC applications to evaluate different hardware and compilers. Examples of such suites include Rodinia [13], SHOC [23], OpenDwarfs [24], Parboil [25], PolyBench [26] and many more. Among these suites, Rodinia is regularly used for evaluating performance on different hardware since it includes OpenMP, CUDA and OpenCL versions of multiple benchmarks that can be used to target a wide variety of hardware. Each of the benchmarks in this suite belongs to one of Berkeley's Dwarfs [27]. Each Berkeley Dwarf represents one class of HPC applications that share a similar compute and memory access pattern. We choose Rodinia so that we can take advantage of the existing OpenMP and CUDA implementations for evaluating CPUs and NVIDIA GPUs. Furthermore, we port and optimize the OpenCL versions for FPGAs based on Intel FPGA SDK for OpenCL to be able to perform a meaningful performance comparison between different hardware architectures and show the strengths and weaknesses of them. We evaluate one or two benchmarks from multiple of the Dwarfs, expecting that our FPGA optimization techniques for each application belonging to a Dwarf can be used as guidelines for optimizing other applications belonging to the same Dwarf.

## 4.2 Methodology

### 4.2.1 Benchmarks

We use the latest v3.1 of the Rodinia benchmark suite to make sure all the latest updates are applied to the CPU and GPU benchmarks. The benchmarks we evaluate in this study are as follows:

**NW:** Needleman-Wunsch (NW) is a Dynamic Programming benchmark that represents a sequence alignment algorithm. Two input strings are organized as the top-most row and left-most column of a 2D matrix. Computation starts from the top-left cell and continues row-wise, computing a score for each cell based on its neighbor scores at the top, left, and top-left positions and a reference value, until the bottom-right cell is reached. This computation pattern results in multiple data dependencies. This benchmark only uses integer values.

**Hotspot:** Hotspot is a Structured Grid benchmark that simulates microprocessor temperature based on a first-order 5-point 2D stencil on a 2D grid and uses single-precision floating-point values. Apart from the center cell and its four immediate neighbors from the *temperature* input, the computation also involves the center cell from a second *power* input. The computation continues iteratively, swapping the input and output buffers after each iteration, until the supplied number of iterations have been processed.

**Hotspot 3D:** Hotspot is also a Structured Grid benchmark and implements the 3D version of Hotspot using a first-order 3D 7-point stencil on a 3D input grid. Similar to Hotspot, this benchmark also uses the center cell from the *power* input, alongside with the center cell and its six immediate neighbors from the *temperature* input in its computation.

**Pathfinder:** Pathfinder is a Dynamic Programming benchmark that attempts to find a path with smallest accumulated weight in a 2D grid. Computation starts from the top row and continues row by row to the bottom, finding the minimum value among the top-right, top, and top-left neighbors and accumulating this value with the current cell. Similar to NW, this access pattern results in data dependencies but in different directions. This benchmark also only uses integer values.

**SRAD:** SRAD is a Structured Grid benchmark used for performing speckle reducing on medical images. Similar to Hotspot, its computation involves stencil computation on a 2D input with single-precision floating-point values. However, SRAD has two stencil passes, is much more compute-intensive, and includes an initial reduction on all of the grid cells.

**LUD:** LU Decomposition (LUD) is a Dense Linear Algebra benchmark that decomposes an arbitrary-sized square matrix to the product of a lower-triangular and an upper-triangular matrix. This benchmark is compute-intensive with multiple instances of single-precision floating-point multiplication, addition and reduction.

### 4.2.2 Optimization Levels

For each benchmark on the FPGA platform, we create a set of NDRange and Single Work-item kernels. For each set, we define three optimization levels:

**None:** The lowest optimization level, i.e. *none*, involves using the original NDRange kernels from Rodinia directly, or a direct Single Work-item port based on either the NDRange OpenCL kernel or the OpenMP implementation of the benchmark. The original NDRange kernel from Rodinia will be used as our FPGA baseline to determine speed-up from our optimizations. To avoid unreasonably slow baselines, we employ the crucial *restrict* (3.2.1.1) and *ivdep* (3.2.1.2) attributes to avoid false dependencies and false loop serialization. This level of optimization shows how much performance can be expected when we only rely on the compiler for optimization.

**Basic:** For the *basic* optimization level, we only apply basic manual and compiler-assisted optimizations (Sections 3.2.1 and 3.2.2) to the unoptimized kernels of each set. This optimization level acts as a representative of the level of performance that can be achieved using a modest amount of effort and by relying only on optimization techniques defined in Intel's documents for programmers with little knowledge of FPGA hardware.

**Advanced:** The *advanced* optimization level involves significant code rewrite in most cases alongside with taking full advantage of the advanced manual and compiler-assisted optimizations (Sections 3.2.3 and 3.2.4). This level of optimization shows how much performance can be expected with a large amount of programming effort and moderate knowledge of the underlying FPGA characteristics.

For all optimization levels, all parameters (block size, SIMD size, unroll factor, benchmark-specific input settings, etc.) are tuned to maximize performance and the best case is chosen. It is worth noting that we avoid the --fpc and --fp-relaxed compiler switches which can reduce area usage of floating-point computations at the cost of breaking compliance with the IEEE-754 standard due to introduction of inaccuracies and rounding errors in the computation.

### 4.2.3 Hardware and Software

We evaluate our benchmarks on two FPGA boards. One contains a Stratix V GX A7 device and the other an Arria 10 GX 1150 device. The newer Arria 10 device has roughly twice the logic, 6% more Block RAMs, and nearly six times more DSPs compared to the Stratix V FPGA. Furthermore, the DSPs in the Arria 10 FPGA have native support for single-precision floating-point operations, giving this FPGA an edge over Stratix V for floating-point computation. Table 4-1 shows the device characteristics of these two FPGAs.

**Table 4-1 FPGA Device Characteristics**

| Board | FPGA | ALM | Register (K) | M20K (Blocks\|Mb) | DSP | External Memory |
|---|---|---|---|---|---|---|
| Terasic DE5-Net | Stratix V GX A7 | 234,720 | 939 | 2,560\|50 | 256 | 2x DDR3-1600 |
| Nallatech 385A | Arria 10 GX 1150 | 427,200 | 1,709 | 2,713\|53 | 1,518 | 2x DDR4-2133 |

To keep the comparison fair, we will compare each FPGA device with a CPU and GPU of its age. Table 4-2 shows a list of the hardware used in our evaluation, and a summary of their characteristics. The peak compute performance numbers reported in this table are for single-precision floating-point computation.

To compile our OpenMP kernels on CPUs, we use GCC v6.3.0 with -O3 flag and Intel C++ Compiler v2018.2 with "-fp-model precise -O3" flags and choose the best run time between the two compilers. The extra flag for ICC is used to disable optimizations that might change the accuracy of floating-point computations. All hyperthreads are used on every CPU in this case. For GPUs we use NVIDIA CUDA v9.1 with "-arch sm_35 -O3" flags. Intel FPGA SDK for OpenCL v16.1.2 is also used for the FPGAs. Due to a bug in this version of Quartus that resulted in routing errors with some of our kernels on Arria 10, we disable "Parallel Synthesis" in the BSP of our Arria 10 board. We use CentOS 6 on our FPGA machines and CentOS 7 on the CPU/GPU machines.

**Table 4-2 Evaluated Hardware and Their Characteristics**

| Type | Device | Peak Memory Bandwidth (GB/s) | Peak Compute Performance (GFLOP/s) | Production Node (nm) | TDP (Watt) | Release Year |
|---|---|---|---|---|---|---|
| FPGA | Stratix V | 25.6 | ~200 | 28 | 40 | 2011 |
|  | Arria 10 | 34.1 | 1,450 | 20 | 70 | 2014 |
| CPU | i7-3930K | 42.7 | 300 | 32 | 130 | 2011 |
|  | E5-2650 v3 | 68.3 | 640 | 22 | 105 | 2014 |
| GPU | Tesla K20X | 249.6 | 3,935 | 28 | 235 | 2012 |
|  | GTX 980 Ti[1] | 340.6 | 6,900 | 28 | 275 | 2015 |

### 4.2.4 Timing and Power Measurement

In this study, we only time the kernel execution and disregard initialization and all data transfers between host and device. Even though this puts the CPUs at a disadvantage, doing so allows us to fairly compare the computational performance of the devices without hampering their performance by the link between the host and the device that is independent of the devices. Furthermore, we expect data transfer between host and device for most HPC applications to be small relative to the total run time or else, there would be little reason to accelerate them using a PCI-E-attached accelerator like an FPGA or a GPU. To maximize our timing accuracy, we use the high-precision clock_gettime() function supported by most major Linux distributions with the CLOCK_MONOTONIC_RAW setting. Furthermore, we make sure our input sizes are big enough so that kernel run time is at least a few seconds in every case to further increase the dependability of our timing and power measurement results. However, in a few cases, even with the largest setting that fit in the 4 GB external memory of the Stratix V board (smallest external memory size among all evaluated devices), run time of the fastest cases went below 1 second, for a minimum of a couple milliseconds for Pathfinder on the GPUs.

Our Stratix V board does not have an on-board power sensor; hence, to estimate the power usage of the board, we run quartus_pow on the placed-and-routed OpenCL kernel, and add 2.34 Watts to the resulting number to account for the power consumption of the two memory modules. We assume that each memory module uses a maximum of 1.17 Watts based on the datasheet of a similar memory model [28]. The Arria 10 board, however, includes a power sensor and the manufacturer provides an API to read the power sensor in C programs. We use the values reported by sensor to measure power consumption on this platform. Similarly, we read the power sensor available on the GPU boards and the CPUs using existing APIs, namely NVIDIA NVML [29] and the Linux MSR driver [30]. For the GPUs and the Arria 10 FPGA, the on-board power sensor is queried once every 10 milliseconds during kernel execution and the reported wattage values are averaged. In two cases (NW and Pathfinder), since the benchmark run times were not long enough to get accurate power measurement on the GPUs, the main computation loop of these benchmarks was wrapped in an extra loop to artificially

---
[1] The GTX 980 Ti GPU used our evaluation is a non-reference model that is shipped with higher core and memory clock compared to the reference model

extend the benchmark run time and allow correct measurement of average power usage. In these cases, run time was measured from the first iteration of the extra loop. Power efficiency in these cases is determined by calculating energy to solution as average power consumption multiplied by the kernel run time. For the CPUs, since the associated MSR register reports energy values, we directly measure energy to solution by subtracting the accumulated energy usage at the beginning of kernel execution from the one at the end. It should be noted that unlike the case for the FPGAs and the GPUs where the power measurement includes the board power, the CPU measurements only include the chip itself and do not include the power consumption of the host memory. This slightly favors the CPUs in power efficiency comparison.

Except a few cases among versions with no or basic optimization on Stratix V where run time was over half an hour, all benchmark configurations on every hardware were repeated five times, and timing and power measurements were averaged.

## 4.3 Results

### 4.3.1 Stratix V

We discuss optimization details and performance improvement with different optimization levels only on the Stratix V FPGA. For Arria 10, only the result of the fastest version of each benchmark is measured, which will be reported in the next section. The source code for all the benchmarks reported in this section is available at https://github.com/zohourih/rodinia_fpga.

**4.3.1.1 NW**

The *original* NDRange kernel from Rodinia implements 2D blocking and takes advantage of diagonal parallelism for this benchmark. For this version we use a block size of 128×128. For the *unoptimized* Single Work-item kernel, we use a straightforward implementation with a doubly-nested loop based on the OpenMP version of the benchmark. Due to load/store dependency caused by the left neighbor being calculated in the previous loop iteration, the compiler fails to pipeline the outer loop, and the inner loop is pipelined with an initiation interval of 328, which is equal to the minimum latency of an external memory write followed by a read.

For the *basic* NDRange version, we set the work-group size (3.2.1.4) and add SIMD and unrolling (3.2.1.5). Setting the work-group size allows the compiler to share the same compute unit between different work-groups to minimize pipeline stalls (*work-group pipelining*), while only one work-group is allowed to occupy a compute unit in the *unoptimized* version due to unknown work-group size. However, since the local buffers need to be further replicated to allow parallel access by the extra work-groups, we are forced to reduce block size to 64×64 in this version. Furthermore, due to the very large number of accesses to the local buffer and numerous barriers in the design, parameter tuning is limited to a SIMD size of two and no unrolling. For the Singe Work-item version of this optimization level, we use one extra register to manually cache the left neighbor and then use this register in the innermost loop (iterating over columns) instead of the external memory access. This allows us to remove dependency to the left neighbor. Since the compiler still detects a false dependency on the external memory buffer, we also add *ivdep* (3.2.1.2) to allow correct pipelining of the inner loop with an initiation interval of one. The outer loop iterating over rows still runs sequentially here due to dependency to top and top-left neighbors updated in the previous row. This dependency is unavoidable in this design. Unrolling cannot be used for the innermost loop since it results in new load/store dependencies.

The characteristics of this benchmark make it clear that a Single Work-item design is more suitable since we could already avoid one of the dependencies in the algorithm by employing an additional register in the Single Work-item version with *basic* optimization. Hence, we choose the Single Work-item model to create the kernel with *advanced* optimization level. In [17], we presented an optimized design for NW that used 1D blocking and took advantage of the shifting pattern of the computation alongside with one block row of extra registers to completely avoid external memory accesses other than to read the initial values on the grid and block boundaries. In this implementation, due to the dependency to the left cell which is computed in the previous iteration, we had to fully unroll the computation over a block row or else the loop-carried dependency prevented pipelining with an initiation interval of one. Because of this, the unroll factor and block size had to be the same, preventing us from using large block sizes to minimize redundant memory accesses on the block boundaries. To avoid this problem, we use a different design here. Fig. 4-1 shows our implementation for this benchmark. For this implementation, instead of computing the cells row-by-row which forces us to unroll the computation over the direction of the loop-carried dependency, we take advantage of *diagonal* parallelism in the algorithm. Our new design uses 1D blocking in the *y* dimension with a block height of *bsize* and a parallelism degree, i.e. number of cells computed per iteration, of *par*. Computation starts from top-left and moves downwards, computing one chunk of columns at a time with a chunk width of *par*. The chunk of columns is processed in a diagonal fashion, with the first diagonal starting from an out-of-bound point and ending on the top-left cell in the grid (yellow color). Then, computation moves downwards, calculating one diagonal with *par* cells (shown with the same color in

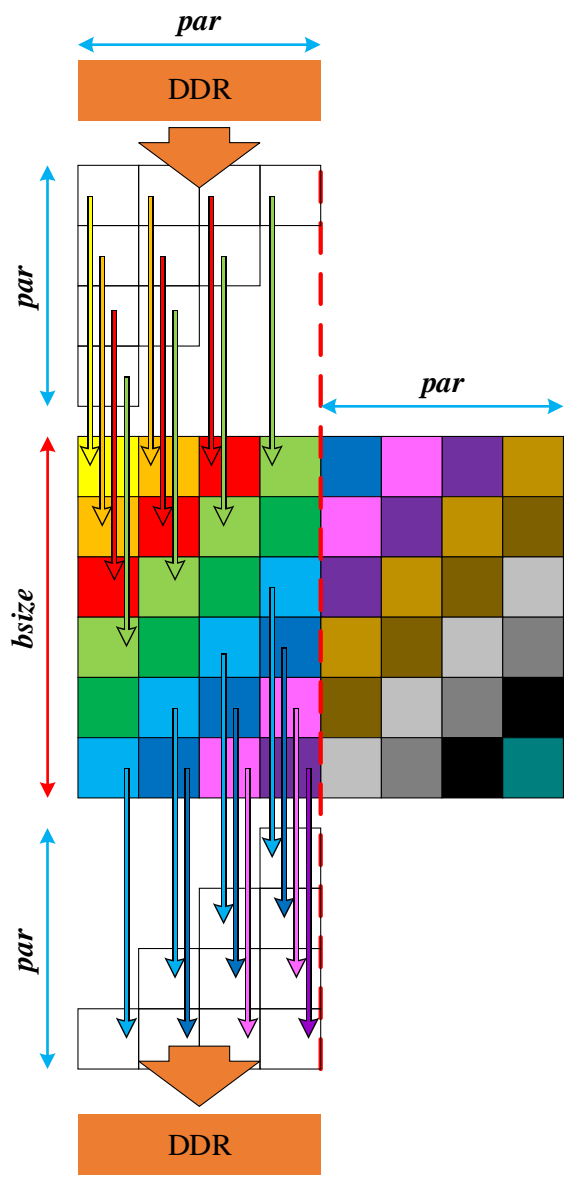

**Figure 4-1 NW implementation**

Fig. 4-1) per loop iteration until the bottom-cell in the diagonal falls on the bottom-left cell in the block (light blue color). For the next diagonal (dark blue color), the cells that would fall inside the next block instead wrap around and compute cells from the next chunk of columns in the current block. When the first cell in the diagonal falls on the block boundary (violet color), the computation of the first chunk of columns is finished and every cell computed after that will be from the second chunk of columns. When all the columns in a block are computed, computation moves to the next block and repeats in the same fashion. To correctly handle the dependencies, each newly-computed cell is buffered on-chip using shift registers (3.2.4.1) for two iterations to resolve the dependency to the top cell in the next diagonal and top-left cell in the diagonal after that. Furthermore, the cells on the right-most column in the current chunk of columns are buffered in a large shift register with the size of *bsize* so that they can be reused in the next chunk of columns to resolve the dependency to the left neighbor. Finally, the blocks are also overlapped by one row to provide the possibility to re-read the cells on the boundary computed in the previous block and handle the top and top-left dependencies in the first row in the new block. Even though this design allows us to separate block size from degree of parallelism, it breaks external memory access coalescing since accesses are diagonal instead of row-wise, resulting in very poor memory performance. To address this issue, we manually insert a set of shift registers between the memory accesses (both read and write) and computation to delay memory accesses and convert diagonal accesses to consecutive coalesceable ones. These shift registers are shown as white cells in Fig. 4-1. For reading, the shift register for the first column in the chunk has a size of *par* and as we move towards the last column in the chunk, the shift registers get smaller by one cell until the last column where the shift register will turn into a single register. For writes, the set of shift registers instead starts from a single register and ends with a shift register of size *par*. In this case, since writes start *par* iterations after reads, the input width is padded by *par* cells to allow the cells in the rightmost chunk of columns to be written back to external memory correctly. Finally, to improve alignment of external memory accesses, we pass the first column of the input that represents one of the strings and is read-only to the kernel using a separate global buffer so that reads from and writes to the main buffer start from the address 0 instead of 1. For this version, we disable the compiler's cache (3.2.3.2), use loop collapse (3.2.4.3) followed by exit condition optimization (3.2.4.4), and perform seed and target $f_{max}$ sweep (3.2.3.5). Our final implementation has three global buffers, one of which is only accessed once per row (read-only buffer for first column) while the other two (main input and output) are accessed every iteration using large vector accesses. Hence, we also perform manual memory banking (3.2.3.1) and put each of the two frequently-accessed buffers in a different bank to maximize external memory performance. A *bsize* of 4096 and *par* size of 64 are used for this version to maximize its performance.

Table 4-3 shows the performance and area utilization of our kernels on the Stratix V FPGA. For this benchmark, we use an input size of 23040× 23040. As expected, the *original* NDRange kernel performs poorly despite its large block size, due to lack of enough parallelism and large amount of pipeline flushes caused by the plethora of barriers in the kernel. The *unoptimized* Single Work-item kernel performs much worse due to high initiation interval caused by the load/store dependency resulting from the same global buffer being used as both input and output. With *basic* optimizations, the performance of the NDRange kernel slightly improves but still the performance is far from competitive. Furthermore, its operating frequency suffers greatly due to inefficient use of local memory buffers and full utilization of FPGA Block RAMs. On the other hand, the very simple optimization used for the *basic* Single Work-item kernel, even without any explicit parallelism in the kernel (no unrolling), manages to outperform the NDRange kernel and achieve an acceptable level of performance compared to the optimization effort. Finally, the *advanced* kernel manages to achieve nearly 40 times higher performance over the *original* NDRange kernel and brings run time down to a level that can compete with other devices. Performance of this benchmark is now limited by the external memory bandwidth to the point that performance difference between a *par* size of 32 and 64 is less than 5%.

Table 4-3 Performance and Area Utilization of NW on Stratix V

| Optimization Level | Kernel Type | Time (s) | Power (W) | Energy (J) | $f_{max}$ (MHz) | Logic | M20K Bits | M20K Blocks | DSP | Speed-up |
|---|---|---|---|---|---|---|---|---|---|---|
| None | NDR | 9.937 | 16.031 | 159.300 | 267.52 | 27% | 16% | 30% | 6% | 1.00 |
| | SWI | 203.864 | 12.998 | 2649.824 | 304.50 | 20% | 5% | 17% | <1% | 0.05 |
| Basic | NDR | 3.999 | 16.643 | 66.555 | 164.20 | 38% | 68% | 100% | 8% | 2.48 |
| | SWI | 2.803 | 12.137 | 34.020 | 191.97 | 19% | 8% | 18% | <1% | 3.55 |
| Advanced | SWI | 0.260 | 19.308 | 5.020 | 218.15 | 53% | 7% | 28% | 2% | 38.22 |

One point to note here is the relatively low operating frequency of the *basic* and *advanced* Single Work-item kernels compared to the rest of our evaluated benchmarks. Even though we can remove the dependency to the computation of the left neighbor using extra registers or shift registers in these kernels, this optimization requires one write to and one read from registers or shift registers that are updated in the previous cycle, resulting in very tight timing requirements. Because of this, the critical path of design is limited by this read-after-write dependency rather than the loop exit condition, rendering the exit condition optimization performed in the *advanced* version ineffective.

#### 4.3.1.2 Hotspot

The *original* NDRange implementation of Hotspot in Rodinia performs 2D spatial blocking by first moving data from external memory to internal memory and computing all the blocked data before writing them back. This implementation also performs temporal blocking; i.e., it computes each cells for multiple consecutive iterations (time-steps) before writing the final output back to external memory. Without setting the work-group size manually, the compiler assumes a work-group size of 256, which limits the size of the 2D block to 16×16 cells for this version. The *pyramid_height* parameter, which controls the degree of temporal parallelism (i.e. number of fused iterations), is set to one in this case since performance does not scale with higher values due to small block size. We create the *unoptimized* Single Work-item kernel based on the OpenMP implementation of the same benchmark in form of a doubly-nested loop.

For the NDRange kernel with *basic* optimization, we manually set the work-group size (3.2.1.4), add SIMD attribute (3.2.1.5) and move calculation of constants outside of the kernel (3.2.2.2). The compiler's failure in coalescing the accesses to the local buffer prevented area scaling beyond a block size of 64×64 in this case. We also used a SIMD size of 16 and *pyramid_height* of 4 to maximize the performance of this kernel. For the Single Work-item kernel of the same optimization level, we move calculation of constants outside of the kernel (3.2.2.2), remove branches on global memory address calculation (3.2.2.3), and unroll the innermost loop (3.2.1.5). The compiler is successful in achieving an initiation interval of one for the loop nest; however, its performance does not scale beyond an unroll factor of two since the compiler fails to coalesce the accesses to global memory and instantiates multiple ports which compete with each other for the limited external memory bandwidth.

For this particular benchmark, we create both an NDRange and a Single Work-item kernel with *advanced* optimization level. This is due to the fact the *original* NDRange kernel which uses temporal blocking can be further tuned to achieve a reasonably-high performance, while at the same time the stencil-based computation of this benchmark also matches very well with Single Work-item kernels.

Multiple optimizations are performed on the NDRange kernel to minimize its local memory usage. This involves making code changes to correctly coalesce accesses to the on-chip buffers to minimize buffer replication (3.2.4.2) and replacing multiple local buffers with private registers. Specifically, since each index in the buffers used to store *power* and computed *temperature* values are only accessed by one work-item, there is no need to define them as shared local buffers and significant Block RAM saving can be obtained by replacing them with one private register per work-item. Furthermore, branches that are used to make sure out-of-bound neighbors of cells on the border fall back on the border cell itself are optimized so that the compiler can correctly coalesce the accesses to the local buffer for reading the neighbors under the presence of SIMD. This is done by using intermediate registers to read both possible values for each case from the local buffer, and then choosing the correct value from the registers instead of the local buffer. Finally, the two write ports to the remaining local buffer, one from external memory to store values for the first fused iteration, and one to write back the output of the current iteration to be used in the next, are merged into one write port. This is done by conditionally writing to a private register instead, and then writing from the private register to the local buffer. Reducing the number of write ports to the local buffer from two to one halves the buffer replication factor on its own. Combined with the rest of the optimizations, Block RAM replication factor and utilization is significantly reduced, allowing us to use larger block sizes or unroll (3.2.1.5) the iteration loop to improve performance. We also add support for non-square blocks to increase freedom in tuning block size. Compared to the version with *basic* optimization, we increase the block size to 128×64 and add an unroll factor of 2. The kernel becomes limited by logic utilization on Stratix V at this point due to lack of native support for floating-point operations in the DSPs of this FPGA which results in a large amount of logic being used to support such operations. However, even with the larger block size and extra unrolling used in this version, Block RAM utilization is reduced compared to the version with *basic* optimization. Furthermore, we perform seed and target $f_{max}$ sweep (3.2.3.5) for this kernel to maximize its operating frequency and experimentally find that performance now scales up to a *pyramid_height* of 6.

For the Single Work-item kernel with *advanced* optimization level, we adopt 1D spatial blocking (but no temporal blocking). We disable the cache (3.2.3.2) and use loop collapse (3.2.4.3), exit condition optimization (3.2.4.4), shift register-based on-chip storage (3.2.4.1), loop unrolling (3.2.1.5) and seed and target $f_{max}$ sweep (3.2.3.5) for this kernel. These optimizations allow us to achieve a design with a very high operating frequency, fully saturate the memory bandwidth using an unroll factor of 16, and use a very large block size of 4096 to minimize the amount of redundant computation, all with a very modest area utilization. The performance of this kernel cannot be improved any further due to saturation of FPGA external memory bandwidth.

Table 4-4 shows the performance and area utilization of the aforementioned kernels on the Stratix V FPGA. For this benchmark, we use an input size of 8000×8000 and an iteration count of 100. The moderate level of optimization in the *original* Rodinia kernel is still not enough for it to perform well on FPGAs, mainly since the compiler is not able to infer correct run-time parameters like work-group size, and no parallelization is performed by default either. Hence, the relatively straightforward *unoptimized* Single Work-item version manages to outperform this kernel. However, *basic* optimizations significantly improve the performance of the NDRange kernel, while the performance of the Single Work-item version hardly improves by such optimizations. The *advanced* Single Work-item kernel achieves very high speed-up over the *unoptimized* versions and very high operating frequency at a very modest area utilization. However, its performance is limited by the external memory bandwidth. On the other hand, the *advanced* NDRange kernel, since it employs temporal blocking, can break away from the limit imposed by the external memory bandwidth and with our careful optimizations, manages to achieve over twice higher performance compared to the *advanced* Single Work-item kernel.

Table 4-4 Performance and Area Utilization of Hotspot on Stratix V

| Optimization Level | Kernel Type | Time (s) | Power (W) | Energy (J) | $f_{max}$ (MHz) | Logic | M20K Bits | M20K Blocks | DSP | Speed-up |
|---|---|---|---|---|---|---|---|---|---|---|
| None | NDR | 45.712 | 13.337 | 609.661 | 303.39 | 22% | 5% | 17% | 12% | 1.00 |
| | SWI | 21.388 | 13.353 | 285.594 | 303.39 | 21% | 10% | 22% | 10% | 2.14 |
| Basic | NDR | 3.276 | 31.561 | 103.394 | 234.96 | 58% | 37% | 78% | 27% | 13.95 |
| | SWI | 14.614 | 13.685 | 199.993 | 255.68 | 24% | 12% | 23% | 4% | 3.13 |
| Advanced | NDR | 1.875 | 28.181 | 52.839 | 206.01 | 78% | 42% | 71% | 52% | 24.38 |
| | SWI | 4.102 | 16.533 | 67.818 | 304.41 | 47% | 5% | 19% | 26% | 11.14 |

We need to emphasize here that the performance difference between the two *advanced* kernels does not show the advantage of the NDRange programming model, but rather, it shows the advantage of temporal blocking for stencil computation. In fact, due to the shifting pattern of stencil computation and the effectiveness of shift register optimization, which is only applicable to the Single Work-item programming model, it is certain that this model should be preferred over NDRange for stencil computation. The relatively high difference between the operating frequencies of the two *advanced* kernels further affirms which kernel model matches better with the underlying hardware. In Chapter 45, we will revisit this benchmark and show that a highly-optimized Single Work-item design with temporal blocking can achieve even higher performance than we achieved here with the NDRange implementation.

**4.3.1.3 Hotspot 3D**

Unlike the 2D version of this benchmark, the *original* NDRange implementation of Hotspot 3D in Rodinia neither employs explicit spatial blocking nor temporal blocking. However, it uses multiple private registers to cache consecutive neighbors to be used by the same thread when traversing the *z* dimension. Similar to Hotspot (2D), we also create the *unoptimized* Single Work-item kernel based on the OpenMP implementation in form of a triply-nested loop.

For the NDRange kernel with *basic* optimization, we set the work-group size (3.2.1.4) and add SIMD (3.2.1.5). The performance of this kernel does not scale beyond a SIMD size of 8 due to unoptimized memory access coalescing. For the Single Work-item kernel with the same optimization level, we remove branches on global memory address calculation (3.2.2.3) and unroll the inner loop (3.2.1.5). Similar to the 2D version, initiation interval of one is achieved for the loop nest, but performance does not scale beyond a SIMD size of four due to memory contention caused by tens of global memory ports competing with each other.

For the *advanced* optimization level, we only create a Single Work-item kernel since it best matches with the shifting pattern of stencil computation. The *original* NDRange implementation of Hotspot 3D in Rodinia is not as optimized the 2D version and hence, creating an *advanced* NDRange implementation based on that will be fruitless. We use the exact same set of optimization as the equivalent kernel for the 2D version in a very similar design. More specifically, we use 2D spatial blocking and collapse a larger loop nest (3.2.4.3) alongside with exit condition optimization (3.2.4.4). This kernel uses a relatively large block size of 512×512 with the cache disabled (3.2.3.2) and an unroll factor of 16 (3.2.1.5). Seed and target $f_{max}$ sweep (3.2.3.5) is also performed to maximize its operating frequency. Again, similar to the 2D case, the performance of this kernel does not scale any further due to saturating the external memory bandwidth.

Table 4-5 shows the performance and area utilization of our Hotspot 3D kernels on the Stratix V FPGA. For this benchmark, we use an input size of 960×960×100 with an iteration count of 100. The *original* OpenCL kernel from Rodinia performs very poorly on FPGAs and probably other hardware due to lack of sufficient optimization. On the particular case of FPGAs, it is outperformed even by the *unoptimized* Single Work-item kernel despite the sheer simplicity of the latter kernel. *Basic* optimizations on the NDRange kernel prove effective but still not enough to overtake even the *unoptimized* Single Work-item kernel. Furthermore, the operating frequency of this kernel drops significantly due to high area utilization. Even though the *unoptimized* Single Work-item kernel achieves relatively good performance, *basic* optimizations on this kernel only yield minor performance improvement due to lack of external memory access coalescing. On the other hand, the kernel with *advanced* optimization achieves

a noticeable jump in performance due to careful optimization of external memory accesses and efficient data caching. The operating frequency of the final kernel could easily be improved to over 300 MHz with slightly smaller blocks (512×256 or 256×256), but the improvement in operating frequency proved to be insufficient to make up for the performance gap caused by more redundant computation. The only reason for the sub-300 MHz operating frequency of this kernel is the placement constraints arisen from the large shift register used for spatial blocking that takes up a large portion of the Block RAM resources.

Table 4-5 Performance and Area Utilization of Hotspot 3D on Stratix V

| Optimization Level | Kernel Type | Time (s) | Power (W) | Energy (J) | $f_{max}$ (MHz) | Logic | M20K Bits | M20K Blocks | DSP | Speed-up |
|---|---|---|---|---|---|---|---|---|---|---|
| None | NDR | 249.164 | 14.991 | 3735.2 | 271.00 | 28% | 11% | 26% | 13% | 1.00 |
|  | SWI | 32.224 | 13.656 | 440.1 | 303.49 | 21% | 13% | 25% | 5% | 7.73 |
| Basic | NDR | 54.834 | 27.813 | 1525.1 | 202.38 | 80% | 31% | 78% | 78% | 4.54 |
|  | SWI | 24.813 | 15.689 | 389.3 | 255.36 | 32% | 21% | 35% | 15% | 10.04 |
| Advanced | SWI | 5.760 | 19.892 | 114.6 | 260.41 | 48% | 37% | 60% | 52% | 43.26 |

### 4.3.1.4 Pathfinder

The *original* NDRange version of this benchmark uses 2D blocking, with the block width being defined at compile-time but the block height being configurable at run-time using the input *pyramid_height* parameter. The block height (*pyramid_height*) controls how many block rows are processed (fused) without writing any data to external memory. As many cells as the block width are loaded from off-chip memory into on-chip memory, and computation iterates over *pyramid_height* rows using only on-chip memory, and then data is written back to off-chip memory. When a complete column of blocks is computed, the next block column starts. Due to the triangular/cone-shaped dependency pattern of the algorithm, the blocks are overlapped by 2 × *pyramid_height* columns to ensure correct output. This implementation very much resembles Rodinia's implementation of Hotspot. Since the compiler limits work-group size to 256 unless it is manually set, we use a block size of 256 for the *unoptimized* NDRange kernel and experimentally tune *pyramid_height*, which achieves the best performance if it is equal to 10. For the *unoptimized* Single Work-item kernel, we wrap the computation in a doubly-nested loop like the OpenMP version of the benchmark, but only keep the loop on columns inside the kernel and put the loop on rows in the host code since it is not pipelineable due to data dependency between computation of consecutive rows. The loop left in the kernel achieves an initiation interval of one.

For the *basic* optimization level, we set the work-group size (3.2.1.4) in the NDRange kernel, unroll the loop on the fused rows, and add SIMD and kernel pipeline replication (3.2.1.5). Specifically, we use a block size of 1024, a SIMD size of 16 and a kernel pipeline replication factor of 2. We also experimentally find a *pyramid_height* of 32 to achieve the best performance in this case. Due to inability of the compiler in correctly coalescing the accesses to the local buffer under the presence of SIMD, unrolling the iteration loop proved to be ineffective and was avoided since it increased the number of write ports to the local buffer, significantly increasing Block RAM replication factor. Furthermore, more pipeline (compute unit) replication was not possible in this kernel due to FPGA area limitation. For the Single Work-item kernel with this optimization level, we move external memory accesses outside of branches and replace them with registers (3.2.2.3) so that the accesses can be correctly coalesced with loop unrolling, and then unroll the loop by a factor of 64 which is the highest value that achieves meaningful performance scaling.

Similar to Hotspot, we create both an *advanced* Single Work-item and an *advanced* NDRange version for this benchmark. On one hand, the NDRange implementation of Pathfinder is very similar to Hotspot and with some extra effort, we expect to be able to significantly improve its performance similar to Hotspot. On the other hand, we also expect it to be possible to efficiently implement Pathfinder using a Single Work-item kernel due to the shifting pattern of the computation, and use shift registers as an efficient cache to minimize external memory accesses while also resolving the loop-carried dependency.

For the *advanced* NDRange version, we use a very similar set of optimizations to that of Hotspot. Specifically, we replace the local *result* buffer with a private register since it is only read and written by the same work-item. Furthermore, all accesses to the remaining *prev* buffer from inside of branches are replaced with temporary registers by first moving both possible values from the local buffer to the temporary register, and then choosing the correct value from the register inside the branch (3.2.4.2). Finally, the two writes to the *prev* buffer are merged into one using extra private registers; this optimization halves the Block RAM replication factor for implementing the buffer on its own. In the end, the number of reads from and writes to the buffer are reduced to the minimum value of 3 and 1, regardless of SIMD size. The large reduction in Block RAM usage in this case allows us to, unlike the *basic* NDRange version, successfully unroll the iteration loop (3.2.1.5) and further increase the block size. Specifically, we increase the block size to 8192, which allows the performance to improve up to a *pyramid_height* of 92, and use a SIMD size of 16 and an unroll factor of 2. Kernel pipeline replication is not used in this case since performance benefits more from unrolling and it is preferred to spend the FPGA resources on this form of parallelism. Seed and target $f_{max}$ sweep (3.2.3.5) is also performed to maximize the operating frequency of this kernel.

Even though Pathfinder has spatial dependencies similar to NW, the dependency is to cells from the previous row and hence, no dependency to the output of the previous iteration exists. Hence, we do not need to follow the same optimization strategy as NW and instead, for the *advanced* Single Work-item version of Pathfinder we use the same design strategy as the NDRange kernel but with an implementation similar to Hotspot and shift registers used as on-chip buffer. Doing so significantly reduces Block RAM usage for this kernel and allows us to increase block size to 32768. The block size can still be increased in this case, but performance improvement from lower redundancy will become minimal and instead, operating frequency will decrease due to complications arisen from placing large shift registers on the FPGA. Moreover, this version uses loop collapse (3.2.4.3) and exit condition optimization (3.2.4.4) alongside with an unroll factor of 32 (3.2.1.5). The cache created by the compiler is also disabled (3.2.3.2) since we perform caching manually using shift registers.

Table 4-6 shows the performance of our kernels in this benchmark with different optimization levels. We use an input size of 1,000,000×1,000 for this benchmark. Decreasing the number of columns and instead increasing the number of rows would have resulted in longer run time and more dependable timing results in this case. However, since most optimizations (SIMD, unrolling, etc.) are performed on the loop on columns, doing so would have resulted in low pipeline efficiency on our FPGA design and less improvement over baseline, and also low occupancy on the CPU and GPU versions, resulting in an unfair comparison. Hence, we chose to use a bigger number of columns instead at the cost of short execution time.

Table 4-6 Performance and Area Utilization of Pathfinder on Stratix V

| Optimization Level | Kernel Type | Time (s) | Power (W) | Energy (J) | $f_{max}$ (MHz) | Logic | M20K Bits | M20K Blocks | DSP | Speed-up |
|---|---|---|---|---|---|---|---|---|---|---|
| None | NDR | 3.918 | 12.901 | 50.546 | 303.39 | 20% | 4% | 16% | 2% | 1.00 |
| | SWI | 3.605 | 12.764 | 46.014 | 304.50 | 20% | 5% | 16% | <1% | 1.09 |
| Basic | NDR | 0.310 | 30.916 | 9.584 | 221.68 | 54% | 35% | 80% | 3% | 12.64 |
| | SWI | 0.749 | 14.469 | 10.837 | 226.03 | 40% | 20% | 32% | <1% | 5.23 |
| Advanced | NDR | 0.188 | 20.716 | 3.895 | 239.69 | 44% | 32% | 55% | 2% | 20.84 |
| | SWI | 0.234 | 15.314 | 3.583 | 278.39 | 34% | 7% | 21% | <1% | 16.74 |

The *original* kernel from Rodinia fails to achieve good performance due to small block size and lack of explicit parallelism, which also leads to poor utilization of external memory bandwidth. Similarly, the *unoptimized* Single Work-item kernel achieves low performance due to poor utilization of external memory bandwidth and lack of efficient data caching. With *basic* optimizations, the performance of both kernel variations improves, but the *basic* NDRange kernel achieves higher performance due to better caching of data. With *advanced* optimizations, the NDRange kernel achieves over 20 times higher performance than baseline, thanks to our careful optimizations on the local memory buffers that make it possible to use much bigger block sizes. The *advanced* Single work-item kernel achieves 25% lower performance compared to its NDRange counterpart but with a much higher operating frequency due to better critical path optimization, and much lower Block RAM utilization despite much bigger block size which also results in better power efficiency. The input buffer in Pathfinder is the only global buffer that is accessed every iteration and the output buffer is only written in the last fused row. Since the FPGA board has two banks, even with interleaving, it is not possible to efficiently saturate the external memory bandwidth when only one buffer exists in the kernel that is accessed every iteration. Apart from that, memory access efficiency is low in this benchmark due to unaligned memory accesses caused by block overlapping. The NDRange kernel in this case seems to achieve better performance due to *work-group pipelining* which can potentially allow better utilization of the memory bandwidth, and this is likely the reason for its higher performance. We expect the performance of the Single Work-item to be improved further by optimizing memory access alignment using padding, as we will discuss in Section 5.3.3, but such optimization is outside the scope of this chapter. It is likely that the operating frequency of the Single Work-item kernel could also be improved further if the number of fused rows is converted from a run-time variable to a compile-time constant, since it would simplify the critical path. In the end, we choose the *advanced* NDRange kernel as the best implementation for this benchmark.

### 4.3.1.5 SRAD

The *original* implementation of SRAD in Rodinia consists of six kernels. Two of these kernels (*compress* and *extract*) perform pre- and post-processing on the input image and only take a very small portion of the run time. Since neither of these kernels are timed on any of the platforms, we move them to the host code in the OpenCL implementation to avoid spending FPGA area on these kernels. In this implementation, the computation starts by reading the input image and calculating the value of each cell multiplied by itself and saving it into another buffer in the *prepare* kernel. In the *reduce* kernel, an additive reduction is performed on the input image and the new buffer from the *prepare* kernel, and two summation outputs are generated. In the *srad* kernel, the first stencil pass, which is a 2D 5-point star-shaped stencil, is performed on the input image and the summation results calculated in the previous kernel are used in the computation. Instead of calculating the addresses of the neighbors inside of the kernel, this implementation calculates the addresses for all cells in the host code and creates four additional buffers with the same size as the input image to store them on the FPGA external memory. Then, in the kernel, the address to access each neighbor in each iteration is read from these buffers, and the neighbor value is then read from the image buffer, resulting in unnecessary indirect memory accesses and extremely poor memory access efficiency. Then, the output of this kernel is stored in five different external memory buffers, one having the same coordinates as the input image and the others each being one column or one row shifted compared to this buffer so that address calculation for accessing neighbors can also be avoided in the next kernel. In the final *srad2* kernel, another stencil pass is performed, this time a 2D 3-point stencil that only uses the *center*, *east* and *south* cells, and the output is stored in a final buffer. The strange design decisions in this implementation make it an unlikely candidate to achieve high performance on any hardware:

- Separating the *prepare* and *reduce* kernels results in unnecessary external memory loads and stores
- The implementation of the *reduce* kernel is extremely inefficient
- Indirect memory accesses in the *srad* and *srad2* kernels to avoid basic address calculation lead to poor memory bandwidth utilization
- Lack of even basic caching in the stencil passes results in a significant amount of redundant external memory accesses.

For this version, we use a work-group size of 256 to align with the limitation imposed by the compiler when work-group size is not manually set. For the *unoptimized* Single Work-item version, we use the same kernel structure as the *original* NDRange implementation, but implement the first two kernels as basic single-loops and each of the two stencil passes as a doubly-nested loop. The *ivdep* pragma (3.2.1.2) is also used to avoid a false dependency in the *srad2* kernel.

For the NDRange kernel with *basic* optimization, the work-group size is manually set for all kernels (3.2.1.4), and SIMD is employed (3.2.1.5) with the exception of the *reduce* kernel in which SIMD cannot be used due to thread-id-dependent branching. Instead, in the *reduce* kernel, full unrolling is used for a simple loop, and partial unrolling is used for other loops (3.2.1.5). After extensive parameter tuning, best performance is achieved by using a work-group size of 256, a SIMD factor of 8 for the *prepare* kernel and 2 for the *srad* and *srad2* kernels, and an unroll factor of 2 for the *reduce* kernel. For the Single Work-item kernel with the same optimization level, the reduction in the *reduce* kernel is optimized using shift registers (3.2.2.1), and the innermost loops of all the kernels are partially unrolled. The latter forces us to add yet another *ivdep* pragma (3.2.1.2) to avoid a new false dependency in the *srad2* kernel. After parameter tuning, best performance for this version is achieved by using an unrolling factor of 8 for both the *prepare* and *reduce* kernels, and 2 for the remaining kernels.

For *advanced* optimization, we choose the Single Work-item kernel type since it can be used to efficiently implement both the reduction operation and the two stencil passes in this benchmark. Considering the suboptimal implementation of this benchmark in Rodinia, a complete code rewrite was required to achieve reasonable performance on our FPGA platform. Specifically, to minimize external memory accesses and maximize local data sharing, we combine all the original kernels into one kernel. The loops of the original *prepare* and *reduce* kernels are combined into one loop, eliminating two global buffers and all accesses associated with them. Also all indirect memory accesses to read neighboring cells are converted to direct addressing, and the four global buffers holding the address of the neighbors from the original implementation are eliminated. Then, the two stencil passes (*srad* and *srad2*) are merged, eliminating five more global buffers that were originally used to pass data from the first pass to the second pass. Over 10x reduction in global memory traffic and usage is achieved like this. The second stencil pass of the computation (*srad2*) can only start when the center, south and east cells are already computed by the first pass (*srad*). This means that if computation starts from the top-left of the grid as is the default case, it is not possible for the second pass to start right after the first cell is computed by the first pass, and some cells need to be buffered until all neighbors are ready to compute the first cell in the second pass. However, if the starting point is changed to bottom-right, since the necessary neighbors for computation of the first cell in the second pass fall outside of the grid boundary and the boundary conditions will be used instead, the second pass can start right after the first pass without any delay. We take advantage of this technique to minimize resource overhead. Similar to Hotspot 2D, we then employ 1D overlapped blocking with shift registers used as on-chip buffers (3.2.4.1). However, since two stencil passes are involved in this benchmark with a dependency between them, it is required that we increase the width of the halo region from one to two cells to correctly handle the dependency. Moreover, loop collapse (3.2.4.3) and exit condition optimization (3.2.4.4) are employed, all loops are partially unrolled (3.2.1.5), the auto-generated cache is disabled (3.2.3.2), and seed and target $f_{max}$ sweep (3.2.3.5) is performed as is the usual case for all of our *advanced* Single Work-item kernels. We also employ manual memory external banking (3.2.3.1) for this benchmark since the final version of our design reduces the total number of global memory buffers to two, with one being read and the other being written every clock cycle. Surprisingly, optimizing this benchmark did not end here. After place and routing the final design, we encountered lower-than-expected operating frequency on both Stratix V and Arria 10. After extensive troubleshooting, it turned out that the version of Intel FPGA SDK for OpenCL we are using has some problem with balancing pipeline stages when a floating-point variable is multiplied by a constant floating-point value. We worked around this problem by converting such multiplications to division without any loss of accuracy. For example, $0.25 \times x$ is replaced by $x/4.0$. The best-performing configuration for this version uses a block size of 4096 and unroll factors of 4 and 16 for the stencil computation and the reduction operations, respectively. No more unrolling is possible here due to DSP limitations on the Stratix V FPGA.

Table 4-7 shows the performance and area utilization of the aforementioned kernel versions. We use an input size of 8000×8000 and 100 iterations for this benchmark. Due to poor design, the *original* kernel from Rodinia performs very poorly on our FPGA. The *unoptimized* Single Work-item kernel, despite being based on the NDRange kernel, achieves higher performance, mainly due to more efficient implementation of the *reduce* kernel. *Basic* optimizations hardly improve the performance of the NDRange kernel due to poor baseline implementation, but the Single Work-item kernel can achieve a reasonable speed-up since the implementation of the reduction operation in this version is close to optimal. Finally, the kernel with *advanced* optimization not only achieves a notable speed-up over baseline, but also very high operating frequency, which shows that this design matches the underlying FPGA architecture very well.

Table 4-7 Performance and Area Utilization of SRAD on Stratix V

| Optimization Level | Kernel Type | Time (s) | Power (W) | Energy (J) | $f_{max}$ (MHz) | Logic | M20K Bits | M20K Blocks | DSP | Speed-up |
|---|---|---|---|---|---|---|---|---|---|---|
| None | NDR | 346.796 | 18.913 | 6558.953 | 248.20 | 47% | 22% | 42% | 26% | 1.00 |
| None | SWI | 276.807 | 16.558 | 4583.370 | 270.56 | 36% | 15% | 33% | 24% | 1.25 |
| Basic | NDR | 265.784 | 24.587 | 6534.831 | 248.57 | 64% | 34% | 78% | 52% | 1.30 |
| Basic | SWI | 42.346 | 20.358 | 862.080 | 251.69 | 48% | 37% | 57% | 46% | 8.19 |
| Advanced | SWI | 9.060 | 18.904 | 171.270 | 304.41 | 57% | 8% | 27% | 87% | 38.28 |

### 4.3.1.6 LUD

The *original* implementation of this benchmark uses 2D square blocking with three kernels. First, the *diameter* kernel computes the top left block in the matrix, then, the *perimeter* kernel computes the remaining blocks in the first block column and block row and then, the remaining square of blocks are processed by the *internal* kernel. In the next step, the starting position of the matrix is moved one block forward in both the *x* and the *y* dimension and the same chain of kernel operations is performed on the new submatrix. In the final round of computation, only the bottom right block will be left and in this case, only the *diameter* kernel processes this block to finish the computation. Every kernel takes full advantage of local memory by avoiding all redundant memory accesses in the block that is being processed. By default, the compiler auto-unrolls some of the loops in the kernel, which results in lower performance since the external memory accesses in the auto-unrolled loops are not consecutive and hence, numerous ports to external memory are created, resulting in a significant amount of contention on the memory bus. We prevent this by forcing an unroll factor of one for these loops. A block size of 16×16 is used for this version since larger values are cannot be used unless work-group size is set manually. For the *unoptimized* Single Work-item implementation, we base our kernel design on the OpenMP version of the benchmark which uses a similar computation pattern to that of the NDRange version. We also use *ivdep* (3.2.1.2) to avoid false dependencies detected by the compiler in the middle loops of the *diagonal* kernel and the outermost loop of the *internal* kernel. Dependencies detected in some other loops were real and hence, *ivdep* was not used for those. Initially, we encountered what seemed to be a functional bug in the compiler using this version. We worked around this issue by making minor modifications in the *internal* kernel to swap the order of the two innermost compute loops, which then allowed us to merge the write-back loop into the compute loop and replace the arrays of *sum* variables with a single variable. A block size of 16×16 was also used in this version since larger block sizes resulted in lower performance.

For the NDRange kernel with *basic* optimization, we manually set the work-group size for each kernel (3.2.1.4) and add SIMD and kernel pipeline replication (3.2.1.5) to all kernels. Furthermore, the loop in the *internal* kernel is fully unrolled, while the loops in the other kernels are partially unrolled with compile-time configurable unroll factors (3.2.1.5). In practice, SIMD could not be used in the *diameter* and *perimeter* kernels due to thread-id-dependent branching. Furthermore, using kernel pipeline replication for the *diameter* kernel was avoided since this kernel is only executed by one work-group. Since run time is dominated first by dense matrix multiplication in the *internal* kernel, and then by the computation in the *perimeter* kernel, while the *diameter* kernel accounts for less than 0.1% of the total run time, we configure the parameters in a way that allocates resources to each kernel based on its portion of the total run time. Based on this, block size is increased to 64×64, the *diameter* kernel is left as it is, while an unroll and kernel pipeline replication factor of 2 is used for the *perimeter* kernel. For the *internal* kernel, a pipeline replication factor of 3 is used on top of full unrolling of the loop in this kernel. These parameters nearly maximize the DSP and Block RAM utilization of the device and higher values cannot be used any more. For the Single Work-item kernel with *basic* optimization, we increase block size to 64×64 and use the shift register-based optimization for floating-point reduction (3.2.2.1) which then allows us to also partially unroll the reduction loops in the *diameter* and the *perimeter* kernels. For the *internal* kernel, the innermost loop is fully, and the middle loop is partially unrolled (3.2.1.5). Furthermore, the loop over blocks for the *perimeter* and *internal* kernels, which are not pipelineable, are moved to the host to save area. Partial unrolling of the innermost loops in the *diagonal* and *perimeter* kernels resulted in slow-down and hence, was avoided. Moreover, apart from fully unrolling the innermost loop of the internal kernel, the middle loop is unrolled by a factor of 2. Higher values could not be used since an unroll factor of 3 resulted in load/store dependencies and a very high initiation interval due to the loop trip count (block width) not being divisible by 3, and a factor of 4 resulted in DSP overutilization on the Stratix V FPGA.

To choose the best kernel type for creating the version with *advanced* optimization, we have to consider two important characteristics of this benchmark: first, the outer loops in the *diagonal* and *perimeter* kernels are not pipelineable due to variable exit condition of the middle or innermost loops. As mentioned in Section 3.1.4, NDRange kernels are preferred in such case since the run-time scheduler can allow a lower average initiation interval compared to sequential execution of the non-pipelineable loops in the equivalent Single Work-item implementation. Second, blocking in this benchmark requires that data transfers between off-chip memory and on-chip memory be separated from the computation, with a complete block being loaded into on-chip memory, computed, and then written back. Hence, no overlapping of computation and memory accesses will exist in a Single Work-item implementation, resulting in poor performance. On the other hand, in an NDRange implementation, each compute unit is shared between multiple work-groups and one work-group can occupy the compute part of the pipeline while another work-group is occupying the memory access part, allowing efficient *work-group pipelining* and overlapping of computation and memory accesses. Hence, we choose NDRange as the best kernel type for this optimization level. Compared to the NDRange kernel with *basic* optimization level, multiple optimizations are employed to minimize local memory ports and replication factor (3.2.4.2) for the *advanced* version. A temporary variable is used as reduction variable instead of the local buffers themselves, to reduce number of ports to the local buffers in the *diagonal* and *perimeter* kernels. The local buffer in the *diameter* kernel and the *dia* buffer in the *perimeter* kernel are split into two buffers, one of which is loaded row-wise and the other, column-wise. Replacing the column-wise accesses to the original buffers with row-wise accesses to the new buffers that are filled in column-wise order allows correct access coalescing under partial unrolling of the loops in these two kernels and significant reduction in Block RAM usage. The *peri_row* buffer in the *perimeter* kernel is also transposed for the same reason. Loads from and writes to external memory are modified in the *perimeter* kernel to remove thread-id-dependent branching. Furthermore, writing back the content of the *peri_row* buffer to external memory is merged into the compute loop to remove one extra read port from this buffer. The same can be done with the write-back of the content of the *peri_col* buffer; however, that would result in an external memory access pattern that is not consecutive based on work-item ID and hence, lowers performance. The write-back for this buffer is kept outside of the compute loop after a barrier so that data can then be written back in a way that accesses are consecutive based on work-item ID. Also common subexpression elimination is performed in all kernels and constant common subexpressions are moved to the host code to minimize logic and DSPs used for integer arithmetic. All these optimizations allow us to increase block size to 96×96, with an unroll factor of 4 for the *diameter* kernel, an unroll factor of 8 and compute unit replication factor of 2 for the *perimeter* kernel, and a SIMD size of 2 for the *internal* kernel. However, fitting this configuration on the FPGA required that we manually perform port sharing on the *diameter* kernel to reduce its Block RAM usage so that more Block RAMs are available to the rest of the kernels. Even though doing so slightly reduced the performance of the *diameter* kernel, the extra performance gained by faster execution of the rest of the kernels made up for the difference. Finally, seed and target $f_{max}$ sweep (3.2.3.5) is performed to maximize the performance of this kernel. Unlike every benchmark discussed so far where we use this compiler-assisted optimization to maximize performance by maximizing the operating frequency of the design, in the particular case of this benchmark, performance increases with operating frequency up to a certain point and after that, it starts decreasing. The reason for this is that with the implemented optimizations, the *internal* kernel nearly saturates the FPGA external memory bandwidth and increasing the operating frequency beyond the saturation point will result in more contention on the memory bus and instead, decreases performance. In the end, we timed the final kernel running at different operating frequencies and chose the fastest one.

Table 4-8 shows the performance of LUD with different optimization levels. We use an input matrix size of 11520×11520, which is divisible by all the block sizes used for every version. The *original* NDRange kernel achieves low performance here due to small block size and lack of explicit parallelism. The *unoptimized* Single Work-item kernel achieves even worse performance due to the non-pipelineable loops and lack of compute and memory access overlapping. With *basic* optimization, the NDRange kernel achieves over two orders of magnitude speed-up, mostly due to full unrolling of the loop in the *internal* kernel that is the most compute-intensive kernel in the benchmark. The performance of the Single Work-item kernel, however, improves only slightly with *basic* optimization since the fundamental problems associated with using this kernel type for this benchmark are still not addressed. Finally, our *advanced* optimizations on the NDRange kernel allow us to maximize the performance of this benchmark on our device with near full utilization of DSP and Block RAM resources. The performance of this benchmark on the Stratix V FPGA is limited by DSP and Block RAM resources.

**Table 4-8 Performance and Area Utilization of LUD on Stratix V**

| Optimization Level | Kernel Type | Time (s) | Power (W) | Energy (J) | $f_{max}$ (MHz) | Logic | M20K Bits | M20K Blocks | DSP | Speed-up |
|---|---|---|---|---|---|---|---|---|---|---|
| None | NDR | 1944.820 | 15.580 | 30300.296 | 262.60 | 30% | 14% | 28% | 13% | 1.00 |
| None | SWI | 2451.187 | 15.885 | 38937.105 | 267.73 | 34% | 12% | 28% | 16% | 0.79 |
| Basic | NDR | 14.800 | 29.712 | 439.738 | 234.57 | 69% | 42% | 95% | 99% | 131.41 |
| Basic | SWI | 1273.347 | 25.667 | 32682.997 | 254.32 | 65% | 24% | 61% | 65% | 1.53 |
| Advanced | NDR | 13.159 | 19.832 | 260.969 | 224.40 | 81% | 50% | 98% | 96% | 147.79 |

It is worth mentioning that one compiler limitation, which applies to all NDRange kernels and still exists even in the latest version of the compiler (v18.0), severely limited our parameter tuning freedom for this benchmark. As mentioned in Section 2.3.2, the compiler automatically pipelines multiple work-groups in the same compute unit for NDRange kernels to maximize pipeline efficiency and minimize the negative performance impact of barriers. However, the number of work-groups that can run simultaneously in a compute unit is automatically decided by the compiler at compile-time, with no way of being influenced by the programmer. In many cases the compiler tries to support tens or even hundreds of work-groups per compute unit, and then replicates all the local buffers in the compute unit by the same number, resulting in significant waste of Block RAMs, while the same performance could probably be achieved using a much smaller number of simultaneous work-groups. In the particular case of the LUD benchmark, the *diameter* kernel does not even require work-group pipelining since it is only executed by one work-group, while the compiler still replicates the local buffers inside of this kernel to support two simultaneous work-groups. Moreover, we had to abandon multiple optimization ideas for the *perimeter* kernel since they resulted in the compiler increasing the degree of work-group pipelining and making the kernel unfittable. If it was possible to directly influence the degree of work-group pipelining, the trade-off between performance and Block RAM utilization could be optimized much more efficiently.

### 4.3.2 Arria 10

#### 4.3.2.1 Changes Compared to Implementations on Stratix V

For NW and Hotspot 3D we use the exact same settings as Stratix V since the small improvement in external memory bandwidth on Arria 10 compared to Stratix V prevents these benchmarks from benefiting from the more resources available in this FPGA. The same applies to Pathfinder; however, we ended up using a smaller block size of 4096 for this benchmark on Arria 10 since bigger block sizes lowered operating frequency by an amount that cancelled out the effect of the bigger block size.

For Hotspot, we decrease the block size to 64×64 but increase the unroll factor to 3 compared to Stratix V. This allows a small performance improvement over using the same configuration as Stratix V despite significant reduction of operating frequency.

For SRAD, we increase the unroll factor for the stencil passes from 4 to 16 on Arria 10, which is made possible by the significant improvement in the number of DSPs and native support for floating-point operations on this device. We also take advantage of single-cycle floating-point accumulation, which is an Arria 10-specific optimization, to eliminate the need for shift register optimization for floating-point reduction (3.2.2.1). However, to our surprise, our implementation on Arria 10 achieved lower operating frequency compared to Stratix V. Further troubleshooting showed that similar to the issue discussed in Section 4.3.1.5, the source of the problem on Arria 10 also seemed to be from the way pipeline stage balancing was performed by the compiler. Specifically, the compiler seemed to implement floating-point division operations inefficiently since removing all such divisions from the kernel increased the operating frequency of the kernel to nearly 350 MHz. Unfortunately, by the time of writing this thesis, we could not find a work-around for this problem. However, even with the lowered operating frequency, the benchmark is memory-bound on Arria 10 and performance improvement is expected to be minimal with higher operating frequency.

For LUD, even though the higher number of DSPs available in Arria 10 compared to Stratix V eliminated one of the two main area bottlenecks, the other area bottleneck, i.e. Block RAM count, is not improved much in the new FPGA and still limits the performance on this device. Furthermore, as mentioned in Section 3.2.3.4, we could not use *flat* compilation for LUD on Arria 10 since an NDRange kernel was being used and regardless of the number of different seeds we tried, we could not prevent timing failure for the peripheral clocks (DDR, PCI-E, etc.). This forced us to use the default PR flow, which resulted in fitting or routing failure for any configuration that utilized more than 95% of the Block RAM resources on the device, further reducing our ability to efficiently use this important resource. To be able to increase the block size to 128×128, we have to decrease compute unit replication for the *perimeter* kernel to one and instead make up for it by increasing the unroll factor to 32 (unrolling has much less Block RAM overhead than pipeline replication). We also increase the unroll factor for the *diameter* kernel to 8 and use a SIMD size of 4 for the *internal* kernel. With *flat* flow, the compute unit replication factor of two could be kept for the *perimeter* kernel with an unroll factor of 16, resulting in 5% higher performance compared to our final configuration with the PR flow, but this configuration had to be discarded due to timing constraints not being met. This shows that using Partial Reconfiguration on Arria 10, apart from the direct disadvantage of lowering operating frequency, can also indirectly result in even more performance disadvantage when area utilization is high by preventing efficient utilization of FPGA resources.

Table 4-9 shows the best-performing results for each benchmark on both Arria 10 and Stratix V, and the resource that is bottlenecking the performance in each case. "BW" in this table refers the external memory bandwidth, while M20K refers to FPGA on-chip memory blocks. The clear trend here is that performance on Arria 10 is limited by its low external memory bandwidth in nearly every benchmark. Because of this, performance improvement in cases where performance was already bottlenecked by this resource on Stratix V shows minimal improvement. Furthermore, power efficiency is lowered in these benchmarks compared to Stratix V due to higher static power consumption and inefficient use of the FPGA area on Arria 10. The only benchmarks that achieve meaningful performance improvement on Arria 10 are SRAD and LUD, which were bound by FPGA resources on Stratix V. SRAD also becomes memory bound with the higher unroll factor used on Arria 10 despite modest area usage. For LUD, minimal improvement in Block RAM count prevents us from trading off more external memory bandwidth by on-chip memory compared to Stratix V and in the end, we

**Table 4-9 Performance and Power Efficiency of All Benchmarks on Stratix V and Arria 10**

| Benchmark | FPGA | Time (s) | Power (W) | Energy (J) | $f_{max}$ (MHz) | Logic | M20K Bits | M20K Blocks | DSP | Bottleneck |
|---|---|---|---|---|---|---|---|---|---|---|
| NW | Stratix V | 0.260 | 19.308 | 5.020 | 218.15 | 53% | 7% | 28% | 2% | BW |
| | Arria 10 | 0.176 | 32.699 | 5.755 | 201.06 | 28% | 8% | 25% | <1% | BW |
| Hotspot | Stratix V | 1.875 | 28.181 | 52.839 | 206.01 | 78% | 42% | 71% | 52% | Logic, BW |
| | Arria 10 | 1.616 | 45.732 | 73.903 | 179.89 | 31% | 44% | 81% | 29% | M20K, BW |
| Hotspot 3D | Stratix V | 5.760 | 19.892 | 114.578 | 260.41 | 48% | 37% | 60% | 52% | BW |
| | Arria 10 | 5.254 | 35.147 | 184.662 | 239.39 | 14% | 36% | 53% | 10% | BW |
| Pathfinder | Stratix V | 0.188 | 20.716 | 3.895 | 239.69 | 44% | 32% | 55% | 2% | BW |
| | Arria 10 | 0.141 | 34.397 | 4.850 | 258.97 | 27% | 19% | 35% | <1% | BW |
| SRAD | Stratix V | 9.060 | 18.904 | 171.270 | 304.41 | 57% | 8% | 27% | 87% | DSP |
| | Arria 10 | 4.721 | 40.889 | 193.037 | 277.33 | 44% | 14% | 27% | 62% | BW |
| LUD | Stratix V | 13.159 | 19.832 | 260.969 | 224.40 | 81% | 50% | 98% | 96% | DSP, M20K |
| | Arria 10 | 5.279 | 46.671 | 246.376 | 240.74 | 33% | 45% | 93% | 41% | M20K, BW |

cannot utilize even half of the DSPs of this device. Even if more Block RAMs were available on Arria 10, performance will not improve much further since the most time consuming part of this benchmark (the *internal* kernel) is already nearly memory-bound.

### 4.3.3 CPUs

Table 4-10 shows the performance and power efficiency of all the benchmarks on both of our evaluated CPUs using both GCC v6.3.0 and ICC 2018.2. The best performance for each benchmark on each CPU has been colored in green. None of the benchmarks are modified other than to add timing and power measurement functions. Most of the CPU benchmarks in Rodinia already take advantage of optimization techniques like loop tiling. We expect that the existing code optimizations coupled with using two state-of-the-art compilers should allow us to achieve a reasonable level of performance on the CPUs and allow fair comparison with the rest of the hardware.

**Table 4-10 Performance and Power Efficiency Results of All Benchmarks on CPUs**

| Benchmark | CPU | Compiler | Time (s) | Power (W) | Energy (J) |
|---|---|---|---|---|---|
| NW | i7-3930k | GCC | 719.651 | 116.691 | 83.977 |
| | | ICC | 744.204 | 115.767 | 86.148 |
| | E5-2650 v3 | GCC | 371.479 | 81.910 | 30.428 |
| | | ICC | 395.222 | 83.746 | 33.090 |
| Hotspot | i7-3930k | GCC | 4056.987 | 126.988 | 515.180 |
| | | ICC | 3331.503 | 127.817 | 425.818 |
| | E5-2650 v3 | GCC | 3149.191 | 87.131 | 274.391 |
| | | ICC | 2659.946 | 87.814 | 233.579 |
| Hotspot 3D | i7-3930k | GCC | 7752.818 | 152.252 | 1180.363 |
| | | ICC | 8806.121 | 151.272 | 1331.353 |
| | E5-2650 v3 | GCC | 6881.140 | 100.302 | 690.168 |
| | | ICC | 6794.439 | 99.955 | 679.140 |
| Pathfinder | i7-3930k | GCC | 306.995 | 133.308 | 40.925 |
| | | ICC | 293.070 | 140.161 | 41.074 |
| | E5-2650 v3 | GCC | 297.511 | 83.687 | 24.896 |
| | | ICC | 309.270 | 86.892 | 26.874 |
| SRAD | i7-3930k | GCC | 41206.358 | 113.265 | 4667.282 |
| | | ICC | 15008.157 | 153.048 | 2296.995 |
| | E5-2650 v3 | GCC | 46510.895 | 58.414 | 2716.417 |
| | | ICC | 11825.654 | 100.860 | 1192.733 |
| LUD | i7-3930k | GCC | 22048.880 | 142.271 | 3136.958 |
| | | ICC | 19396.328 | 133.585 | 2591.064 |
| | E5-2650 v3 | GCC | 17896.558 | 94.115 | 1684.335 |
| | | ICC | 14326.216 | 88.891 | 1273.477 |

Based on our results, in most cases ICC outperforms GCC by a large margin. Also other than Pathfinder which does not seem to scale well with multi-threading, the newer CPU is faster than the old one in every benchmark. However, this CPU is at best twice faster than the old one, despite being of a much newer generation and having four (6 vs. 10) more cores.

### 4.3.4 GPUs

Table 4-11 shows performance and power efficiency of all of our benchmarks on both of our evaluated GPUs. Default block size was increased to 32×32 for Hotspot, and *pyramid_height* was tuned for Hotspot and Pathfinder on each GPU. Changing the default parameters were also attempted in other benchmarks, but were discarded since they did not improve performance. No further modifications were made in any of the benchmarks other than adding timing and power measurement functions. All of the CUDA versions of the benchmarks in Rodinia have already gone through a good degree of optimization and we believe that coupled with NVIDIA's most recent CUDA toolkit and compiler, the performance results we have obtained here are a good representative of the capabilities of the GPUs.

**Table 4-11 Performance and Power Efficiency Results of All Benchmarks on GPUs**

| Benchmark | GPU | Time (s) | Power (W) | Energy (J) |
|---|---|---|---|---|
| NW | K20X | 270.587 | 102.184 | 27.649 |
| | 980 Ti | 133.116 | 132.465 | 17.633 |
| Hotspot | K20X | 823.476 | 132.297 | 108.943 |
| | 980 Ti | 1161.366 | 152.340 | 176.921 |
| Hotspot 3D | K20X | 2893.110 | 118.531 | 342.922 |
| | 980 Ti | 1393.586 | 174.916 | 243.748 |
| Pathfinder | K20X | 50.200 | 138.755 | 6.965 |
| | 980 Ti | 21.503 | 219.690 | 4.724 |
| SRAD | K20X | 3758.656 | 145.440 | 546.660 |
| | 980 Ti | 2374.360 | 222.598 | 528.516 |
| LUD | K20X | 4884.329 | 134.892 | 658.856 |
| | 980 Ti | 1292.572 | 237.113 | 306.458 |

Based on our results, the newer 980 Ti GPU outperforms its older counterpart by two times or more in nearly every benchmark; the only exception is Hotspot were 980 Ti actually achieves lower performance. Since Hotspot relies heavily on caching, this performance regression could be caused by differences in the memory and cache hierarchy of these two GPUs which are from different generations.

### 4.3.5 Comparison

Fig. 4-2 shows the performance and power efficiency of all of our benchmarks on all the evaluated hardware. Our results show that the FPGAs can outperform their same-generation CPUs in every case while achieving up to 16.7 times higher power efficiency. Compared to the GPUs, however, the results are different. Other than the NW benchmark in which the Stratix V FPGA can narrowly overtake its same-generation GPU, in no other case can any of the FPGAs outperform their same-generation GPUs. Furthermore, the newer Arria 10 FPGA is outperformed by even the older K20X GPU in every benchmark other than NW; though the difference tends to be smaller in the more compute-intensive SRAD and LUD benchmarks. Still, the FPGAs have a clear power efficiency advantage over the GPUs to the point that the aged Stratix V FPGA can achieve better power efficiency than the much newer GTX 980 Ti GPU in every benchmark. The largest power efficiency advantage is observed in the NW benchmark were the Stratix V FPGA is 5.6 times more power efficient than it same-generation GPU. Unfortunately, power efficiency improvements on Arria 10 compared to Stratix V are minimal to none since in none of the benchmarks we can efficiently use the resources of this newer FPGA due the external memory bandwidth bottleneck. LUD is the only benchmark in which Arria 10 achieves better power efficiency than Stratix V.

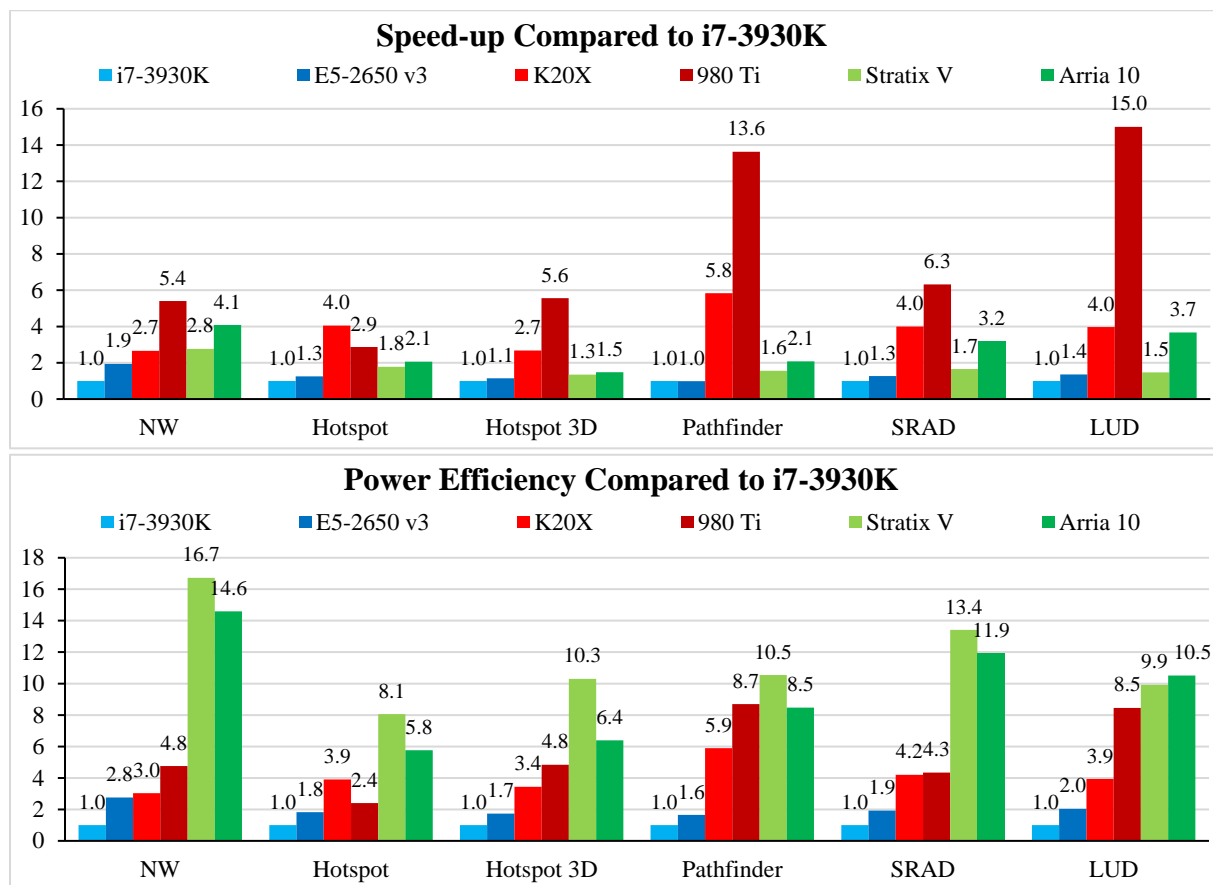

Figure 4-2 Performance and Power Efficiency Comparison Between Different Hardware

The smallest performance difference between the Arria 10 FPGA and the 980 Ti GPU is observed in the NW and Hotspot benchmarks. For the case of NW, the gap between the FPGAs and GPUs is minimized since our FPGA design can efficiently handle the complex loop-carried dependency of this benchmark while fully pipelining the design with an initiation interval of one. Our FPGA implementation of this benchmark is near-optimal and its performance is effectively limited by the external memory bandwidth of the FPGAs. However, the GPU implementation of Rodinia is sub-optimal and a more optimized implementation could likely achieve better performance on the GPUs and increase the gap. On the other hand, our FPGA implementation and Rodinia's GPU implementation for Hotspot are based on the same algorithm, but temporal blocking achieves better scaling on FPGAs compared to GPUs. Hence, the performance gap between these devices could potentially become even smaller if temporal blocking is employed more efficiently on the FPGAs. Performance of this benchmark scales up to a degree of temporal parallelism (*pyramid_height*) of 6 on the FPGAs due to large block size, while on the GPUs the performance only scales up to a degree of 2 or 3 due to relatively smaller block size. Hence, we conclude that stencil computation is likely one of the computation patterns that could be efficiently accelerated on FPGAs, and despite the large external memory bandwidth gap between FPGAs and GPUs, it could be possible to achieve competitive performance on FPGAs compared to GPUs for this computation pattern if temporal blocking is efficiently utilized. Dynamic programming applications like NW and Pathfinder are also good candidates for acceleration on FPGAs. However, as we shown in this chapter, these types of applications quickly reach the limit of the external memory bandwidth on FPGAs and cannot benefit from temporal blocking since they are generally not iterative. Hence, an optimized implementation of such benchmarks on a GPU will likely always be faster than an optimized implementation on an FPGA due to their large gap in external memory bandwidth. For the more compute-bound benchmarks, the performance difference between FPGAs and their same-generation GPUs gets larger since the large gap in the peak compute performance of these devices cannot be easily filled. Even though FPGAs allow us to create custom pipelines for each application to achieve higher computational efficiency compared to GPUs, it is generally not enough to fill the large compute performance gap unless a specific application achieves very low computational efficiency on GPUs (less than 10%).

In the end, we should emphasize that the goal of this chapter was *not* to achieve the best performance for each benchmark on the FPGAs, but rather, to determine effective FPGA-based optimizations for each benchmark and optimize each to a degree that would allow fair comparison with the CPU and GPU platforms. We expect that there could still be room to further optimize some of benchmark on the FPGAs. For example, all of the stencil benchmarks could benefit from temporal blocking (some of which we will revisit in the next chapter), or LUD could benefit from a systolic array-based implementation which could match much more efficiently with the underlying FPGA hardware. Reducing the width of datatypes is also an important FPGA-specific optimization that we did not study here; this optimization could significantly reduce area and memory bandwidth usage on these devices for some benchmarks (especially integer ones) and lead to noticeable performance improvements. However, if such optimizations are performed for the FPGAs, the optimization level of the FPGA kernels might go beyond the CPU and GPU kernels and prevent fair comparison between the devices.

## 4.4 Related Work

In [24], the authors present OpenDwarfs, a multi-platform benchmark suite written in OpenCL that can target different hardware including FPGAs. They implement and evaluate multiple benchmarks on a range of CPU and GPU devices and one Xilinx Virtex-6 FPGA, and report performance results and detailed comparison. For converting the OpenCL kernels to synthesizable HDL code, they use SOpenCL [31]. Our work differs from theirs in this respect that we use newer hardware and a much more mature OpenCL compiler for our FPGA platform which is made by the FPGA manufacturer. Furthermore, they explore very few optimizations on each hardware and their FPGA optimizations are limited to basic loop unrolling, while we use CPU and GPU implementations that are already optimized to a reasonable degree and also manually apply a range of FPGA-specific optimizations to our FPGA kernels to allow fair comparison with the other hardware. In [32], the same authors present a visual programming framework that can automatically generate and tune code targeting Intel FPGA SDK for OpenCL. Their framework supports generation of both NDRange and Single Work-item kernels and can automatically apply optimizations related to data parallelism (Section 3.2.1.5), shift register optimization for floating-point reduction (Section 3.2.2.1), *restrict* keyword (Section 3.2.1.1), aligned DMA transfers, and general shift register inference (Section 3.2.4.1). However, the framework cannot automatically detect patterns that could benefit from shift register inference in the application and hence, this optimization needs to be guided by the programmer. The authors then implement three benchmarks (Electrostatic Surface Potential Calculation, Gene Sequence Search and Time-domain FIR Filter) on an Intel Stratix V FPGA, and discuss the effect of using the NDRange or Single Work-item kernel type, basic data parallelism optimizations, and shift registers as on-chip memory.

In [33], the authors of the Rodinia benchmark suite present a preliminary study using three benchmarks, namely NW, DES and Gaussian Elimination, and compare performance in terms of number of clock cycles between a CPU, a GPU and an FPGA. This work uses VHDL to implement the benchmarks on the FPGA and does not discuss platform-specific optimizations in detail. The Xilinx Virtex-II FPGA they use is also relatively old compared to the other hardware used in their evaluation. In contrast, we use two modern generations of FPGAs and compare each to a CPU and GPU of its own generation. Furthermore, we use a mature HLS tool to implement the benchmarks on the FPGAs that significantly reduces development effort, and take advantage of multiple FPGA-specific optimizations to allow a fair comparison with the already-optimized CPU and GPU implementations.

In [34], the authors present a framework for converting C/C++ code that is annotated with OpenACC directives to OpenCL targeting Intel FPGA SDK for OpenCL. They add equivalent functions to OpenACC to support basic OpenCL functions like host/device data transfer, setting kernel arguments, kernel invocation, etc. Furthermore, the framework automatically enables aligned DMA transfers, and provides a set of pragmas to support on-chip channels, loop unrolling, SIMD and kernel pipeline replication. All of these functions and pragmas are directly mapped to their equivalent in OpenCL host and kernel code to be used with Intel FPGAs. One shortcoming of this work is that the resulting OpenCL kernel uses the NDRange programming model, which is not the preferred model on FPGAs. Furthermore, all optimizations and parameter tuning are still left to the programmer. They perform basic parameter tuning on FPGAs for multiple benchmarks, and present performance comparison with CPU, GPU and Xeon Phi platforms using the same OpenACC code, which is not necessarily optimized for either of the hardware. We, however, take advantage of many basic and advanced optimization techniques on FPGAs and provide a more dependable performance comparison since our CPU and GPU benchmarks are already optimized to a reasonable degree. Very recently, the authors extended their work in [35] by adding support for multiple new optimizations to their framework that can significantly reduce programming effort. Specifically, they add support for Single Work-item kernels, shift register optimization for floating-point reduction (Section 3.2.2.1), Loop collapsing (Section 3.2.4.3), shift register inference for local data storage (Section 3.2.4.1) and branch-variant code motion optimization. They also provide support for both of the shift register-based optimizations being coupled with unrolling. Some analysis is provided on the effect of each optimization on a set of benchmarks and performance and power efficiency comparison between a Stratix V D5 FPGA, a Xeon CPU and an NVIDIA GPU is reported. One shortcoming of the work is that when unrolling is coupled with optimization of floating-point reduction operations, they directly pass the unroll pragma to the OpenCL kernel and let the OpenCL compiler unroll the loop. As mentioned in Section 3.2.2.1, this method of unrolling would require increasing the size of the shift register and can result in significant area overhead for large unroll factors. In contrast, if the loop is unrolled using the method we proposed in Section 3.2.2.1, a larger shift register will not be required and the extra area overhead can be avoided. In addition, since their compiler does not yet support loop blocking/tiling using OpenACC directives, using the automatic shift register inference will require manual loop blocking by the programmer since this optimization cannot be applied if the loop bounds are not known at compile-time.

In [36], the authors use Xilinx SDAccel to implement a set of benchmarks (K-Nearest Neighbor, Monte Carlo Method and Bitonic Sort) on a Xilinx Virtex-7 FPGA, and compare performance and power efficiency with two GPU platforms. A set of basic FPGA-based optimizations are performed and better power performance in some cases and better power efficiency in most is reported compared to the GPUs. However, multiple shortcomings reduce the dependability of their results. The GPUs used in the study are relatively low-end and not comparable with the FPGA platform. On top of that, the majority of the GPU kernels are not optimized, which gives an unfair advantage to the FPGA. In addition, the input sizes are very small in every case and benchmark run times are below 1 ms or even 1 μs in most cases, and in one case the input is saved on the FPGA on-chip memory instead of external memory, giving the FPGA platform even more unfair advantage over the GPUs. Power consumption comparison also puts the GPU board against the FPGA chip alone, without considering the FPGA external memory. In contrast, we keep the comparison as fair as possible by using same-generation devices, employing kernels on each device that have gone through a reasonable amount of optimization, and using large input sizes that allow dependable run time measurements.

In [37], the authors present an OpenCL-based benchmark suite for FPGAs targeting Intel FPGA SDK for OpenCL. They discuss nine benchmarks extracted from Rodinia [13], OpenDwarfs [24], Intel's OpenCL examples and other sources, and compile each with multiple performance parameters for a total of over 8000 different configurations. They also attempt to mathematically model the relationship between design parameters and performance. This work only uses the NDRange programming model, which is not the preferred programming model on FPGAs, and only takes advantage of basic compiler-assisted optimization techniques. They conclude that the relationship between performance and design parameters is difficult to model mathematically due to complex interactions between such parameters.

In [38], the authors present a framework that uses information from both OpenCL host and kernel code to automate certain optimizations that Intel FPGA SDK for OpenCL Offline Compiler cannot perform on its own due to lack of knowledge of the host code. Specifically, they provide the means of automatically adding the *restrict* keyword to kernel buffers (Section 3.2.1.1) if no pointer aliasing exists in the host code, and converting NDRange kernels to Single Work-item by wrapping each region between two barriers in a multiply-nested loop using local and global work size information extracted from the host code. Moreover, they automate shift register inference for optimizing floating-point reduction (Section 3.2.2.1). Their work is an early step in automating optimizations that we performed manually here.

In [39], the authors perform a study similar to ours but target Xilinx tools and FPGAs. They port multiple benchmarks from the Rodinia benchmark suite [13] and optimize each using a set of general optimizations targeting Xilinx Vivado HLS [5] and SDAccel [7]. Speed-up over baseline, and performance and power efficiency comparison between a Xilinx Virtex-7 FPGA and an NVIDIA Tesla K40c GPU are reported, with the FPGA being faster than the GPU in some kernels. They use sequential implementations that do not even have pipelining enabled as FPGA baseline, resulting in over 4,000 times speed-up after optimization for some benchmarks. We, however, use baselines that are optimized for GPUs rather than such unoptimized codes that do not represent real-world scenarios to avoid such unrealistic speed-up ratios. Furthermore, their timing results seem to suggest that some of the benchmarks were running for a few days (e.g. 50 hours for LUD). Based on the more detailed version of the publication available at [40], it seems the reported timing values are actually in µs, and the timing unit is incorrectly reported as seconds in [39]. Having this in mind, the run time for all of the benchmarks in their case is lower than 200 ms, and go as low as 48 µs in case of NW; such short run times that can be heavily affected by profiling overhead cannot be considered as basis for dependable performance comparison. In fact, the three kernels in which they report better performance on the FPGA compared to the GPU are among the shortest ones. Moreover, as mentioned in Section 4.2.4, such short run times will not allow correct power readings on the GPU either as is also evident in their reported power consumption for the GPU. Only two of their evaluated benchmarks use more than 100 Watts on the GPU, and most are under 80 Watts (less than one third of the GPU's TDP) and go as low as 53.7 Watts in case of NW; such values represent the idle power of the GPU rather than power usage during kernel execution. They also do not report how FPGA power consumption is measured and whether it includes FPGA external memory or not. In contrast, apart from general optimizations, our work also includes benchmark-specific transformations in every case and our results are highly dependable since we make sure to use big input sizes to allow correct time and power measurement. In addition, we achieve good performance and better power efficiency on the FPGAs compared to GPUs in every case, while lack of benchmark-specific optimizations in their case results in very low performance in some benchmarks that we successfully optimize (e.g. LUD and Hotspot) and lower power efficiency compared to the GPU in multiple benchmarks.

In [41], the authors present another study similar to ours that also targets Xilinx FPGAs and tools. This work presents a wide range of HLS-based optimization techniques for FPGAs, some of which we also covered in this chapter, and includes code examples for many of them. They apply the optimization techniques to three applications, namely a Jacobi 2D stencil, General Matrix Multiplication (GEMM) and an N-body code, and report speed-up achieved with each level of optimization. This work is analogues to ours on the Xilinx platform but lacks performance comparison with other devices.

## 4.5 Publication Errata

The publication in WRC'16 [22] presented a very preliminary version of the results presented here with only four benchmarks and limited advanced optimizations. The current results are updated and largely different from the numbers reported in that publication.

Compared to the publication at SC'16 [17]:

- Performance model has been improved by splitting initiation interval into compile-time and run-time initiation interval and discussing them separately.
- Multiple new optimization techniques have been added and all optimizations have been discussed in a more organized manner.
- All the results presented in this chapter are updated with more optimizations and parameter tuning in nearly every case, resulting in better performance and power efficiency.
- All kernels have been compiled using a newer version of Quartus and Intel FPGA SDK for OpenCL Offline Compiler.
- Hotspot 3D has been added to the benchmarks but CFD has been removed.
- Arria 10 results have been reported for every benchmark and performance and power efficiency comparison between Stratix V and Arria 10 has been added.
- Performance of Hotspot on FPGAs has significantly improved compared to other hardware since the input settings used in the publication were too small, resulting in the input completely fitting in the CPU and GPU caches and giving them an unfair advantage.
- In the publication, it is incorrectly assumed than loop unrolling increases the pipeline depth by the unroll factor; the pipeline depth does increase with unrolling, but not by the unroll factor. This issue has been corrected in this document.
- Newer versions of ICC and CUDA have been used to maximize the performance of the CPUs and GPUs. Furthermore, "-fp-model precise" is added to ICC compilation parameters to make sure that accuracy of floating-point computations is the same on all hardware.

## 4.6 Conclusion

In this chapter, we ported a subset of the Rodinia benchmark suite for FPGAs as a representative of typical HPC workloads. We showed that even though the original NDRange kernels designed for GPUs generally perform poorly on FPGAs, and basic guidelines from Intel's documents hardly improve their performance, advanced FPGA-specific optimizations are effective on both NDRange and Single Work-item kernels, allowing us to achieve at least one order of magnitude performance improvement over the baseline on FPGAs for every benchmark. Furthermore, we showed that in most cases the Single Work-item programming model matches better with the underlying FPGA architecture, allowing us to take better advantage of the unique features of these devices.

Our results showed that FPGAs have a performance and power efficiency advantage over their same-generation CPUs in every case. However, compared to GPUs, it is generally not possible to achieve better performance due to the large gap in external memory bandwidth and compute performance between current-generation FPGAs and GPUs. The main bottleneck of performance for current-generation FPGAs is their low external memory bandwidth, resulting in memory-bound performance for nearly every benchmark on the new Arria 10 FPGA. Despite these limitations, FPGAs can still achieve higher power efficiency compared to not only their same-generation GPUs (up to 5.6 times) but also newer-generation ones.

Among the evaluated benchmarks, stencil-based applications showed better matching with the FPGA architecture than the rest and we expect that by taking full advantage of the FPGA-specific shift register buffers and temporal blocking, we should be able to minimize the performance gap between FPGAs and GPUs for this type of computation. Based on this conclusion, we extend our work by implementing a highly-optimized stencil accelerator on FPGAs, which will be discussed in the next chapter.

# 5 High-Performance Stencil Computation on FPGAs Using OpenCL

In this chapter, we will first introduce stencil computation and discuss its importance in HPC, and then we will review related work and show the advantage of our work against existing solutions. In the next step, we will present our high-performance FPGA-based accelerator for stencil computation and its associated performance model. Finally, we will discuss our results including performance projection for upcoming FPGAs and comparison with other FPGA work and highly-optimized implementations on other hardware. Unlike the previous chapter where our comparison involved moderately-optimized (but not optimal) implementations on different hardware, in this chapter we will thrive to compare highly-optimized implementations on each hardware. The contents of this chapter have been partially published in [42] and [43].

## 5.1 Background

### 5.1.1 Stencil Computation

Stencils are one of the most important computation patterns in HPC that are used in weather, wave, seismic and fluid simulations, image processing and convolutional neural networks. This computation pattern involves iteratively traversing a multi-dimensional grid, and calculating the weighted sum of a set of coefficients multiplied by the value of each cell and its neighbors. The pattern of the stencil determines which neighbors, and to what distance from the center cell, are involved in the computation. The maximum distance between the neighbors and the center cell is called the stencil *radius*. Alternatively, a stencil with a radius of $r$ is also called an *r-order* stencil[1]. Fig. 5-1 shows an example of first-order 2D and 3D *star-shaped* stencils.

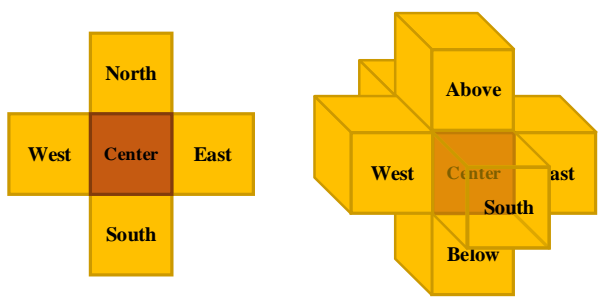

**Figure 5-1 First-order 2D and 3D stencils**

Due to the high byte to FLOP ratio of stencil computation, applications involving this computation pattern are generally memory-bound on most hardware. However, this computation pattern exhibits good spatial and temporal locality, allowing significant reduction of required external memory bandwidth by employing *spatial* and *temporal blocking*.

---

[1] In some scientific publications, a stencil with a radius of $r$ is called a *2r-order* stencil; i.e. what we call a first-order stencil is called second-order

### 5.1.2 Spatial Blocking

In typical stencil computation, since the neighboring cells are reused regularly in the computation, the grid is *blocked* (tiled) in multiple dimensions, allowing full spatial reuse inside of the block and significant reduction in redundant external memory accesses. The only redundant accesses in this case will happen on the boundaries of the blocks. This technique is called *spatial blocking*. On CPUs, spatial blocking is generally implemented using loop tiling. On GPUs, spatial blocking can be implemented by transferring and keeping data on local memory including scratchpad, registers and caches. We will discuss how we implement spatial blocking on FPGAs in Section 5.3.1.

### 5.1.3 Temporal Blocking

Even with spatial blocking, many stencils, especially low-order ones, will still be memory-bound on most hardware. Stencil computation is usually iterative, with an iteration or *time* loop encompassing the loops traversing the spatial domain. Due to read-after-write dependency between the different iterations of the time loop, this loop effectively runs sequentially by default, with data being fully written back to external memory before the next iteration of the loop can start. To further reduce external memory accesses for stencil computation, it is possible to add *temporal blocking* on top of spatial blocking so that multiple iterations of the time loop are calculated on chip, before results are written back to external memory. Our implementation of temporal blocking on FPGAs is discussed in Section 5.3.2. We will show that FPGAs have multiple architectural advantages compared to other hardware that allow them to achieve better performance scaling than such hardware with temporal blocking.

## 5.2 Related Work

Due to their importance, there is a large body of work discussing different implementations of stencil computation on different hardware. One of the most prominent algorithms for stencil computation was presented by Intel in 2010 [44]. In this paper, the *3.5D blocking* technique for 3D stencils is introduced, which involves 2.5D blocking in space and 1D blocking in time. 2.5D spatial blocking refers to blocking two spatial dimensions out of three and *streaming* the last dimension. This is in contrast to classical 4D blocking, which blocks every three spatial dimension. This implementation uses square blocks and overlapped temporal blocking which involves redundantly computing the *halo* regions. Using a mathematical model, this paper shows how blocking one less dimension in 3.5D blocking allows having multiple times bigger blocks compared to 4D blocking with the same amount of on-chip memory, reducing the ratio of redundant memory accesses to block size and consequently, allowing better scaling and speed-up with temporal blocking. Other block shapes have also been proposed in literature, including diamond [45] and hexagonal blocks [46]. Such blocking types are used to reduce or eliminate *redundant computation* by skewing the block shape in different dimensions; however, all such block shapes still have a moderate amount of *redundant memory accesses*. Since such blocking techniques have only been evaluated with 4D blocking so far, they cannot improve the ratio of redundant memory accesses to block size compared to 3.5D rectangular blocking due much smaller block size and hence, generally fall short of the latter blocking technique with respect to performance. Combining such blocking schemes with 3.5D blocking could prove worthwhile, but will likely require significant engineering effort.

For Xeon and Xeon Phi processors, Yount proposed a technique called "Vector Folding" for stencil computation that is suitable for wide-vector architectures [47]. This implementation was further refined and made available to public as the state-of-the-art YASK framework [48]. We will use this framework for evaluation on Xeon and Xeon Phi platforms.

On GPUs, one of the most highly-optimized implementations of 3D stencil computation was proposed by Maruyama et al. [49]. This work uses 3.5D blocking as proposed in [44] and applies multiple GPU-specific optimizations for newer NVIDIA GPUs. Despite the fact that this work is now 4 years old, it still achieves the highest performance for first-order 3D stencil computation on a single GPU reported to this day. We will use the publicly-available implementation from this work to evaluate the performance of first-order 3D stencil computation on GPUs. [50] uses 4D mixed hexagonal and classical tiling and reports performance for multiple 2D stencils. We will use the results reported in this work to compare our implementation on FPGAs with GPUs for first-order 2D stencil computation.

On FPGAs, implementations of stencil computation can be grouped into two categories: the first category use thread-based designs that use implementations similar to what is used on GPUs, while the second category use deep-pipelined designs with shift registers as on-chip buffer. Recent examples of the first category include [51, 52, 53, 54]. The major shortcoming of such implementations is that they do not use shift registers as on-chip buffers (Section 3.2.4.1) and hence, miss one of the most important advantages of FPGAs for stencil computation. Furthermore, such work usually block all of the spatial dimensions rather than streaming one of them as outlined in [44], missing even more room for optimization. In Section 5.3, we will show how we take advantage of both shift registers and 3.5D blocking to maximize performance.

Among the second category, there are multiple recent examples in literature where temporal blocking has been employed in a deep-pipeline design to achieve high performance [55, 56, 57, 58, 59]. However, most of such work only use temporal blocking and avoid spatial blocking; i.e. they stream all of the spatial dimensions and only block the time dimension. The advantage of doing so is that halo regions are eliminated and no redundant computation or memory accesses will exist in the computation, allowing linear performance scaling with temporal blocking. However, avoiding spatial blocking comes at the cost of limiting row size for 2D stencils (to a few thousand cells), and plane size for 3D (to 128×128 or lower), relative to the size of the FPGA on-chip memory. Such limitation is unacceptable in HPC where input grids for stencil computation are generally in the order of tens of thousands of cells in each dimension [60, 61, 62]. In fact, the inputs are generally so large that they are spatially split over thousands of nodes in a world-class supercomputer, with the problem size per node still being larger than what is supported by such work. This restriction will become even more limiting for high-order stencils due to the higher on-chip memory requirement of such stencils. One major advantage of our work is that we combine spatial and temporal blocking in a deep-pipelined design to achieve high performance without restricting input size, as we will discuss in Section 5.3. Among the few examples that use a similar design strategy as us, [63] presents a configurable VHDL template for stencil computation with both spatial and temporal blocking, but their design is only evaluated using a very small input that is not applicable to real-world scenarios. In [64], the authors use the built-in stencil library from MaxCompiler [65], called MaxGenFD [66], to create a framework that can dynamically distribute stencil computation over multiple FPGAs. This work is only applicable to the specific environment of software and hardware provided by Maxeler Technologies, and focuses on efficient scheduling rather than maximizing performance. Finally, in [67], the authors present an HLS-based implementation of Jacobi 2D targeting Xilinx Vivado HLS and SDAccel which uses both spatial and temporal blocking and employs multiple HLS-based optimizations. They evaluate their implementation on a Kintex UltraScale KU115 device and report performance and power efficiency.

All the related work discussed so far only discuss first-order stencils; however, many scientific applications *require* high-order stencil computation. Three out of the nine nominees for the ACM Gordon Bell award in the last two years, including both winners, accelerated computations that involved high-order stencil computation [68, 69, 70]; this clearly shows the importance of accelerating such stencils. For Xeon and Xeon Phi processors, the YASK framework [48] already supports high-order stencils. On GPUs, [71] is one of the few publications that discusses the general optimization of high-order stencils. This implementation uses 2.5D spatial blocking (but lacks temporal blocking) and an "in-plane" method that computes the stencil plane-by-plane in form of a partial sum. This allows better memory coalescing and alignment compared to previous implementations, at the cost of extra computation (due to partial summing). We will compare the performance of our high-order stencil implementation on FPGAs with this work. On FPGAs, publications involving high-order stencil computation are scarce. In [72], Shafiq et al. implement first to fourth-order 3D star-shaped stencils on a Virtex-4 LX200 FPGA. They only use spatial blocking with a cache-like on-chip storage, and use DMA to stream the data from the host to the FPGA, rather than streaming it from the FPGA on-board memory. Also in [73], the authors implement a first-order 3D cubic and a third-order 3D star-shaped stencil using a design that combines spatial and temporal blocking similar to ours. They use MaxCompiler and evaluate their stencils on two Virtex-5 LX330 FPGAs.

## 5.3 Implementation

Our design needs to achieve multiple goals:

- It should support spatial blocking so that any input size, as long as it fits in the FPGA external memory, can be accelerated.
- Considering the low external memory bandwidth of current FPGA boards, it should support temporal blocking so that reasonable performance can be achieved.
- It should be parameterized so that stencil radius and performance parameters can be changed with ease, and the design can be easily scaled over bigger FPGAs.

To realize these goals, we combine spatial and temporal blocking in a fashion similar to what is described in [44]. We stream one of the spatial dimensions but block the rest (i.e. 1.5D spatial blocking for 2D stencils, and 2.5D for 3D stencils). Such design requires a complex multiply-nested loop to support all the index and block variables. As explained in Sections 3.2.4.3 and 3.2.4.4, nested loops incur extra area overhead and lower operating frequency on FPGAs. Furthermore, overlapped blocking also lowers external memory access efficiency due to access alignment issues. We employ the necessary FPGA-specific optimizations to tackle these issues (Section 5.3.3).

We use the Single Work-item programming model due to its advantages over the NDRange model (Section 3.1.4) for stencil computation, specially the fact that shift registers can only be inferred in this model. The outline of our multi-kernel FPGA-based stencil accelerator is depicted in Fig. 5-2. Our design consists of a *read*, a *write*, and a *compute* kernel. The first two kernels are the only ones that have an interface to external memory, and all memory reads and writes are handled by these two kernels. The *compute* kernel is replicated into multiple Processing Elements (PEs) and data is streamed from external memory by the *read* kernel, through the *compute* PEs, and finally written back to external memory by the *write* kernel. These kernels are connected using on-chip channels. Stencil radius and performance parameters are parameterized in our OpenCL design, providing us with extra freedom in terms of targeting different stencil orders and maximizing performance for a given FPGA.

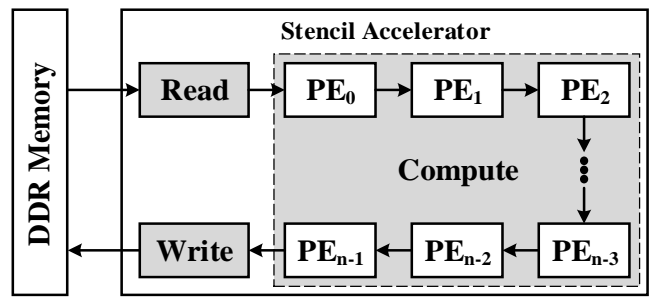

**Figure 5-2 Overview of stencil accelerator**

### 5.3.1 Spatial Blocking on FPGAs

Fig. 5-3 shows 1.5D spatial blocking for 2D stencils, and 2.5D spatial blocking for 3D stencils. This figure is drawn for a similarly-sized 2D and 3D input and the dimension that is *streamed* is shown in green. Computation starts from top left and moves forward in the *x* direction. When the border of the spatial block is reached, computation moves to the next row in the block. For 2D stencils, computation is streamed until the bottom of the input in *y* direction and then moves to the next spatial block. For 3D stencils, computation continues until the bottom of the spatial block in *y* direction, and then computation is streamed plane by plane until the last plane from the input is reached. When the spatial block is fully computed, computation moves to the next spatial block in the same row. Spatial blocks are processed row by row until the entire input grid has been covered.

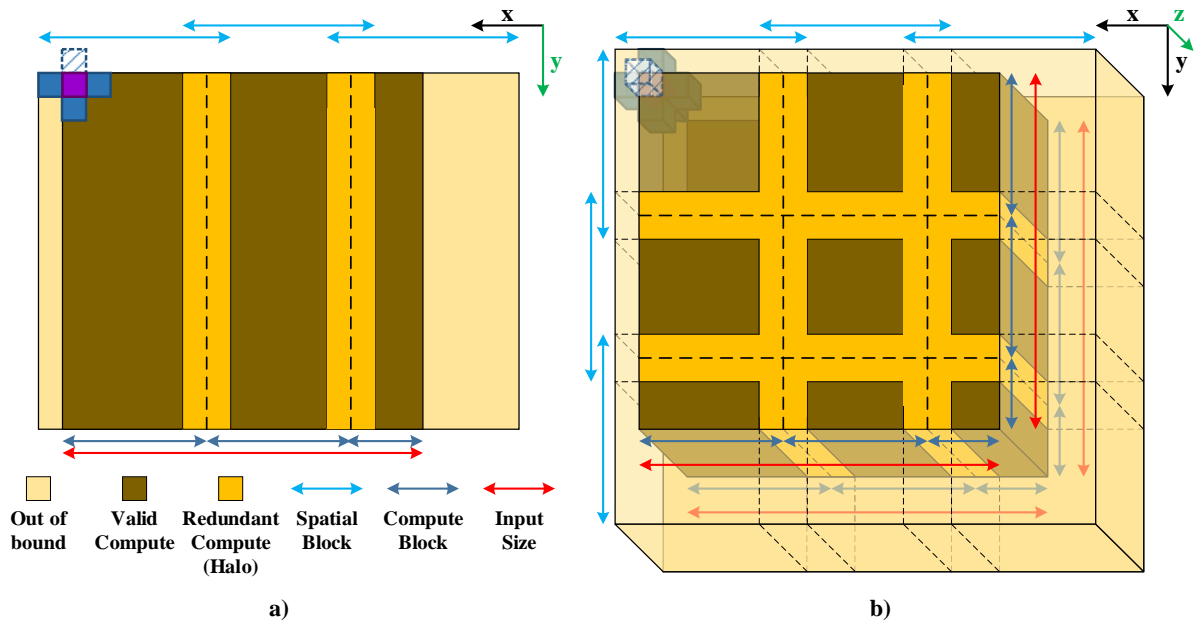

**Figure 5-3 Spatial blocking in a) 2D and b) 3D stencils**

In our implementation, the spatial blocks are overlapped so that no data exchange is required between them. This technique is called *overlapped blocking (tiling)*. The overlapped sections of the blocks are called *halos* or *ghost zones*. For 2D stencils, each block has two columns of halos, one on each block border, and a width equal to the stencil radius. For 3D stencils, two rows and two columns of halos exist on the borders of each block, again each being as wide as the stencil radius. We define the section of each spatial block that excludes the halo regions as the *compute block*. To keep our loops regular and allow correct pipelining without breaking the static shift register addressing, if the dimensions of an input do not align with the size of our spatial block in that dimension (as also shown in Fig. 5-3), extra out-of-bound computation is performed until the borders of the last spatial block are reached. This out-of-bound computation could incur significant overhead for small inputs; however, this overhead will be negligible for inputs that are much larger than the spatial block size.

We use shift registers as on-chip buffers for spatial blocking (Section 3.2.4.1); this storage type perfectly aligns with the shifting pattern of stencil computation. This technique is widely used on FPGAs [56, 57, 58]; however, it is not applicable to CPUs, Xeon Phi and GPUs due to lack of hardware support for shift registers. Fig. 5-4 shows how data in a given spatial block is cached in a shift registers. For 2D stencils, grid cells need to be cached starting from the *North* neighbor down to the *South* neighbor including all block rows in between. For 3D stencils, cells are cached from the *Above* neighbor down to the *Below* neighbor including all rows and planes between them. For a stencil of radius *rad*, the shift register needs to be filled with *rad* rows if the stencil is 2D, and *rad* planes if it is 3D, before the computation can start. As the computation progresses, one new cell is loaded from external memory into the tail of the shift register (*Bottom* neighbor of next cell in 2D, and *Below* neighbor of next cell in 3D stencils) every iteration, and the cell at the head of the shift register is discarded. In total, two spatial block rows and two spatial block planes need to be cached for first-order 2D and 3D stencils, respectively. In contrast, on other hardware, a full rectangle for 2D and a full rectangular cuboid for 3D stencils needs to be cached in on-chip memory. In other words, **for a fixed spatial block size, the required amount of on-chip memory on FPGAs is one block row and one block plane less than other hardware, for 2D and 3D stencils, respectively. Hence, the availability of shift registers on FPGAs is one of the advantages of this platform for stencil computation compared to other hardware.**

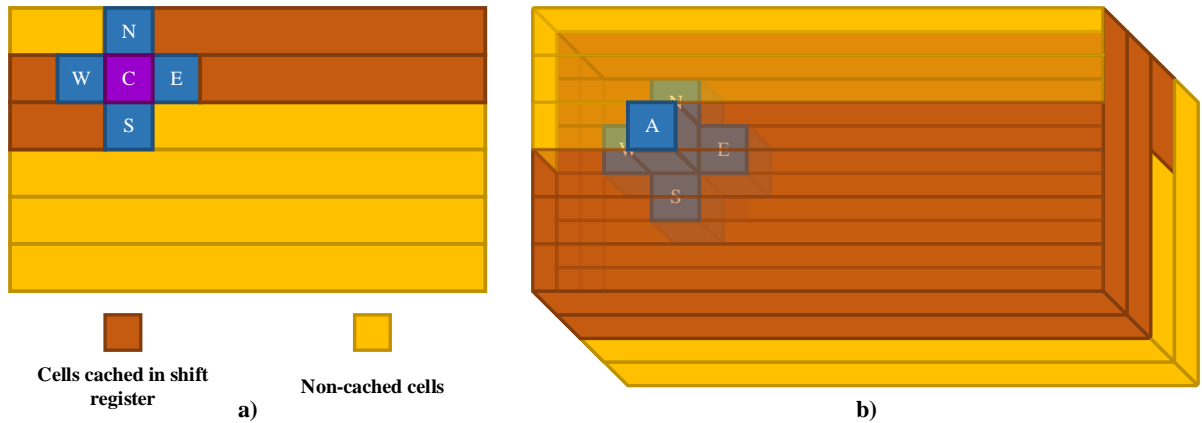

**Figure 5-4 Shift register for spatial blocking in a) 2D stencils and b) 3D stencils**

To parallelize the computation spatially and allow efficient use of the FPGA external memory bandwidth by memory access coalescing, we vectorize the computation in each spatial block by loop unrolling (Section 3.2.1.5). In this case, for a vector size of $par_{vec}$, $par_{vec}$ cells are loaded into the shift register per iteration and the same number are discarded every iteration. Furthermore, the size of the shift register is also increased by the same amount. Fig. 5-5 shows how vectorization is applied to the computation of a 2D stencil.

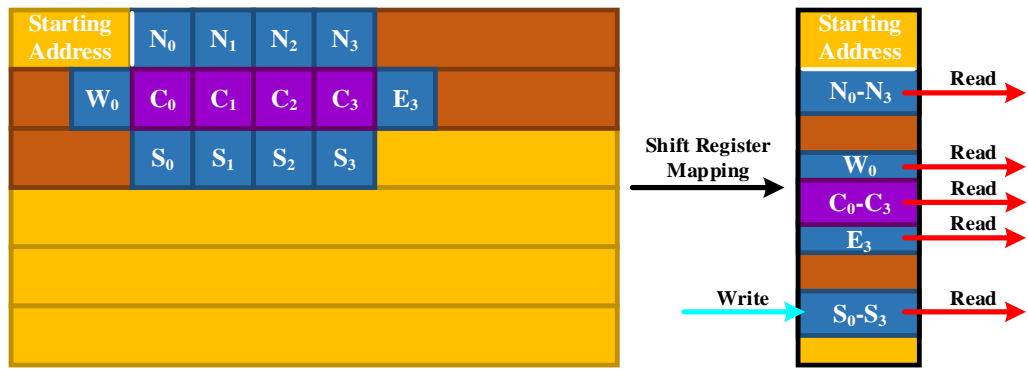

**Figure 5-5 Spatial blocking with vectorization**

If we denote the size of the spatial block in each dimension as $bsize_{\{x|y\}}$ and the stencil radius as *rad*, the size of the shift register required in our implementation is as follows:

$$size = \begin{cases} 2 \times rad \times bsize_x + par_{vec} & 2D \\ 2 \times rad \times bsize_x \times bsize_y + par_{vec} & 3D \end{cases} \quad (5\text{-}1)$$

In practice, multiple accesses are required per iteration to the shift register to fetch all of the neighbors, with the size of the vectorized access being larger than the width of the Block RAM ports. Hence, the compiler will interleave data if the shift register spans over enough Block RAMs, and will replicate the shift register if it does not, to provide enough ports for all the parallel accesses. Since spatial blocking eliminates all redundant external memory accesses per block, we also disable the cache that is generated by the compiler to save Block RAMs (Section 3.2.3.2).

### 5.3.2 Temporal Blocking on FPGAs

To implement temporal blocking, we map each parallel temporal iteration to a PE, and connect the PEs using on-chip channels (FIFOs). To achieve this design pattern, we use the *autorun* kernel type (Section 3.2.3.3) to replicate the *compute* kernel into as many PEs as the *degree of temporal parallelism* (i.e. the number of iterations from the time loop that are computed in parallel). In our design, each PE computes the same spatial block of a different (but consecutive) time-step, and the intermediate data is passed between the PEs via the on-chip channels. Since the computation of a given PE can start only after the first output of the previous PE has been generated, the computation in each PE is always *rad* rows, for 2D, and *rad* planes, for 3D stencils, behind its previous PE. This delay is required so that the necessary cells for the computation of the first new output are loaded into the shift register in each PE.

Due to the read-after-write dependency between the iterations in the time dimension, it is required that we increase the width of the halo regions proportional to the degree of the temporal parallelism. Hence, even though the block size is the same for all the PEs, the amount of valid computation decreases as we move towards the last PE. This is shown in Fig. 5-6 for 2D and 3D stencils.

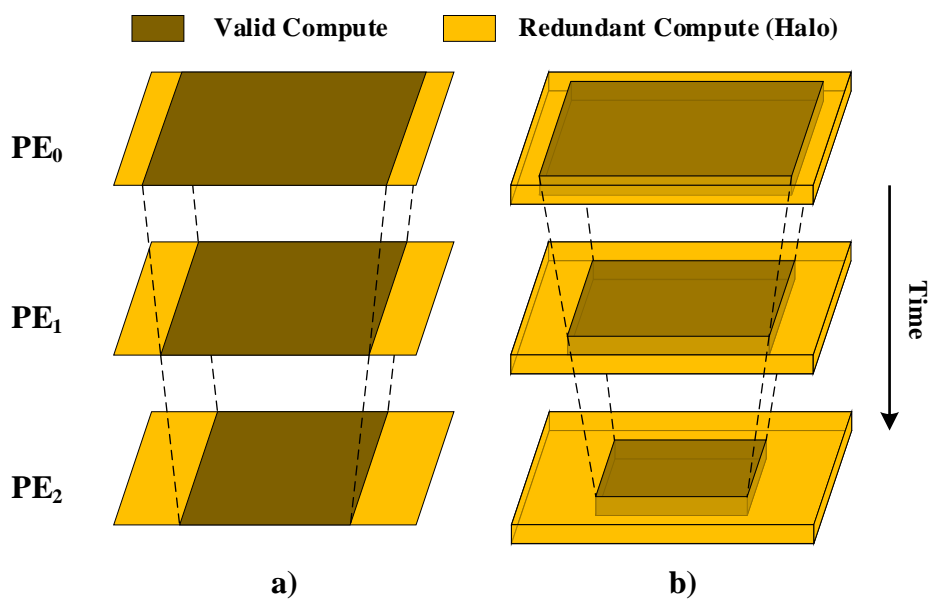

**Figure 5-6 Temporal blocking for a) 2D stencils and b) 3D stencils**

In this case, for a degree of temporal parallelism of $par_{time}$, the width of each halo region (in cells) in the last PE will be equal to:

$$size_{halo} = rad \times par_{time} \tag{5-2}$$

Halo processing results in *thread divergence* on GPUs, since the threads processing the halo regions and the threads performing the valid computation go through different paths. Alleviating this problem requires complex Warp Specialization optimizations [49]. However, in a deep-pipelined FPGA design, thread divergence does not exist since all flow control statements (if, switch/case, etc.) are eliminated by implementing both paths and multiplexing out the correct result. This eliminates flow divergence, at the cost of an area overhead. In our design, we also minimize this area overhead by avoiding control flow statements on computation and redundantly computing all halo regions, and only avoiding write back of invalid outputs to external memory. **Lack of thread divergence and the need for Warp Specialization is another advantage of FPGAs compared to GPUs for stencil computation.**

On GPUs, local memory is scattered across different Streaming Multiprocessors (SMs). Even though GPUs have a large amount of local memory available in total, the amount of local memory per SM is relatively small (less than 0.5 MB). Since each warp is scheduled onto one SM and can only utilize the local memory of that SM, the spatial block size on GPUs will be limited to the size of the local memory per SM. However, no such restriction exists on FPGAs and the programmer even has the flexibility of using all the local memory (~6.6 MB on Arria 10) for implementing only one spatial block. Because of this, the spatial block size on an FPGA can be multiple times larger than a GPU with the same total local memory size. **This shows the third and most important advantage of FPGAs over GPUs for stencil computation: bigger spatial block size on FPGAs allows a lower ratio of redundant to valid memory accesses per spatial block for a fixed degree of temporal parallelism, resulting in better scaling with temporal blocking on these devices compared to GPUs.**

### 5.3.3 FPGA-specific Optimizations

Multiple advanced manual optimizations are implemented in our design to maximize the performance of our stencil accelerator:

**Loop collapse:** As explained in Section 3.2.4.3, to avoid the area overhead of loop nesting, we manually collapse the loop nest in our code into one loop, and apply the necessary updates to the index and block variables inside the collapsed loop.

**Exit condition optimization:** We also apply the exit condition optimization explained in Section 3.2.4.4 on top of loop collapsing to improve the operating frequency of our design. This optimization allows us to increase operating frequency from ~200 MHz to over 300 MHz.

**Padding:** Our observations show that accesses to the FPGA external memory must be 512-bit-aligned or else, the access will be split into two smaller accesses, resulting in significant waste of memory bandwidth. To keep our loops regular and allow correct pipelining, the first spatial block starts from a point that is $size_{halo}$ cells to the left of the input grid (Fig. 5-3). However, valid memory accesses start from the actual starting point of the grid, which is also the beginning of the first compute block. Because of this, unless $size_{halo}$ is a multiple of 512 bits, the starting memory access and all accesses after that within the same block will not be aligned. Furthermore, due to overlapping of spatial blocks, even if the start of the first compute block is 512-bit-aligned and $bsize_{\{x|y\}}$ are a multiple of 512 bits, depending on the halo size, the starting point (and hence, all accesses) in the next spatial blocks might not be 512-bit-aligned.

Based on Eq. (5-2), for first-order stencils ($rad = 1$) with single-precision floating-point values, $par_{time}$ must be a multiple of 16 for the starting point of the first compute block to be 512-bit-aligned; however, the requirement could be less restricting for higher-order stencils; e.g. a multiple of 4 is enough for fourth-order stencils. On the other hand, for the rest of the accesses to be also aligned, the distance between the starting points of each two consecutive spatial blocks must be a multiple of 512 bits; this distance, which is equal to the size of the compute block ($csize_{\{x|y\}}$), is equal to:

$$csize_{\{x|y\}} = bsize_{\{x|y\}} - 2 \times size_{halo} \tag{5-3}$$

In this case, if $bsize_{\{x|y\}}$ is divisible by 512 bits, and $size_{halo}$ is divisible by 256 bits, $csize_{\{x|y\}}$ will be divisible by 512 bits. For first-order stencils, this means that $par_{time}$ should be a multiple of 8. Finally, the dimensions of the input ($dim_{\{x|y\}}$) that are blocked (i.e. excluding the dimension that is streamed) should also be a multiple of 512 bits to allow all accesses to be 512-bit-aligned and enable highest-possible performance. To summarize, the following requirements need to be met to achieve best memory throughput, and consequently, best computational performance, in our design:

$$\begin{cases} (rad \times par_{time} \times size_{cell}) \bmod 512b = 0 \\ (bsize_{\{x|y\}} \times size_{cell}) \bmod 512b = 0 \\ (dim_{\{x|y\}} \times size_{cell}) \bmod 512b = 0 \end{cases} \tag{5-4}$$

In this case, $size_{cell}$ is the size of each cell in bytes. For single-precision floating-point grid cells, Eq. (5-4) can be simplified as:

$$\begin{cases} (rad \times par_{time}) \bmod 16 = 0 \\ bsize_{\{x|y\}} \bmod 16 = 0 \\ dim_{\{x|y\}} \bmod 16 = 0 \end{cases} \tag{5-5}$$

Among the three requirements, the second one usually holds since the spatial block size is a large power of two. The third case can also be handled by padding the rows and columns of the input by a few bytes. However, the first case will limit our parameter tuning, since it restricts maximum performance to certain values of $par_{time}$. To alleviate this issue, we pad the device buffers holding the input(s) and output of the computation by $(rad \times par_{time}) \bmod 16$ indexes. This padding guarantees that the starting point of the first compute block is always 512-bit-aligned and hence, only the requirement for the distance between the starting points of each two consecutive spatial blocks will need to be considered. Such padding effectively relaxes our requirements from Eq. (5-5) to:

$$\begin{cases} (rad \times par_{time}) \bmod 8 = 0 \\ bsize_{\{x|y\}} \bmod 16 = 0 \\ dim_{\{x|y\}} \bmod 16 = 0 \end{cases} \tag{5-6}$$

Apart from increasing our freedom for parameter tuning and allowing maximum performance for cases where all requirements are met, including up to 30% performance improvement for cases where $rad \times par_{time}$ is a multiple of 8 but not 16, the padding also improves performance for cases where the first requirement in Eq. (5-5) is not held by 10-15%.

Among the advanced compiler-assisted optimizations, we take advantage of the following optimizations in our design:

**Disabling cache:** as explained in Section 5.3.1, we disable the cache (Section 3.2.3.2) that is automatically generated by Intel FPGA SDK for OpenCL Offline Compiler since spatial blocking eliminates all redundant accesses per spatial block, and keeping the caches will only waste Block RAMs.

*autorun* **Kernels:** as explained in Section 5.3.2, we define the *compute* kernel in our design as *autorun* (Section 3.2.3.3) to be able to efficiently replicate it once per parallel time-step and achieve efficient floor-planning and high operating frequency with tens of PEs.

**Flat compilation:** we use flat compilation (Section 3.2.3.4) on our Arria 10 device to minimize the possibility of placement and routing failures with high area usage, and maximize operating frequency. In our experience, many of our best-performing kernels even failed to fit with the default PR flow.

**Target $f_{max}$ and seed sweep:** to maximize the operating frequency, and consequently, performance of our best-performing configuration for each benchmark, we take advantage of both target $f_{max}$ and seed sweeping (Section 3.2.3.5).

### 5.3.4 Support for High-order Stencils

To support first-order and high-order stencils in the same kernel, it was required that we parameterize the stencil radius in our kernel. Because of this, different parameters and code segments that relied on the stencil radius were parameterized based on this value:

- Shift register size and address for accesses to the shift register
- Comparison statements on block and index variables, and also the global index
- Boundary conditions; since this could not be efficiently realized using unrolled loops and branches, we created a code generator to generate the boundary conditions and insert them into the base kernel

## 5.4 Performance Model

Our implementation includes multiple parameters that affect its performance. Since FPGA placement and routing is a resource intensive (up to 50 GB of memory is required per Arria 10 compilation) and time consuming (up to 24 hours per Arria 10 compilation) operation, it is impossible to choose the best parameters by exhaustively searching the parameter search space. Hence, we devise a performance model that allows us to quickly choose the best performance parameters based on stencil and FPGA characteristics and minimize performance tuning time.

In our design, four parameters affect performance:

- **$bsize_{\{x|y\}}$**: The bigger the spatial block size is, the lower the ratio of the redundant to valid computation will be and hence, better performance scaling can be achieved with temporal blocking. This value is constrained by Block RAM resources on the FPGA.
- **$par_{vec}$**: A larger vector size will improve both memory and compute throughput. However, choosing a vector size larger than the value that saturates the external memory bandwidth will not improve performance any further. This value is mostly constrained by logic and DSP resources.
- **$par_{time}$**: A higher degree of temporal parallelism will improve compute throughput with a fixed amount of memory bandwidth. This value is constrained by logic, DSP and Block RAM resources. For a fixed spatial block size, since amount of redundant computation increases with respect to the degree of temporal parallelism (Eq. (5-2)), there will be a limit to the extent performance can improve with more temporal parallelism.
- **$f_{max}$**: As long as memory bandwidth is not saturated, higher operating frequency improves both memory and compute performance; however, this value is not predictable and cannot be tuned other than by sweeping the target $f_{max}$ and seed.

These parameters create multiple area trade-offs:

- **Block RAM:** The Block RAM utilization depends on $par_{time}$ and $bsize_{\{x|y\}}$.
- **DSP:** DSP utilization depends on $par_{time}$ and $par_{vec}$.
- **Logic:** Logic resources (LUTs and registers) do not generally become a bottleneck on the Intel Arria 10 device except for very high values of $par_{time}$ (100+). On Stratix V, however, logic usage can become a bottleneck due to lack of support for floating-point operations in the DSP of these devices which forces parts of each floating-point multiplication and all of each floating-point addition operation to be implemented using logic resources. Furthermore, high logic utilization (above 75%) can significantly limit operating frequency due to placement and routing congestion.

Table 5-1 shows the name and description of the parameters we use in our model. Iterative stencil computation is generally memory-bound on most hardware due to high FLOP-to-byte ratio. Hence, in our model we assume the computation is memory bound and that the latency of external memory accesses is hidden by the deep pipeline. Furthermore, even though filling and emptying the array of PEs once per iteration block incurs extra overhead since only half the memory bandwidth (either only reading or writing) is used in these periods, we ignore this overhead since the amount of data that is transferred in these phases is less than 1% of the total input size. To predict run time and performance, we need to accurately count the total amount

of data transferred between the FPGA and its external memory and model the external memory bandwidth.

**Table 5-1 Model Parameters**

| Parameter | Description | Unit |
|---|---|---|
| $rad$ | Stencil radius | Cells |
| $par_{vec}$ | Compute vector size (width) | N/A |
| $par_{time}$ | Degree of temporal parallelism (Number of parallel time-steps) | N/A |
| $f_{max}$ | Kernel operating frequency | Hz |
| $f_{mem}$ | External memory operating frequency | Hz |
| $size_{cell}$ | Size of each grid cell | Bytes |
| $size_{input}$ | Number of cells in input grid | Cells |
| $size_{bus}$ | Width of the bus to each external memory bank | Bits |
| $num_{read}$ | External memory reads per cell update | N/A |
| $num_{write}$ | External memory writes per cell update | N/A |
| $num_{acc}$ | External memory accesses per cell update | N/A |
| $num_{bank}$ | Number of external memory banks | N/A |
| $bsize_{\{x|y\}}$ | Size of spatial block in each dimension | Cells |
| $csize_{\{x|y\}}$ | Size of compute block in each dimension | Cells |
| $dim_{\{x|y|z\}}$ | Input size in each dimension | Cells |
| $bnum_{\{x|y\}}$ | Number of spatial/compute blocks in each dimension | Cells |
| $trav_{\{x|y\}}$ | Number of traversed cells in each dimension | Cells |
| $th_{mem}$ | Effective/utilized memory throughput | GB/s[1] |
| $th_{max}$ | Peak memory throughput | GB/s |
| $iter$ | Number of iterations | N/A |

External memory throughput depends on multiple parameters. One parameter is the vector width ($par_{vec}$) since it determines the size of the memory ports between the kernel and the memory interface. Another parameter is the number of accesses to external memory which determines the number of memory ports. The next parameter is the kernel operating frequency ($f_{max}$) which determines the rate of memory requests per second. In case of both the Intel Stratix V and Arria 10 FPGAs, the memory controller operates at $1/8$ the clock of the memory modules installed on the boards (200 and 266 MHz for Stratix V and Arria 10, respectively). Furthermore, each external memory bank is connected to the FPGAs via a 64-bit bus ($size_{bus}$). This means that for a kernel running at the same operating frequency as the FPGA memory controller, one 512-bit access per cycle per external memory bank is required to saturate the external memory bandwidth. However, since a set of FIFOs are also implemented between the kernel and the memory interface to allow them to run at different operating frequencies, if the kernel is running at a higher frequency than the memory controller, the memory bandwidth can still improve or be saturated with smaller accesses. Considering these points, external memory bandwidth can be modelled as follows:

$$num_{acc} = num_{read} + num_{write} \tag{5-7}$$

$$th_{max} = \frac{num_{banks} \times size_{bus} \times f_{mem}}{8 \times 10^9} \tag{5-8}$$

$$th_{mem} = \min\left(th_{max}, \frac{f_{max} \times par_{vec} \times num_{acc} \times size_{cell}}{10^9}\right) \tag{5-9}$$

---

[1] All throughput numbers reported in this document are in GB/s = $10^9$ B/s, and not GiB/s = $2^{30}$ B/s

Here, since we employ spatial blocking to eliminate all redundant external memory accesses per spatial block, $num_{read}$ and $num_{write}$ do not depend on stencil shape or size and are equal to the number of input and output buffers that are accessed once per iteration.

To calculate the amount of data transferred between the FPGA and its external memory, first we calculate the total number of cells that are processed, including the redundant and out-of-bound ones. As seen in Fig. 5-3, with overlapped blocking, the compute blocks will be consecutive in the last PE. Hence, each dimension of the input is traversed up to an index where the traversed size is a multiple of the compute block size. Based on this, the number of spatial/compute blocks in each dimension can be calculated as:

$$bnum_{\{x|y\}} = \left\lceil \frac{dim_{\{x|y\}}}{csize_{\{x|y\}}} \right\rceil \tag{5-10}$$

Consequently, the total number of processed cells is:

$$t_{cell} = \begin{cases} bnum_x \times bsize_x \times dim_y, & 2D \\ bnum_x \times bsize_x \times bnum_y \times bsize_y \times dim_z, & 3D \end{cases} \tag{5-11}$$

Out-of-bound external memory reads and writes are skipped in our implementation, and redundant writes to halo regions are also avoided. Based on this, the number of cells that are read from external memory will be equal to $t_{cell}$ minus the out-of-bound reads multiplied by the number of reads per cell update (i.e. number of input buffers). The amount of data that is written to external memory per output buffer will instead be exactly equal to the input size ($t_{write} = size_{input}$). The number of reads for 2D and 3D stencils can be calculated as follows:

$$trav_{\{x|y\}} = bnum_{\{x|y\}} \times csize_{\{x|y\}} + 2 \times size_{halo} \tag{5-12}$$

$$t_{read_{2D}} = \left(t_{cell} - (trav_x - dim_x) \times dim_y\right) \times num_{read} \tag{5-13}$$

$$t_{read\_3D} = t_{cell} - (trav_x \times trav_y - dim_x \times dim_y) \times dim_z$$
$$- \Big((bnum_x - 1 + bnum_y - 1) \times (2 \times size_{halo}) \times size_{halo}$$
$$+ \big((trav_x - size_{halo} - dim_x) \times (bnum_y - 1) \quad (5\text{-}14)$$
$$+ (trav_y - size_{halo} - dim_y) \times (bnum_x - 1)\big)$$
$$\times (2 \times size_{halo})\Big) \times dim_z$$

Finally, we calculate run time (s) and computation throughput (GB/s) as:

$$run\_time = \frac{\left\lceil \frac{iter}{par_{time}} \right\rceil \times (t_{read} + t_{write}) \times size_{cell}}{10^9 \times th_{mem}} \quad (5\text{-}15)$$

$$throughput = \frac{num_{acc} \times size_{input} \times size_{cell} \times iter}{10^9 \times run\_time} \quad (5\text{-}16)$$

Throughput from Eq. (5-16) can be converted to compute performance (GFLOP/s) and number of cells processed per second (GCell/s) by using the byte-to-FLOP and byte per cell update ratios of the stencil.

We can also calculate the ratio of redundant memory accesses to total, which shows how much of the external memory bandwidth is wasted due to overlapped blocking, as follows:

$$redundancy = \frac{num_{read} \times t_{read} + num_{write} \times t_{write}}{num_{acc} \times size_{input}} \quad (5\text{-}17)$$

## 5.5 Methodology

### 5.5.1 Benchmarks

For our evaluation, we use Hotspot 2D and 3D from the Rodinia benchmark suite [13] and Diffusion 2D and 3D from [49]. All of these stencils are first-order stencils. To evaluate high-order stencils, we extend Diffusion 2D and 3D to higher orders up to fourth as a representative of high-order star-shaped stencils.

Table 5-2 shows the equation and characteristics of our evaluated stencils. In this table, $f_x$ refers to the value of the cell in position $x$, $c_x$ refers to the coefficient of this cell, *rad* refers to stencil order, $x$ is a member of the set {Center, West, East, South, North, Below, Above}, and the set {$x,i$} refers to the $i^{th}$ neighbor cell in the direction of $x$. All our stencils use single-precision floating-point values. All values except $TEMP_{AMB}$ which is a compile-time constant, are passed to the kernel as variables and can be changed at run-time without kernel recompilation.

**Table 5-2 Stencil Characteristics**

| Benchmark | Equation | FLOP Per Cell Update | Bytes Per Cell Update | $\frac{\text{Byte}}{\text{FLOP}}$ | $\frac{\text{FLOP}}{\text{Byte}}$ |
|---|---|---|---|---|---|
| Diffusion 2D | $c_c \times f_c + \sum_{i=1}^{rad}(c_w \times f_{w,i} + c_e \times f_{e,i} + c_s \times f_{s,i} + c_n \times f_{n,i})$ | $8 \times rad + 1$ | 8 | $\xRightarrow{rad=1} 0.889$ $\xRightarrow{rad=2} 0.470$ $\xRightarrow{rad=3} 0.320$ $\xRightarrow{rad=4} 0.242$ | $\xRightarrow{rad=1} 1.125$ $\xRightarrow{rad=2} 2.125$ $\xRightarrow{rad=3} 3.125$ $\xRightarrow{rad=4} 4.125$ |
| Diffusion 3D | $c_c \times f_c + \sum_{i=1}^{rad}(c_w \times f_{w,i} + c_e \times f_{e,i} + c_s \times f_{s,i} + c_n \times f_{n,i} + c_b \times f_{b,i} + c_a \times f_{a,i})$ | $12 \times rad + 1$ | 8 | $\xRightarrow{rad=1} 0.615$ $\xRightarrow{rad=2} 0.320$ $\xRightarrow{rad=3} 0.216$ $\xRightarrow{rad=4} 0.163$ | $\xRightarrow{rad=1} 1.625$ $\xRightarrow{rad=2} 3.124$ $\xRightarrow{rad=3} 4.625$ $\xRightarrow{rad=4} 6.125$ |
| Hotspot 2D | $f_c + sdc \times (power_c + (f_n + f_s - 2.0 \times f_c) \times R_{y\_1} + (f_e + f_w - 2.0 \times f_c) \times R_{x\_1} + (TEMP_{AMB} - f_c) \times R_{z\_1})$ | 15 | 12 | 0.800 | 1.250 |
| Hotspot 3D | $c_c \times f_c + c_n \times f_n + c_s \times f_s + c_e \times f_e + c_w \times f_w + c_a \times f_a + c_b \times f_b + sdc \times power_c + c_a \times TEMP_{AMB}$ | 17 | 12 | 0.706 | 1.417 |

Among the first-order stencils, Hotspot has higher arithmetic intensity compared to Diffusion, and has two input buffers instead of one. This requires an extra shift register to cache the second input (*power* input) of Hotspot; though this buffer will be smaller than the one used for the main input since only the *Center* cell is required to be cached. For the high-order implementation of Diffusion stencils, the coefficient for all the neighbors in a given direction is fixed; however, since we do not allow reordering of floating-point operations, the coefficient is not shared and hence, $4 \times rad + 1$ and $6 \times rad + 1$ floating-point multiplications (FMUL) and $4 \times rad$ and $6 \times rad$ floating-point addition operations (FADD) are required per cell update, for Diffusion 2D and 3D, respectively. Optimizing this implementation is equal to optimizing the worst-case scenario where all the coefficients for all of the neighboring cells are different. Finally, all out-of-bound neighbors for boundary cells in our implementation fall back on the boundary cell itself which requires complex branches to implement.

The "Byte per Cell Update" column in Table 5-2 shows the amount of data that needs to be read from or written to the FPGA external memory for each cell update, with the assumption of full spatial reuse (no redundant memory accesses) but no temporal blocking. This value is equal to $num_{acc} \times size_{cell}$. The byte-to-FLOP ratio of the stencils shows that first-order stencils are highly memory-bound; however, the ratio decreases for higher-order stencils, making such stencils less memory-bound than first-order ones.

### 5.5.2 Hardware Setup

We evaluate our stencils on the same FPGAs as the ones used in the previous chapter (Section 4.2.3). On top of that, we will project the performance of our stencils for the upcoming

Stratix 10 MX 2100 and GX 2800 devices. The characteristics of these FPGAs are shown in Table 5-3. The yellow rows show unreleased boards/devices.

**Table 5-3 FPGA Device Characteristics**

| Board | FPGA | ALM | Register (K) | M20K (Blocks\|Mb) | DSP | External Memory |
|---|---|---|---|---|---|---|
| Terasic DE5-Net | Stratix V GX A7 | 234,720 | 939 | 2,560\| 50 | 256 | 2x DDR3-1600 |
| Nallatech 385A | Arria 10 GX 1150 | 427,200 | 1,709 | 2,713\| 53 | 1,518 | 2x DDR4-2133 |
| Bittware S10VM4 [74] | Stratix 10 MX 2100 | 702,720 | 2,811 | 6,847\|134 | 3,960 | 4-tile HBM2 |
| Nallatech 520C [75] | Stratix 10 GX 2800 | 933,120 | 3,732 | 11,721\|229 | 5,760 | 4x DDR4-2400 |

For first-order 2D stencil comparison, we compare our implementation of Diffusion 2D on FPGAs with the results published recently in [50]. Since the Jacobi 2D stencil used in this work uses shared coefficients, we estimate its performance if the coefficients had not been shared by scaling their reported performance by the difference in the FLOP per cell update of the two cases. Specifically, their stencil has a FLOP per cell update of 5 [46], while without sharing coefficients similar to our stencil, this value will increase to 9. Hence, we scale their results by a ratio of $9/5$. Furthermore, by analyzing their results we can see that the performance difference between their two evaluated GPUs is very close to the ratio of the external memory bandwidth of these GPUs. Hence, to estimate how their implementation performs on newer hardware, we linearly extrapolate their results for newer GPUs based on the ratio of improvement in external memory bandwidth. Since their implementation is not available to public at the time of writing this thesis, it is not possible to directly measure performance on such hardware. We also implement Diffusion 2D using the state-of the-art YASK framework [48] for evaluation on a 12-core Intel Xeon E5-2650 v4 CPU, and a 64-core Intel Xeon Phi 7210F processor. The Xeon processor is accompanied by quad-channel DDR4 memory operating at 2400 MHz. The Xeon Phi processor is set to operate in flat mode and *numactl* is used to set the faster MCDRAM memory as the preferred memory. All hyperthreads are used on both processors. It is worth noting that boundary conditions in YASK are different from our implementation. In this framework, the allocated grid is bigger than the input grid so that out-of-bound neighbors can also be read from external memory. This results in extra memory accesses, but allows correct vectorization on grid boundaries. In our implementation, all out-of-bound neighbors fall back on the grid cell that is on the border, instead, which avoids some extra external memory accesses at the cost of extra area usage for implementing branches.

For first-order 3D stencil comparison, we compare our implementation of Diffusion 3D with the highly-optimized GPU implementation from [49] on multiple high-end NVIDIA GPUs. We use the publicly-available code from this work to directly measure performance on our GPUs. To keep the comparison fair, we disable ECC on all the GPUs that support it. Similar to the 2D case, we also implement Diffusion 3D using the YASK framework for evaluation on the Xeon and Xeon Phi processors.

For the rest of the first order stencils, we avoid comparison with CPUs and GPUs due to lack of a highly-optimized implementation. Even though Rodinia has OpenMP and CUDA implementations for both Hotspot 2D and 3D, these implementations are not well-optimized to the point that our FPGA implementation on Arria 10 achieves over twice higher performance compared to the NVIDIA Tesla P100 GPU. We used these suboptimal implementations for comparison in the previous chapter because our FPGA implementations in that chapter had a similar level of optimization; however, the level of optimization on the FPGAs in this chapter is beyond those implementations. Hence, it is not possible to perform a fair comparison between our FPGA implementation in this chapter and Rodinia's OpenMP and CUDA implementations anymore.

For high-order 2D stencil comparison, we could not find a general implementation on GPUs that could be used for comparison. For evaluation on Xeon and Xeon Phi, we implemented Diffusion 2D using YASK. For high-order 3D stencil comparison, we compare our results from high-order Diffusion 3D with the implementation from [71] on GPUs. This work also uses shared coefficients and hence, the FLOP per cell update for their stencil is lower than ours. Similar to the case of the first-order 2D stencil, we assume their reported cell updates per second will be the same if coefficients were not shared and estimate compute performance for the stencil without shared coefficients by adjusting its FLOP per cell update ratio. We use the best results from this work that were obtained on an NVIDIA GTX 580. Furthermore, we linearly extrapolate their results for newer GPUs based on the ratio of improvement in the theoretical external memory bandwidth of these devices compared to GTX 580. We also implement the same stencils using YASK for comparison with Xeon and Xeon Phi.

Table 5-4 shows the characteristics of all the hardware used in our evaluation. Peak compute performance is for single-precision floating-point operations with each FMA operation being counted as two FLOPs. The byte-to-FLOP ratio shows the ratio of the external memory bandwidth of the device to its compute performance.

In stencil computation, if temporal blocking is *not* used, computation will be memory-bound if the byte-to-FLOP ratio of the device is lower than the byte-to-FLOP ratio of the stencil. Comparing the byte-to-FLOP ratios from Table 5-2 and 5-4 shows that *without* temporal blocking, all of our evaluated stencils will be memory-bound on all of our evaluated hardware, even for the less-memory-bound high-order stencils. Comparing the byte-to-FLOP ratios of the different hardware in Table 5-4 shows that the Arria 10 GX 1150 and the Stratix 10 GX 2800 platforms are by far the most *bandwidth-starved* hardware. This implies that these platforms will be the least suitable platforms for stencil computation; however, we will show that due to effectiveness of temporal blocking on FPGAs, it is possible to overcome the memory-bound nature of stencil computation on these platforms and achieve comparable performance to that of devices with much higher byte-to-FLOP ratio.

Table 5-4 Hardware Characteristics

| Type | Device | Peak Compute Performance (GFLOP/s) | Peak Memory Bandwidth (GB/s) | $\frac{\text{Byte}}{\text{FLOP}}$ | $\frac{\text{FLOP}}{\text{Byte}}$ | On-chip[1] Memory (MiB) | Transistors (Billion) | Node (nm) | TDP (Watt) | Year |
|---|---|---|---|---|---|---|---|---|---|---|
| FPGA | Stratix V GX A7 | ~200 | 26.5 | 0.133 | 7.5 | 6.3 + 0.9 = 7.2 | 3.8 | 28 | 40 | 2011 |
| FPGA | Arria 10 GX 1150 | 1,450[2] | 34.1 | 0.024 | 42.5 | 6.6 + 1.6 = 8.2 | 5.3 | 20 | 70 | 2014 |
| FPGA | Stratix 10 MX 2100 | 5,940[2] | 512.0 | 0.081 | 11.6 | 16.7 + 1.4 = 18.1 | ~20 | 14 | 150[3] | 2018 |
| FPGA | Stratix 10 GX 2800 | 8,640[2] | 76.8 | 0.008 | 112.5 | 28.6 + 1.9 = 30.5 | ~30 | 14 | 200[3] | 2018 |
| CPU | Xeon E5-2650 v4 | 700 | 76.8 | 0.110 | 9.1 | 30 + 3 + 0.4 = 33.4 | 4.7 | 14 | 105 | 2016 |
| CPU | Xeon Phi 7210F | 5,325 | 400.0 | 0.075 | 13.3 | 32 + 2 = 34 | 8 | 14 | 235 | 2016 |
| GPU | GTX 580 | 1,600 | 192.4 | 0.122 | 8.2 | 1 + 2 + 0.8 = 3.8 | 3 | 40 | 244 | 2010 |
| GPU | Tesla K40c | 4,300 | 288.4 | 0.067 | 14.9 | 1 + 3.8 + 1.5 = 6.3 | 7.1 | 28 | 235 | 2013 |
| GPU | GTX 980 | 5,000 | 224.4 | 0.045 | 22.2 | 1.5 + 4 + 2 = 7.5 | 5.2 | 28 | 165 | 2014 |
| GPU | GTX Titan X | 6,700 | 336.6 | 0.050 | 19.9 | 2.2 + 6 + 3 = 11.2 | 8 | 28 | 250 | 2015 |
| GPU | GTX 980 Ti | 6,900 | 336.6 | 0.049 | 20.5 | 2 + 5.5 + 3 = 10.5 | 8 | 28 | 275 | 2015 |
| GPU | Tesla P100 PCI-E | 9,500 | 720.9 | 0.078 | 12.9 | 3.5 + 14 + 4 = 31.5 | 15.3 | 16 | 250 | 2016 |
| GPU | Tesla V100 SXM2 | 15,700 | 897.0 | 0.060 | 16.554 | 7.5 + 20 + 6 = 33.5 | 21.1 | 12 | 300 | 2017 |

## 5.5.3 Software Setup

All of our systems use CentOS v6 or v7 as operating system. We use GCC v5.4.0 for compiling our OpenCL host codes, and Intel Quartus and FPGA SDK for OpenCL Offline Compiler v16.1.2 for compiling the kernel codes. We avoided newer versions of Quartus (v17.0, v17.1 and v18.0) since they reliably resulted in lower performance (20-30% lower) and higher area utilization (5-10% more Block RAMs) for the same kernels compared to v16.1.2. On the Xeon and Xeon Phi processors, we use Intel C/C++ Compiler v2018.1 and YASK's built-in compilation settings. For the GPUs, we use CUDA v8.0 for the older ones (pre-Tesla P100) and CUDA v9.0 for the newer ones (Tesla P100 and V100) with "-arch sm_35 -O3" flags.

---

[1] FPGA: M20K + MLAB, Xeon: L3 + L2 + L1D, Xeon Phi: L2 + L1D, GPU: Shared/L1 + Register + L2

[2] Assuming full DSP utilization with FMA operations running at 480 MHz for Arria 10 and 750 MHz for Stratix 10 MX 2100 and GX 2800

[3] Stratix 10 GX 2800 TDP has been estimated based on the results reported in [76]. Stratix 10 MX 2100 TDP has been scaled based on its smaller size.

### 5.5.4 Performance and Power Measurement

Similar to our evaluation in the previous chapter, we only time the kernel computation and ignore the initialization and host to device transfers for all platforms. For the FPGA and GPU, we use the same high-precision timer introduced in Section 4.2.4 and for the Xeon and Xeon Phi platforms, we use timing values reported by YASK's built-in timer. In all cases, all runs are repeated five times and all values are averaged.

For power measurement, similar to the previous chapter, we sample the on-board sensors for the Arria 10 FPGA and all the GPUs once every 10 milliseconds. For the Stratix V FPGA, we follow a similar approach as the previous chapter, with the exception that since our implementation in this chapter is better optimized compared to the previous chapter, we increase the FPGA toggle rate to 25% for estimation (default is 12.5%). For the Xeon and Xeon Phi processors, we instrument YASK with our power measurement function based on the MSR driver [30], which starts and ends with YASK's built-in timer. For estimating the power usage of the first-order 2D implementation from [50] and high-order 3D implementation from [71] on GPUs, we use the same ratio of measured power to TDP as what we measure in practice for the first-order 3D implementation from [49] (~75%). For estimating the power usage of the Stratix 10 FPGAs, we use the results reported in [76].

We calculate performance in number of cells updated per second (GCell/s) as follows:

$$\frac{run\_time}{size_{input} \times iter} \tag{5-18}$$

We calculate computation performance (GFLOP/s) and throughput (GB/s) by multiplying the GCell/s value by the FLOP and byte per cell update values of the stencil (Table 5-2), respectively. In this case, redundant computation and memory accesses are *not* included in the reported performance values.

### 5.5.5 Benchmark Settings

For the FPGA benchmarks, to minimize out-of-bound computation in the last spatial block and show the maximum potential of these devices, we choose $dim_{\{x|y\}}$ to be a multiple of $csize_{\{x|y\}}$. For 2D stencils on Stratix V and Arria 10, the input dimensions are the closest multiple of $csize_x$ to 16000 cells, and for 3D stencils, they are between 490 and 850 cells. At least 1 GB of FPGA external memory is used in every case. Number of iterations is also set to 1000 in all cases. This results in a minimum run time of 3 seconds for 2D, and 11 seconds for the 3D stencils. For performance projection on Stratix 10, we increase the size of the input dimensions to the closest multiple of $csize_x$ to 32000 cells for 2D, and the closest multiple of $csize_{\{x|y\}}$ to 2000 cells for 3D, and set the number of iterations to 5000. The exact dimension sizes for every case are reported in Section 5.7. We observed less than 50 ms of variation in our FPGA executions.

For Xeon and Xeon Phi, we experimentally found the best-performing input sizes by trying multiple different values. On the Xeon processor, the best results were obtained with an input size of 16384×16384 and 768×768×768, for 2D and 3D stencils, respectively. For the Xeon Phi processor, input sizes of 32768×32768 and 768×768×768 achieved the best performance.

All benchmarks used 1000 iterations, for a minimum run time of 53 seconds on the Xeon, and 20 seconds on the Xeon Phi processor.

For the GPU implementation from [49], we again tried multiple input sizes and used the best performance that was achieved with an input size of 512×512×512 for Diffusion 3D. It is worth noting that this implementation does not support input sizes that are not a multiple of the spatial block size.

## 5.6 Performance Tuning

### 5.6.1 Xeon and Xeon Phi

The YASK framework includes a built-in performance tuner that runs automatically at the beginning of execution and chooses the best block size based on input characteristics and the given hardware before running the actual benchmark. We use the standard flow of this framework and allow it to choose the best block size for each benchmark run. YASK also supports temporal blocking; however, after trying multiple cases, we could not achieve a meaningful performance improvement with temporal blocking on any of the hardware. Based on the author's recent work [77], temporal blocking in YASK is useful only when a Xeon Phi processor is set to *cache* mode and an input that is larger than the MCDRAM is used. In this case, temporal blocking will allow performance to reach a level close (but not higher) to the case where all the input can fit on the MCDRAM, minimizing the negative effect of the slower but larger DDR memory. Since in our benchmarks the inputs completely fit in the MCDRAM, maximum performance can be achieved out of the box and enabling temporal blocking in YASK does not improve the performance any further.

### 5.6.2 GPU

The implementation from [49] uses a fixed degree of temporal parallelism of two. However, the block sizes and the number of thread blocks in the *z* dimension can be tuned in this implementation. For every one of our evaluated GPUs, we separately tuned these parameters and chose the best-performing one. On all GPUs, the best block size was 32×8. However, the best number of threads blocks varied between 1 and 16 depending on the GPU.

### 5.6.3 FPGA

To tune the performance parameters of our FPGA design, we first calculate the total degree of parallelism based on the number of DSPs on the FPGA and the number of DSPs required for one cell update. The number of DSPs required for one cell update depends on the computational characteristics of the stencil and can be extracted from the compiler's area report. This report is generated in a few minutes when the first stage of OpenCL compilation is completed. This number can also be directly calculated based on the stencil equation; however, sometimes certain restrictions in the way the pipeline is implemented by the compiler and the ordering of the operations might increase the number of required DSPs.

Table 5-5 shows the expected number of DSPs for each cell update based on the stencil equation, and the number implemented by the compiler on the Arria 10 FPGA. Each DSP in this FPGA can support one FADD, one FMUL or one FMA operation. Furthermore, the compiler does not use DSPs to implement multiplications between a floating-point number and a constant value. For Diffusion 2D and 3D, all multiplications and their succeeding additions can be fused into an FMA operation, and one FMUL will be required for the last multiplication in the chain that is not followed by an addition. For Hotspot 2D and 3D, the operations are more complex and the implemented DSP usage is higher than what we expect.

**Table 5-5 Number of DSPs Required for One Cell Update on Arria 10**

| Benchmark | FADD | FMUL | FMA | Expected DSP Usage | Implemented DSP Usage |
|---|---|---|---|---|---|
| Diffusion 2D | 0 | 1 | $4 \times rad$ | $par_{time} \times par_{vec} \times (4 \times rad + 1)$ | $par_{time} \times par_{vec} \times (4 \times rad + 1)$ |
| Diffusion 3D | 0 | 1 | $6 \times rad$ | $par_{time} \times par_{vec} \times (6 \times rad + 1)$ | $par_{time} \times par_{vec} \times (6 \times rad + 1)$ |
| Hotspot 2D | 5 | 0 | 4 | $par_{time} \times par_{vec} \times 9$ | $par_{time} \times par_{vec} \times 10$ |
| Hotspot 3D | 0 | 0 | 8 | $par_{time} \times par_{vec} \times 8$ | $par_{time} \times (par_{vec} \times 9 + 1)$ |

In the next step, based on the number of DSPs available on the FPGA, we can determine the total degree of parallelism. For example, for the Intel Arria 10 GX 1150 device which has 1518 DSPs we have:

$$par_{total} = \begin{cases} \left\lfloor \dfrac{1518}{c_{2D} + 4 \times rad + 1} \right\rfloor & \text{Diffusion 2D} \\ \left\lfloor \dfrac{1518}{c_{3D} + 6 \times rad + 1} \right\rfloor & \text{Diffusion 3D} \end{cases} \quad (5\text{-}19)$$

In Eq. (5-19), $c_{2D}$ and $c_{3D}$ denote the number of DSPs required in the *read* and *write* kernels for address calculation, which are equal to 4 and 8, respectively. In the next step we have:

$$par_{time} \times par_{vec} \leq par_{total} \quad (5\text{-}20)$$

To choose the pairs of $par_{time}$ and $par_{vec}$ that satisfy Eq. (5-20), we need to consider the following restrictions:

- $par_{vec}$ must be a power of two since regardless of the vector size, the compiler always infers memory ports with a width that is a power of two words and the extra words will be masked out in the kernel, resulting in significant waste of memory bandwidth.
- Values of $par_{time}$ that satisfy Eq. (5-6) are preferred.

In the next step, we need to determine viable configurations for the spatial block. We consider the following restrictions:

- $bsize_{\{x|y\}}$ are restricted to powers of two. This allows updating the block variables using an efficient *bit masking* operation with very low area overhead. Other values can also be supported using conditional branching, at the cost of 5-20 MHz of lower operating frequency and slightly higher area overhead.
- $bsize_x$ must be divisible by $par_{vec}$ to keep the computation loop regular.

Modelling Block RAM utilization is not straightforward since the exact way the compiler replicates shift registers, interleaves data, and allocates ports is unknown. Furthermore, the FPGA *mapping* process involves complex Block RAM packing optimizations and mapping of smaller buffers to distributed memory – a process that is near-impossible to model. Hence, we experimentally find the range of $bsize_{\{x|y\}}$ values that could fit on the FPGAs by performing a few example compilations and taking advantage of the resource estimation provided by Intel FPGA SDK for OpenCL Offline Compiler. Based on our findings, $bsize_x = 4096$ is suitable for every case of the 2D stencils on Stratix V and Arria 10, while the block size for 3D stencils can vary between 128×128 and 512×512 depending on *rad* and $par_{time}$.

In the next step, we insert all candidate configurations in our model, and with the assumption of a fixed operating frequency for a fixed stencil, extract the top two or three configurations that are expected to achieve the highest performance. Then, we place and route these configurations and measure their performance on the board. To eliminate the effect of variabilities in $f_{max}$, we normalize the measured performance values for a fixed operating frequency to choose the best-performing configuration. Finally, we sweep the target $f_{max}$ and seed (Section 3.2.3.5) on the chosen configuration to maximize its operating frequency.

For high-order stencils, from Eq. (5-1) and Table 5-5 we can see that both Block RAM utilization (shift register size) and DSP usage will increase relative to the increase in stencil radius. Intuitively, to optimize performance parameters for high-order stencils, one direct solution would be to just divide the $par_{time}$ value of the best configuration for the first-order stencil by the radius of the high-order stencil. In this case, the DSP and Block RAM utilization is expected to stay roughly the same between the different stencil orders. Furthermore, number of cells updated per second (GCell/s) will drop relative to the stencil radius in this case, while the compute performance (GFLOP/s) will stay relatively constant since FLOP per cell update increases relative to the stencil radius.

## 5.7 Results

### 5.7.1 FPGA Results

**5.7.1.1 First-order stencils:**

Table 5-6 shows the configuration and performance of the first order stencils we evaluated, on Stratix V and Arria 10. The estimated performance is calculated based on our performance model, described in Section 5.4, and adjusted for the post-place-and-route operating frequency of the kernel. For each stencil on each FPGA, the highest estimated performance is marked in yellow. Configurations that do not satisfy the requirement on $par_{time}$ for best memory access

alignment (Eq. (5-6)) are marked with blue hachures. In cases where such configuration is chosen as the fastest by our model, the estimated performance is marked in hachured yellow, while the best-performing configuration that does satisfy this requirement is marked in solid yellow. Furthermore, the highest performance measured on each board for each stencil is marked in green, and the resource bottleneck for this configuration is marked in red. Since the OpenCL flow uses the maximum possible $f_{max}$ that meets timing and can be generated by the PLLs on the FPGA, the $f_{max}$ can be an irregular value. Model accuracy also refers to the ratio of the performance measured on the board, to the performance estimated by our model. The hachured cells in this column show cases where accuracy could be potentially inflated since the model assume memory bandwidth is saturated with the associated configurations (Eq. (5-9)), while bandwidth is not saturated in practice. These cases will be further explained in Section 5.7.2.

**Table 5-6 Configuration and Performance of First-order Stencils on FPGAs**

| Benchmark | Device | bsize | $par_{time}$ | $par_{vec}$ | Input Size | Estimated Perf. (GB/s) | Measured Perf. (GB/s\|GFLOP/s\|GCell/s) | $f_{max}$ (MHz) | Logic | Memory (Bits\|Blocks) | DSP | Power (Watt) | Model Accuracy |
|---|---|---|---|---|---|---|---|---|---|---|---|---|---|
| Diffusion 2D | S-V | 4096 | 6 | 8 | 16336×16336 | 116.141 | 87.616\|098.568\|10.952 | 303.39 | 61% | 9%\| 33% | 95% | 26.517 | 75.4% |
| | | 4096 | 12 | 4 | 16288×16288 | 115.360 | 100.505\|113.068\|12.563 | 303.49 | 64% | 14%\| 40% | 95% | 27.889 | 87.1% |
| | | 4096 | 24 | 2 | 16192×16192 | 110.894 | 96.257\|108.289\|12.032 | 292.39 | 71% | 22%\| 52% | 95% | 30.491 | 86.8% |
| | A-10 | 4096 | 36 | 8 | 16096×16096 | 766.918 | 662.655\|745.487\|82.832 | 337.78 | 56% | 38%\| 83% | 95% | 65.516 | 86.4% |
| | | 4096 | 72 | 4 | 15808×15808 | 690.137 | 589.604\|663.305\|73.701 | 306.06 | 70% | 65%\|100% | 95% | 64.245 | 85.4% |
| Hitspot 2D | S-V | 4096 | 6 | 8 | 16336×16336 | 153.068 | 110.426\|138.033\| 9.202 | 272.47 | 91% | 13%\| 43% | 77% | 33.654 | 72.1% |
| | | 4096 | 12 | 4 | 16288×16288 | 131.977 | 115.081\|143.851\| 9.590 | 231.64 | 95% | 21%\| 53% | 77% | 36.103 | 87.2% |
| | A-10 | 4096 | 18 | 8 | 16240×16240 | 543.622 | 406.091\|507.614\|33.841 | 318.52 | 44% | 30%\| 46% | 95% | 46.218 | 74.7% |
| | | 4096 | 36 | 4 | 16096×16096 | 566.361 | 490.599\|613.249\|40.883 | 333.33 | 46% | 53%\| 86% | 95% | 50.349 | 86.6% |
| | | 4096 | 72 | 2 | 15808×15808 | 535.303 | 459.483\|574.354\|38.290 | 317.95 | 67% | 90%\|100% | 95% | 53.209 | 85.8% |
| Diffusion 3D | Stratix V | 512×256 | 4 | 8 | 504×744×504 | 64.874 | 54.457\|088.493\| 6.807 | 256.14 | 64% | 68%\|100% | 91% | 32.397 | 83.9% |
| | | 256×256 | 4 | 8 | 744×744×744 | 74.194 | 62.105\|100.921\| 7.763 | 296.12 | 60% | 36%\| 67% | 91% | 29.379 | 83.7% |
| | | 256×256 | 5 | 8 | 738×738×738 | 60.533 | 40.939\| 66.526\| 5.117 | 194.36 | 73% | 44%\| 81% | 100% | 23.316 | 67.6% |
| | Arria 10 | 256×256 | 12 | 16 | 696×696×696 | 378.345 | 232.378\|377.614\|29.047 | 285.71 | 60% | 94%\|100% | 89% | 64.409 | 61.4% |
| | | 256×128 | 20 | 8 | 648×704×648 | 298.799 | 194.321\|315.772\|24.290 | 300.00 | 50% | 81%\|100% | 74% | 63.637 | 65.0% |
| | | 256×128 | 24 | 8 | 832×720×832 | 326.680 | 202.701\|329.389\|25.338 | 300.00 | 70% | 94%\|100% | 89% | 72.432 | 62.0% |
| Hotspot 3D | S-V | 256×256 | 4 | 8 | 496×496×496 | 97.522 | 65.844\| 93.279\| 5.487 | 259.47 | 80% | 68%\|100% | 100% | 37.044 | 67.5% |
| | | 256×128 | 8 | 4 | 720×560×720 | 91.077 | 72.355\|102.503\| 6.030 | 263.08 | 84% | 68%\|100% | 100% | 37.972 | 79.4% |
| | A-10 | 256×128 | 8 | 16 | 720×560×720 | 245.569 | 193.334\|273.890\|16.111 | 250.98 | 46% | 67%\|100% | 77% | 53.450 | 78.7% |
| | | 256×128 | 10 | 16 | 708×540×708 | 298.144 | 206.387\|292.382\|17.199 | 261.91 | 60% | 81%\|100% | 96% | 59.970 | 69.2% |
| | | 128×128 | 20 | 8 | 528×528×528 | 373.169 | 232.858\|329.882\|19.405 | 311.11 | 63% | 81%\|100% | 97% | 69.573 | 62.4% |

Based on the results, we achieve two times or higher throughput (GB/s) for 2D stencils, versus 3D, on both FPGAs. This difference is expected since 3D stencils require one extra dimension to be blocked and hence, the spatial block size ($bsize_{\{x|y\}}$) will be smaller, increasing the amount of redundant memory accesses with temporal blocking and consequently, lowering performance scalability with temporal blocking. For 2D stencils, however, since $bsize_x$ is relatively large, redundancy is minimized and we can scale performance up to tens of parallel time steps and achieve close-to-linear scaling with temporal parallelism. This difference brings us to a very important conclusion: **For 3D stencils it is better to spend FPGA resources to support a larger vector size, rather than more temporal parallelism, since scaling with temporal parallelism has high overhead due to small block size and large amount of redundant memory accesses, while better scaling can be achieved with vectorization. For 2D stencils, however, it is more efficient to spend FPGA resources on increasing temporal parallelism, rather than vector size. This is due to the fact that the latter achieves close-to-linear performance scaling due to large block size, while performance scaling with the former depends on the behavior of the memory controller and in our experience, scaling is sub-linear except for very small vector sizes (up to four).** Still, higher degree of temporal parallelism will result in higher logic utilization and consequently, more routing complications and lower $f_{max}$. Because of this, using the highest $par_{time}$ with a $par_{vec} = 1$ will not necessarily result in the highest performance for 2D stencils.

For the 2D stencils on Stratix V, Hotspot 2D achieves higher throughput (GB/s) than Diffusion 2D despite lower $par_{time}$. This is because the higher $num_{acc}$ in Hotspot allows better utilization of the memory bandwidth with the narrow vector size. The difference becomes even larger if we compare compute performance (GFLOP/s), since Hotspot 2D also has a higher FLOP-to-byte ratio. It is not possible to fully utilize the DSPs on Stratix V for Hotspot 2D since this stencil has a high number of floating-point additions and subtractions that are not natively supported by the DSPs on this device and hence, performance scaling is constrained by logic utilization. On Arria 10, however, throughput (GB/s) is 35% higher in Diffusion 2D compared to Hotspot 2D since both are constrained by DSP utilization on this FPGA, while the much lower compute intensity (FLOP per cell update) of Diffusion 2D allows using a twice wider vector at the same $par_{time}$. This is enough to offset the better memory bandwidth utilization of Hotspot 2D that results from its higher $num_{acc}$. This 35% throughput difference is exactly equal to the ratio of $num_{acc} \times par_{vec} \times f_{max}$ between these two stencils. However, due to the higher FLOP-to-byte ratio of Hotspot 2D, the difference in the compute performance (GFLOP/s) of these two stencils is smaller.

For the 3D stencils on Stratix V, similar to the 2D case, Hotspot 3D achieves higher throughput (GB/s) than Diffusion 3D despite the same total degree of parallelism ($par_{total}$) and lower $f_{max}$. This is again due to higher $num_{acc}$ in Hotspot 3D which allows better utilization of memory bandwidth. However, unlike the 2D case, compute performance (GFLOP/s) of the two 3D stencils is similar since this time, Diffusion 3D has the higher FLOP-to-byte ratio. Hotspot 3D achieves lower $f_{max}$ here due to over 80% logic utilization and 100% Block RAM and DSP utilization. On Arria 10, the computation throughput (GB/s) of the 3D stencils is close. On this device, Diffusion 3D benefits from the higher total degree of parallelism and bigger $bsize_{\{x|y\}}$, while Hotspot 3D benefits from higher $num_{acc}$ and $f_{max}$. $bsize_{\{x|y\}}$ in Hotspot 3D is smaller since two input buffers need to be cached in this stencil.

Comparison of performance numbers between Stratix V and Arria 10 show that Arria 10 achieves between 3 to 7 times higher compute performance (GFLOP/s) and 1.7 to 3 times higher power efficiency for a fixed stencil compared to Stratix V. These results show that unlike the preliminary evaluation in the previous chapter, our much better optimized design in this chapter is much more successful in taking advantage of the higher computational capabilities of the Arria 10 FPGA. A small part of the large performance difference between Stratix V and Arria 10 comes from the 33% higher memory bandwidth of Arria 10 compared to Stratix V. However, the main reason for this large difference is the much higher computational capability of Arria 10, enabled by the close-to-6-times improvement in number of DSPs on this device and their native support for floating-point operations. Block RAM improvement from Stratix V to Arria 10 is minor (only 6%) but apart from logic, many Block RAMs are also used on the Stratix V FPGA to support floating-point operations due to lack of native support for such operations in the DSPs of this device. On the other hand, this Block RAM overhead does not exist on Arria 10, allowing us to better take advantage of the limited amount of Block RAMs for implementing shift registers on this newer device. Despite all this, the 3D stencils still become Block RAM-bound on Arria 10, resulting in smaller improvement from Stratix V to Arria 10 for these stencils compared to 2D ones.

As shown in Table 5-6, we achieve an $f_{max}$ of over 300 MHz in cases that routing is not constrained by area utilization. This shows that our implementation maps well to the underlying FPGA architecture, and that we have been successful in optimizing the critical path. The $f_{max}$ we achieve with our design is relatively higher compared to similar designs on FPGAs. Since 2D stencils have less dimension variables, their critical path is shorter compared to 3D stencils and hence, $f_{max}$ is higher. It should be noted that, as explained in Section 5.6.3, target $f_{max}$ and seed sweep is only done for the candidate that achieves the highest normalized performance and hence, the $f_{max}$ values reported in Table 5-6 for configurations other than the best-performing ones are *not* necessarily the highest-achievable $f_{max}$ for these configurations.

As a final note on power consumption, in most cases our design uses close to or even over the 70-Watt TDP of the Arria 10 board. This further asserts that we are pushing the boundaries of performance on this device.

### 5.7.1.2 High-order stencils:

Table 5-7 shows the configuration and performance of Diffusion 2D and 3D from first to fourth order. Only the best-performing configuration is reported for each stencil here. The hachured rows show high-order 3D cases on the Stratix V FPGA that the compiler failed to compile and crashed during the OpenCL to Verilog conversion. The numbers included in these rows are estimated. This issue seems to be only limited to Quartus Prime Standard that is used for Stratix V, and the same kernels compile correctly with much bigger configurations using Quartus Prime Pro that is used for Arria 10.

For 2D stencils, based on the compiler's area reports, Block RAM usage per PE increases proportional to stencil radius as we expected. This allowed us to keep $bsize_x$ the same in all cases since $par_{time}$ needed to be reduced to adjust for the higher compute intensity of the higher-order stencils. However, to get the best configuration for high-order ones, rather than dividing the degree of temporal parallelism of the first-order stencil by stencil radius as predicted in Section 5.6.3, we found other configurations that allowed us to better utilize the DSPs available on each device. For 3D stencils, however, with a fixed spatial block size, the Block RAM utilization per PE increased by a factor of 2.5-3x when doubling the stencil radius, which is in contrast to what we expected. This forced us to reduce $bsize_{\{x|y\}}$ from 256×256 to 256×128 on Arria 10 for high-order stencils, despite lower $par_{time}$. We believe that the extra Block RAM usage is either due to some shortcoming in the OpenCL compiler when inferring large shift registers, or some device limitation that requires more Block RAMs than we predicted to provide enough ports for all the parallel accesses to the shift register. This issue could also be the main contributing factor to compilation failure for high-order 3D stencils on the Stratix V device. Here, the best configuration for the high-order 3D stencils on Arria 10 could be obtained by dividing the $par_{time}$ value used for the first-order stencil by the radius of the high-order stencils. On Stratix V, a similar strategy would apply for all orders except third since $par_{time}$ for first-order is not divisible by three.

**Table 5-7 Configuration and Performance of High-order Stencils on FPGAs**

| Stencil | Device | rad | bsize | $par_{time}$ | $par_{vec}$ | Input Size | Estimated Perf. (GB/s) | Measured Perf. (GB/s\|GFLOP/s\|GCell/s) | $f_{max}$ (MHz) | Logic | Memory (Bits\|Blocks) | DSP | Power (Watt) | Model Accuracy |
|---|---|---|---|---|---|---|---|---|---|---|---|---|---|---|
| Diffusion 2D | Stratix V | 1 | 4096 | 12 | 4 | 16288×16288 | 115.360 | 100.505\|113.068\|12.563 | 303.49 | 64% | 14%\| 40% | 95% | 27.889 | 87.1% |
| | | 2 | 4096 | 6 | 4 | 16288×16288 | 58.006 | 50.534\|107.385\| 6.317 | 303.39 | 60% | 14%\| 37% | 86% | 26.494 | 87.1% |
| | | 3 | 4096 | 4 | 4 | 16288×16288 | 38.890 | 33.879\|105.872\| 4.235 | 304.50 | 59% | 14%\| 36% | 83% | 25.928 | 87.1% |
| | | 4 | 4096 | 7 | 2 | 16160×16160 | 33.791 | 29.290\|120.821\| 3.661 | 303.58 | 67% | 29%\| 55% | 95% | 29.955 | 86.7% |
| | Arria 10 | 1 | 4096 | 36 | 8 | 16096×16096 | 766.918 | 662.655\|745.487\|82.832 | 337.78 | 56% | 38%\| 83% | 95% | 65.516 | 86.4% |
| | | 2 | 4096 | 42 | 4 | 15712×15712 | 422.848 | 359.817\|764.611\|44.977 | 322.22 | 64% | 75%\|100% | 100% | 67.819 | 85.1% |
| | | 3 | 4096 | 28 | 4 | 15712×15712 | 264.700 | 225.226\|703.831\|28.153 | 302.56 | 57% | 75%\|100% | 96% | 64.387 | 85.1% |
| | | 4 | 4096 | 22 | 4 | 15680×15680 | 205.240 | 174.399\|719.396\|21.800 | 300.00 | 60% | 78%\|100% | 99% | 66.977 | 85.0% |
| Diffusion 3D | Stratix V | 1 | 256×256 | 4 | 8 | 744×744×744 | 74.194 | 62.105\|100.921\| 7.763 | 296.12 | 60% | 36%\| 67% | 91% | 29.379 | 83.7% |
| | | 2 | 256×256 | 2 | 8 | 744×744×744 | 32.488 | 27.171\| 84.909\| 3.396 | 259.33 | 58% | 52%\| 89% | 84% | 31.378 | 83.6% |
| | | 3 | 256×128 | 4 | 2 | 696×624×696 | 14.069 | | 250.00 | | | 63% | | |
| | | 4 | 256×256 | 1 | 8 | 744×744×744 | 15.660 | | 250.00 | | | 81% | | |
| | Arria 10 | 1 | 256×256 | 12 | 16 | 696×696×696 | 378.345 | 232.378\|377.614\|29.047 | 285.71 | 60% | 94%\|100% | 89% | 64.409 | 61.4% |
| | | 2 | 256×128 | 6 | 16 | 696×728×696 | 176.622 | 97.930\|306.031\|12.241 | 262.75 | 44% | 73%\| 87% | 83% | 58.293 | 55.4% |
| | | 3 | 256×128 | 4 | 16 | 696×728×696 | 114.538 | 63.963\|295.829\| 7.995 | 255.07 | 44% | 81%\| 99% | 81% | 60.160 | 55.8% |
| | | 4 | 256×128 | 3 | 16 | 696×728×696 | 81.563 | 44.615\|273.267\| 5.577 | 242.67 | 47% | 85%\|100% | 80% | 60.354 | 54.7% |

In terms of operating frequency, we expected $f_{max}$ to be only affected by whether the stencil is 2D or 3D, since the design critical path is determined by the number of index and block variables (Section 3.2.4.4). Our expectation is confirmed by the results of Diffusion 2D on the Stratix V device. However, on Arria 10, new *device-dependent* critical paths appear due to device placement and routing complications resulting from chaining tens of Block RAMs to implement large shift registers, which reduce $f_{max}$ as stencil radius increases. Similarly, for 3D stencils, $f_{max}$ decreases for high-order cases. As an extra disadvantage, for high-order 3D stencils on Arria 10 (second to fourth), we cannot achieve an $f_{max}$ above the operating frequency of the memory controller (266 MHz), which also results in lowered peak memory bandwidth.

In terms of computation throughput (GB/s), we can see that in every case except third and fourth-order Diffusion 3D on Stratix V, temporal blocking is still effective, allowing us to achieve an effective throughput higher than the available external memory bandwidth. For 2D stencils, we expect temporal blocking to be still effective even for radiuses higher than four on Arria 10, but likely not on Stratix V. For 3D stencils, due to high Block RAM and DSP requirements, fifth and sixth-order stencils will be limited to two parallel temporal blocks on Arria 10, and for higher values, temporal blocking will be unusable. On Stratix V, temporal blocking will be ineffective for third-order Diffusion 3D and above. Higher performance for such stencils on these devices will only be possible with faster external memory.

In terms of compute performance (GFLOP/s), we achieve similar performance for high-order stencils to that of the first-order ones. In fact, in multiple cases we achieve slightly higher performance (GFLOP/s) with a high-order stencil compared to a lower-order one. The reason for this slight increase is that for certain stencil orders we can find sets of $par_{time}$ and $par_{vec}$ that better utilize the DSPs. In addition, the lower $par_{time}$ used for high-order stencils reduces the amount of redundant memory accesses, further increasing performance. For other cases, performance can slightly decrease due to lower DSP utilization or lower operating frequency. For 3D stencils on Arria 10, even though compute performance is similar for second to fourth-order, there is a gap between first and second-order. This is due to three reasons: lower DSP utilization, smaller spatial block size, and lower operating frequency. Here, we see one major problem of accelerating 3D stencils on FPGAs: due to high FLOP per cell update in high-order 3D stencils, and the restrictions we need to put on our design parameters ($par_{time}$ and $par_{vec}$) to achieve high performance, the number of DSPs used for each PE reaches a few hundred. Because of this, many DSPs are left unused since they cannot be used to accommodate one additional PE.

In terms of updated cells per second (GCell/s), for 2D stencils, performance decreases proportional to the stencil radius. This aligns with what we predicted in Section 5.6.3. For 3D stencils, this relationship is valid for second to fourth-order, but first-order is over twice faster than second-order. The reason for this difference is the same as above.

**Overall, our results show that by tuning the different performance parameters, our design can be scaled to occupy the majority of FPGA area and achieve very high performance, regardless of stencil shape or size.**

### 5.7.2 Model Accuracy

Model accuracy in Tables 5-6 and 5-7 show how much of our predicted performance is achieved in practice on the board. In reality, there is no source of inaccuracy in the way we calculate the amount of data transferred between the FPGA and its external memory. Adding counters to the kernels and counting the number of reads and writes also showed that our model matches run-time counting and is completely accurate. **Hence, the only possible source of lowered model accuracy is if the memory controller and memory interface are not behaving as we expect. In other words, this value shows the efficiency of the memory controller and memory interface, rather than the accuracy of our model.**

For 2D stencils, in every case in which all the requirements from Eq. (5-6) are met, the model accuracy shows very small variations and stays between 85 and 87%. For cases where $par_{time} \times rad$ is a multiple of two, accuracy is lowered by 12-15%, and for cases where this value is an odd number, accuracy drops by 20%. This completely aligns with what we expected as described in Section 5.3.3 when presenting the padding optimization, and clearly shows that the accuracy value is independent of our model and directly follows the behavior of the memory controller.

For 3D stencils, in multiple cases, the accuracy reaches 80-85%, which means that it is also possible to achieve the same level of *memory controller efficiency* as 2D stencils for 3D cases; however, in most cases, accuracy is between 55 and 60%, which shows that the memory controller behaves very erratically with 3D stencils. The main reason for this is that for 3D stencils, we have to rely on large vector sizes. Even though we assume that memory throughput increases linearly with vector size as long as the peak throughput has not been reached (Eq. (5-9)), in reality, this is only valid for small vector sizes up to four due to inefficiencies in the memory controller. For large vector sizes (8 and 16), the memory controller does not seem to be able to handle the accesses efficiently and hence, memory controller efficiency is lowered. This problem will be further pronounced when we consider this fact that redundancy only exists in external memory reads (and not writes), and hence, external memory writes which happen less frequently than reads are more likely to be stalled due to bus conflict. Such stalls will propagate all the way to the top of the pipeline, resulting in even more bandwidth waste for large vectors and significant reduction in pipeline throughput. Profiling the kernels using Intel FPGA Dynamic Profiler for OpenCL showed that the average burst size in our design is always lower than $par_{vec}$, and hardly goes beyond 8 words. This implies that for large vector sizes, many accesses are being split into smaller ones by the memory controller at run-time, resulting in significant waste of memory bandwidth and lowered efficiency. On the other hand, experimenting with manual banking (Section 3.2.3.1) for the Diffusion 3D stencil showed that for cases where $(par_{time} \times rad) \bmod 8 = 0$, which means padding (Section 5.3.3) is not required, and $par_{vec} = 16$, over 80% of the estimated performance can be achieved. This further asserts that the lowered model accuracy is due to inefficiencies in the memory controller since the manual banking optimization does not change anything in our model or the kernel. Trying to align our parameter tuning with this more restrictive requirement would significantly reduce our tuning freedom and in practice, we could not actually achieve higher performance for Diffusion 3D by following this more strict requirement, since we had to lower the total degree of parallelism to achieve such configurations and this cancelled out the improvement from higher efficiency. These issues have little to no effect on 2D stencils since a small vector size is preferred for such stencils; however, the same lowered efficiency as 3D stencils could be observed for experimental 2D kernels with large vector sizes. This observation asserts that model accuracy does not depend on whether the stencil is 2D or 3D and only depends on the vector size. We do not expect this situation to improve without major improvements in the FPGA's memory controller/interface.

Among the different configurations for each stencil in Table 5-6, if we eliminate configurations that do not satisfy Eq. (5-6) (blue hachures) where our model might choose the best configuration incorrectly (yellow hachures), the model allows us to predict the best configuration correctly in every case except one. This is the case of Hotspot 3D on Stratix V where, even though both configurations satisfy Eq. (5-6), one case achieves significantly higher efficiency. This difference is again caused by the memory controller's lower efficiency in handling the configuration with the larger vector size of 8. It is possible to improve our model by considering the extra performance drop for cases where the requirements of Eq. (5-6) are not satisfied. However, this performance drop is caused by inefficiencies in the memory controller, which might be eventually fixed in future versions of Intel FPGA SDK for OpenCL or with the introduction of new FPGAs with improved memory controllers.

As a final note, the cases hachured in orange in Table 5-6 show configurations where the effective throughput from Eq. (5-9) is expected to be capped by $th_{max}$ due to saturation of the external memory bandwidth. However, due to inefficiencies in the memory controller as explained above, memory bandwidth is not saturated in practice and hence, memory access efficiency is artificially inflated. If we eliminate the capping factor of $th_{max}$ from Eq. (5-9) for such cases, we would be able to obtain a more accurate estimation of memory access efficiency as long as the predicted uncapped bandwidth is not too far from the capped one. For example, for the fastest configuration of first-order Diffusion 3D on Arria 10, if we eliminate the bandwidth cap from our model, model accuracy will be reduced to 57.2% which is likely a more accurate estimation of the memory controller efficiency in this case. For cases where the difference between the capped and uncapped bandwidth is too high (e.g. the first two configurations of Hotspot 3D on Arria 10), removing the cap will result in an abnormally-low memory controller efficiency since the external memory bandwidth is saturated also in practice for these cases.

### 5.7.3 Performance Projection for Stratix 10

To evaluate the potential of the upcoming Stratix 10 FPGA family, we use our model to predict the performance of our evaluated stencils on two devices of this family. One is the GX 2800 device that is coupled with conventional DDR4 memory and has an extremely low byte-to-FLOP ratio (Table 5-4), and the other is the MX 2100 device that is accompanied by HBM2 memory and has a byte-to-FLOP ratio comparable to modern GPUs.

For parameter tuning, we follows the following restrictions for performance projection on Stratix 10:

$par_{time}$: We restrict $par_{time}$ to values that satisfy Eq. (5-6).

$par_{vec}$: On Stratix 10 GX 2800, since DDR4 memory is used, we restrict $par_{vec}$ to powers of two similar to Stratix V and Arria 10. However, since MX 2100 is accompanied by HBM2 memory, we expect a different restriction to be required. Based on [78], the interface between the FPGA and memory in case of the Stratix 10 MX series will consist of multiple 128-bit channels. Hence, we will assume that the memory ports between the kernel and memory interface will be restricted to values that are a multiple of four words (128 bits) and hence, restrict $par_{vec}$ to values that are a multiple of four instead of powers of two.

**$bsize_{\{x|y\}}$:** For $bsize_{\{x|y\}}$, we follow the same restrictions as Stratix V and Arria 10 which were discussed in Section 5.6.3. However, we relax the requirement for $bsize_x$ (but not $bsize_y$) being a power of two since this value needs to be a multiple of $par_{vec}$, and $par_{vec}$ is not necessarily a power of two on the MX 2100 device. Furthermore, considering the large size of on-chip memory on the Stratix 10 device family, restricting both $bsize_x$ and $bsize_y$ to powers of two will limit our parameter tuning and could lead to Block RAM underutilization.

To be able to determine which configurations would fit on these devices based on their available resources, we extrapolate resource utilization for each stencil based on the configuration and area utilization of the same stencil on Arria 10. Specifically:

**DSP usage:** Since the DSPs in Stratix 10 are similar to Arria 10, the operations that are supported per DSP are the same. Hence, we use the formulas from Table 5-5 and the configuration parameters to estimate DSP usage on Stratix 10.

**Block RAM usage:** For Block RAM utilization, we extrapolate utilization based on the most area-consuming configurations on Arria 10 from Table 5-6 and 5-7 (which are not necessarily the fastest configurations) and configuration parameters. In this case, we need to consider two factors. One is the total size in Block RAMs that is required, which depends on $bsize_{\{x|y\}}$ and $par_{time}$, and the other is the number of ports that are required, which depends on $par_{time}$, $par_{vec}$ and number of accesses to the shift register for each cell update. For the first case, both memory bits and blocks are extrapolated, but we assume Block RAM overutilization only if expected memory bits utilization goes above 95%. If number of blocks goes over 100%, we assume it will be exactly 100%. For the second case, we calculate the minimum number of memory blocks that are required to provide enough ports for all the parallel accesses. If this case predicts overutilization of blocks, we will discard the candidate configuration. Then, for the final Block RAM utilization, we consider the maximum value between what is calculated in the first case and the second case. We also consider Block RAMs occupied by the OpenCL BSP. On our Arria 10 board, 12% of the Block RAMs are occupied by the BSP. We assume this value will be reduced to 10% on Stratix 10 since it is a larger FPGA. We predict bits and blocks utilization on Stratix 10 as follows:

$$bits_{used_{s10}} = \left\lceil \frac{par_{time_{a10}} \times \prod_{i \in \{x,y\}} bsize_{i_{a10}}}{par_{time_{s10}} \times \prod_{i \in \{x,y\}} bsize_{i_{s10}}} \times \frac{bits_{total_{a10}}}{bits_{total_{s10}}} \\ \times \left(bits_{used_{a10}} - 12\right) \right\rceil + 10 \tag{5-21}$$

$$blocks_{size_{used_{s10}}} = \min\left(\left\lceil \frac{par_{time_{a10}} \times \prod_{i \in \{x,y\}} bsize_{i_{a10}}}{par_{time_{s10}} \times \prod_{i \in \{x,y\}} bsize_{i_{s10}}} \times \frac{blocks_{total_{a10}}}{blocks_{total_{s10}}} \\ \times \left(blocks_{used_{a10}} - 12\right) \right\rceil + 10, 100 \right) \tag{5-22}$$

$$blocks_{ports_{used_{s10}}} = \left\lceil \frac{par_{time} \times par_{vec} \times num_{read\_local}}{blocks_{total_{s10}}} \right\rceil + 10 \qquad (5\text{-}23)$$

$$blocks_{used_{s10}} = \max(blocks_{size_{used_{s10}}}, blocks_{ports_{used_{s10}}}) \qquad (5\text{-}24)$$

In Eq. (5-21) to (5-24), *a10* and *s10* refer to Arria 10 and Stratix 10, respectively, and all the *bits* and *blocks* values are in percentage. $num_{read\_local}$ also refers to the number of reads from the shift register per cell update. Since each Block RAM only has two ports, one of which is used for writing to the shift register, a minimum of one Block RAM per read from shift register is required to satisfy all accesses in parallel. Eq. (5-25) shows the value of $num_{read\_local}$ for all of our evaluated stencils:

$$num_{read\_local} = \begin{cases} rad \times 4 + 1, & Diffusion\ 2D \\ rad \times 6 + 1, & Diffusion\ 3D \\ 6, & Hotspot\ 2D \\ 8, & Hotspot\ 3D \end{cases} \qquad (5\text{-}25)$$

For Diffusion 2D and 3D, the number of reads from the shift register per cell update is equal to the number of neighbors involved in the computation. For Hotspot 2D and 3D, apart from the neighbors, one extra read per cell update is also required for the *power* shift register.

**Logic usage:** Modelling logic utilization on FPGAs is not straightforward. Apart from that, as we shown in Section 5.7.1, logic utilization was never a bottleneck in our design for any of the configurations on the Arria 10 FPGA. Hence, we assume that this resource will not be a bottleneck on the upcoming Stratix 10 FPGAs either, except for cases with very high number of PEs (200+). We will avoid such configurations to improve the dependability of our estimation.

Finally, we estimate $f_{max}$ and memory controller efficiency (model accuracy) as follows:

$\boldsymbol{f_{max}}$: Designs on the Stratix 10 family are expected to reach an $f_{max}$ of up to 1 GHz, enabled by the latest 14 nm manufacturing node and HyperFlex technology [79]. The extended register insertion and re-timing capabilities offered by HyperFlex are expected to improve $f_{max}$ in case of routing congestion caused by placement and routing restrictions or high area utilization. However, when $f_{max}$ is instead limited by the critical path in the design (which will be the case in many complex designs), HyperFlex will have limited effect. For the specific case of stencil computation, as discussed in Section 3.2.4.4, the critical path of the design will be the chain of operations that update the dimension and block variables. Hence, we expect limited $f_{max}$ improvement with HyperFlex on Stratix 10 compared to Arria 10 for cases where $f_{max}$ is limited by the design rather than the device. On the other hand, for cases where $f_{max}$ is limited by the device, we expect HyperFlex to be very effective in improving $f_{max}$. This specifically applies to the case of high-order stencils where $f_{max}$ decreases on Arria 10 as stencil radius increases due to placement restrictions resulting from the large shift registers; we expect HyperFlex on Stratix 10 to eliminate this problem. Considering these points, we only assume a conservative 100-MHz increase in $f_{max}$ compared to the highest values obtained on Arria 10, mostly resulting from the smaller production node on Stratix 10, and base our estimations on an $f_{max}$ of 450 MHz for 2D stencils, and 400 MHz for 3D, regardless of radius.

**Memory controller efficiency (model accuracy):** Even though we expect improvements in the memory controller of the Stratix 10 FPGAs and consequently, higher efficiency, we base our estimations on the efficiency values we measured on Arria 10. Specifically, we assume an efficiency of 85% for 2D stencils, and 60% for 3D, and use these values as correction factors to adjust our predicted performance.

Table 5-8 shows our projected performance results for all of our evaluated stencils on the Stratix 10 MX 2100 and GX 2800 devices. In this table, resources that bottleneck the performance are marked in red. Apart from area bottlenecks, the amount and percentage of the utilized memory bandwidth is also included in this table to allow us to more accurately determine the source of the bottleneck. The "Total Redundancy" column also refers to the ratio of redundant memory accesses to total, equal to the amount of the wasted memory bandwidth due to overlapped blocking, calculated from Eq. (5-17).

**Table 5-8 Performance Projection Results for Stratix 10**

| Device | Stencil | rad | bsize | $par_{time}$ | $par_{vec}$ | $f_{max}$ (MHz) | Memory Controller Efficiency | Estimated Performance (GB/s\|GFLOP/s\|GCell/s) | Total Redundancy | Utilized Memory Bandwidth (GB/s\|%) | Memory (Bits\|Blocks) | DSP |
|---|---|---|---|---|---|---|---|---|---|---|---|---|
| Stratix 10 MX 2100 | Diffusion 2D | 1 | 16320 | 8 | 96 | 450 | 85% | 2349.504\|2643.192\|293.688 | 0.02% | 345.6\| 68% | 20%\| 67% | 97% |
| | | 2 | 16308 | 4 | 108 | 450 | 85% | 1321.596\|2808.390\|165.199 | 0.02% | 388.8\| 76% | 20%\| 67% | 98% |
| | | 3 | 16340 | 4 | 76 | 450 | 85% | 929.898\|2905.931\|116.237 | 0.04% | 273.6\| 53% | 25%\| 68% | 100% |
| | | 4 | 16356 | 2 | 116 | 450 | 85% | 709.746\|2927.703\| 88.718 | 0.02% | 417.6\| 82% | 20%\| 68% | 100% |
| | Diffusion 3D | 1 | 980×512 | 4 | 140 | 400 | 60% | 1066.630\|1733.274\|133.329 | 0.80% | 448.0\| 88% | 94%\|100% | 99% |
| | | 2 | 592×512 | 2 | 148 | 400 | 60% | 562.053\|1756.415\| 70.257 | 1.12% | 473.6\| 93% | 85%\|100% | 97% |
| | | 3 | 364×256 | 4 | 52 | 400 | 60% | 370.432\|1713.247\| 46.304 | 7.81% | 166.4\| 33% | 88%\|100% | 100% |
| | | 4 | 468×512 | 1 | 156 | 400 | 60% | 295.679\|1811.032\| 36.960 | 1.30% | 499.2\| 98% | 81%\| 96% | 99% |
| | Hotspot 2D | 1 | 16368 | 8 | 48 | 450 | 85% | 1761.412\|2201.764\|146.784 | 0.07% | 259.2\| 51% | 24%\| 44% | 97% |
| | Hotspot 3D | 1 | 972×256 | 4 | 108 | 400 | 60% | 1202.747\|1703.891\|100.229 | 2.17% | 512.0\|100% | 94%\|100% | 98% |
| Stratix 10 GX 2800 | Diffusion 2D | 1 | 8192 | 140 | 8 | 450 | 85% | 3355.470\|3774.903\|419.434 | 1.33% | 28.8\| 38% | 59%\| 90% | 97% |
| | | 2 | 8192 | 78 | 8 | 450 | 85% | 1855.527\|3942.994\|231.941 | 1.48% | 28.8\| 38% | 65%\| 86% | 98% |
| | | 3 | 8192 | 52 | 8 | 450 | 85% | 1243.394\|3885.607\|155.424 | 1.48% | 28.8\| 38% | 65%\| 86% | 94% |
| | | 4 | 16384 | 21 | 16 | 450 | 85% | 1021.622\|4214.190\|127.703 | 0.26% | 57.6\| 75% | 69%\| 88% | 99% |
| | Diffusion 3D | 1 | 544×256 | 24 | 32 | 400 | 60% | 960.545\|1560.886\|120.068 | 14.77% | 76.8\|100% | 91%\| 97% | 93% |
| | | 2 | 352×256 | 12 | 32 | 400 | 60% | 466.036\|1456.362\| 58.254 | 18.56% | 76.8\|100% | 88%\|100% | 87% |
| | | 3 | 320×256 | 8 | 32 | 400 | 60% | 308.438\|1426.525\| 38.555 | 19.52% | 76.8\|100% | 90%\|100% | 85% |
| | | 4 | 256×256 | 7 | 32 | 400 | 60% | 251.012\|1537.451\| 31.377 | 28.38% | 76.8\|100% | 89%\|100% | 97% |
| | Hotspot 2D | 1 | 8192 | 140 | 4 | 450 | 85% | 2505.663\|3132.079\|208.805 | 1.77% | 21.6\| 28% | 81%\| 90% | 97% |
| | Hotspot 3D | 1 | 272×256 | 24 | 16 | 400 | 60% | 853.364\|1208.933\| 71.114 | 29.18% | 76.8\|100% | 92%\|100% | 61% |

Even with a conservative estimation, we expect the larger GX 2800 device to achieve up to 3.7 TFLOP/s of compute performance for first-order Diffusion 2D, and even higher performance for higher orders. Furthermore, this device is expected to achieve over 3.1 TFLOP/s for Hotspot 2D. For all these cases, the DSP count is expected to be the performance bottleneck. Due to high number of floating-point additions in Hotspot 2D, which leave the multipliers in the DSPs unoccupied, this stencil achieves lower *DSP occupancy* and hence, lower compute performance compared to first-order Diffusion 2D despite similar DSP utilization. **These results show that having a large FPGA can offset low external memory bandwidth for 2D stencil computation thanks to temporal blocking. We expect the GX 2800 device to be able to easily out-perform its same-generation GPUs for 2D stencil computation.** For 3D stencils, this device achieves over 1.4 TFLOP/s for Diffusion 3D, and 1.2 TFLOP/s for Hotspot 3D. In every case, the low memory bandwidth quickly bottlenecks performance and due to small block size, temporal parallelism has diminishing returns. For Diffusion 3D, we expect to be able to scale the performance up to a point where the majority of the DSPs are used regardless of the radius; however, for Hotspot 3D this will not be possible since this benchmark requires caching the extra *power* input which significantly increases Block RAM requirement and reduces block size. Further reducing the block size for this stencil compared to the configuration reported in Table 5-8, and increasing $par_{time}$ instead, will result in negative performance scaling since the amount of redundant memory accesses will then cancel out the performance improvement from the higher $par_{time}$.

For the smaller but higher-bandwidth MX 2100 FPGA, we expect between 2.6 and 3 TFLOP/s for Diffusion 2D of different orders, and 2.2 TFLOP/s for Hotspot 2D. This is significantly lower than the predicted performance for the GX 2800 device, since we can easily scale performance of 2D stencils with temporal parallelism due to large block size and hence, the larger area available on the GX 2800 device and its higher DSP count gives this device a considerable edge over the smaller MX 2100 device. Unlike Stratix V and Arria 10 where we prefer smaller $par_{vec}$ and larger $par_{time}$ due to better scaling with temporal parallelism, on the MX 2100 device we assume linear performance scaling with vectorization as long as the utilized memory bandwidth is far from the peak. Hence, we try to utilize a reasonable portion of the memory bandwidth and avoid the added redundancy of extra temporal parallelism. The main bottleneck of 2D stencil computation on this device is DSP usage. On the other hand, we predict the MX 2100 device to be more efficient in 3D stencil computation than the GX 2800 device, achieving over 1.7 TFLOP/s for Diffusion 3D of different orders and Hotspot 3D. The higher memory bandwidth of this device allows us to rely more on vectorization and instead, reduce the degree of temporal parallelism, which would in turn also reduce the wasted performance due to redundant memory accesses. Because of this, the MX 2100 device can achieve higher performance than the GX 2800 device for 3D stencil computation, despite being relatively smaller. **This shows the advantage of higher memory bandwidth over having a bigger FPGA for 3D stencil computation. We expect this device to provide competitive performance in 3D stencil computation compared to its same-generation GPUs.** Our estimated results show that the more balanced byte-to-FLOP ratio of the MX 2100 device will allow us to simultaneously have high area and high memory bandwidth utilization. The only exception is the case of third-order Diffusion 3D where the restrictions imposed by Eq. (5-6) force us to use a smaller $par_{vec}$ and larger $par_{time}$ to avoid the lowered memory access efficiency, at the cost of underutilizing external memory bandwidth and higher redundant memory accesses. For fourth-order Diffusion 3D, since the byte-to-FLOP ratio of the stencil without temporal blocking (0.163) is close to the byte-to-FLOP ratio of the device at the predicted $f_{max}$ (0.162), device resources can be fully utilized and maximum performance can be achieved *without* temporal blocking and with minimum amount of redundancy. **This observation emphasizes the importance of having a balanced byte-to-FLOP ratio for HPC accelerators.**

It is worth noting that even though the MX 2100 device is faster than the GX 2800 device for 3D stencil computation, the performance difference is small (10%). This means that the GX device can still be a good candidate for 3D computation (while being a great one for 2D), and that the freedom offered by our design due to different ways of achieving parallelism can even overcome the extremely low byte-to-FLOP ratio of the GX 2800 device.

### 5.7.4 Comparison with Other Hardware

#### 5.7.4.1 First-order 2D stencil

Fig. 5-7 shows the performance and power efficiency of all of our evaluated hardware for first-order Diffusion 2D. Estimated results for Stratix 10 MX 2100 and GX 2800 are also included as hachured bars. Power efficiency for these devices has been calculated by estimating power usage based on the results reported in [76]. Using their results, power usage of the Stratix 10 GX 2800 device can be estimated to be around 150 Watts at 400-450 MHz. Furthermore, we assume a typical power consumption of 125 Watts for the smaller MX 2100 device. It should be noted that the TDP values reported in Table 5-4 are for the peak operating frequency of 800-900 MHz on these devices. Results reported by [50] and estimated results for newer GPUs based on these results are also included in Fig. 5-7 with all the estimated values being hachured. The *Roofline* [80] performance in this figure shows the maximum achievable performance on each device with full utilization of the peak external memory bandwidth and full spatial reuse (no redundant memory accesses) but *without temporal blocking*. This value is equal to the FLOP-to-byte ratio of the stencil (1.125) multiplied by the peak external memory bandwidth of the device (Table 5-4) and can help determine the effectiveness of temporal blocking on each device.

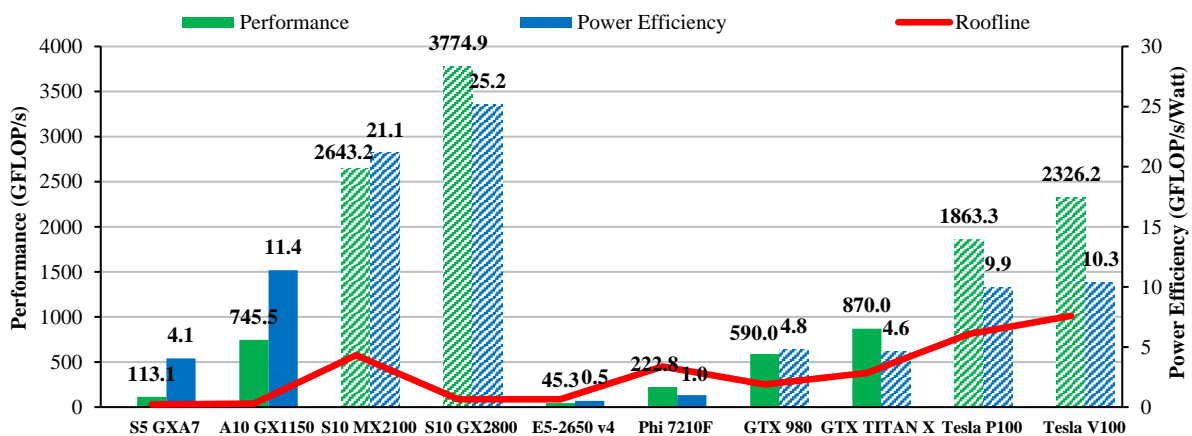

**Figure 5-7 Performance results for first-order 2D stencil computation on all hardware**

Based on our results, the aged Stratix V FPGA outperforms the modern Xeon and achieves half the performance of the Xeon Phi device. Moreover, it achieves better power efficiency than both of these devices and gets close to the estimated power efficiency of the older GPUs. The Arria 10 FPGA also outperforms the GTX 980 GPU and is expected to be able to achieve better power efficiency compared to even the state-of-the-art Tesla V100 GPU. Our estimations also show that the Stratix 10 MX device will likely outperform its same-generation GPUs for first-order 2D stencil computation, while the GX device will probably be even faster than next-generation GPUs. Both of these devices are also expected to offer a level of power efficiency in 2D stencil computation that could remain unchallenged for multiple year to come.

Comparing the roofline performance on each device with the achieved or estimated performance on that device shows an important trend. **For 2D stencil computation, temporal blocking achieves great scaling on FPGAs, allowing these devices to achieve tens of times higher performance than the limit imposed by their external memory bandwidth if temporal blocking is not used. GPUs also achieve modest scaling with temporal blocking (~2.3x) for 2D stencil computation, but far from the level of scaling achieved on FPGAs. However, it is not possible to overcome the limit of external memory bandwidth on the Xeon and Xeon Phi devices using YASK since temporal blocking does not scale on these devices. This trend clearly shows the advantage of FPGAs for 2D stencil computation.**

#### 5.7.4.2 First-order 3D stencil

Fig. 5-8 shows the performance and power efficiency of all of our evaluated hardware for first-order Diffusion 3D. Similar to the previous comparison, estimated results for Stratix 10 MX 2100 and GX 2800 are also included as hachured bars with the same estimated power usage. The Roofline performance has also been calculated in the same way as the 2D case but with the FLOP-to-byte ratio of the 3D stencil (1.625).

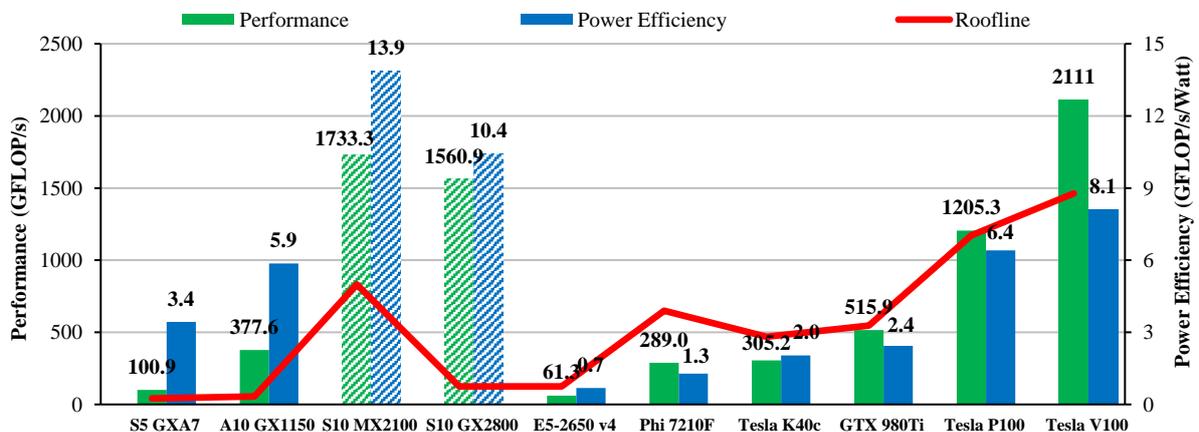

**Figure 5-8 Performance results for first-order 3D stencil computation on all hardware**

Based on our results, the old Stratix V FPGA can beat the modern Xeon in performance, and beat the Xeon Phi and all the GPUs up to 980 Ti in power efficiency. The newer Arria 10 FPGA can further beat the Xeon Phi and the Tesla K40c in performance, and reach a level of power efficiency close to that of the modern Tesla P100. Furthermore, we estimate that the upcoming Stratix 10 devices would be able to achieve higher performance than Tesla P100, and higher power efficiency compared to the state-of-the-art Tesla V100 GPU. It is worth noting that the input size used for the GPUs here is smaller than every other hardware (512×512×512 on GPUs vs. 696×696×696 and higher on other hardware), and for bigger input sizes, even if all the dimensions are a multiple of 512 cells, the GPUs loose up to 20% of their performance. This loss of performance is likely caused by lower cache hit rate on these devices for bigger input sizes. In contrast, as long as the input dimensions are divisible by the dimensions of the compute block, the performance of our FPGA implementation remains nearly constant regardless of how big the input is.

Even though neither Stratix V nor Arria 10 can reach the same level of performance as the more modern GPUs in 3D stencil computation, the same advantage as the 2D case can also be seen here. **Temporal blocking for 3D stencil computation also scales well on FPGAs, while it achieves limited scaling on GPUs (less than 2D), and no scaling on Xeon and Xeon Phi processors (same as 2D). This allows FPGAs to achieve multiple times higher compute throughput than their external memory bandwidth also in 3D stencil computation – something that is not possible on the other hardware.** To put things into perspective, the implementation from [49] achieves highest performance with a block size of only 32×8, which is limited by the size of on-chip memory per SM rather than the total size per GPU. This effectively prevents temporal blocking to scale beyond a degree of temporal parallelism of two for 3D stencil computation (which is already used in this implementation), since **scalability of temporal blocking is directly proportional to the ratio of the size of halos to the block size**. On the other hand, we can use much bigger block sizes on FPGAs (Table 5-6 and 5-7) since, unlike GPUs, we have the freedom to even use all of the on-chip memory to implement one spatial block on these devices. Furthermore, the on-chip memory saving from using shift registers on FPGAs allows us to further increase the gap between the block size on FPGAs and to other hardware, allowing performance scaling up to hundreds of parallel temporal blocks for 2D stencil and tens of such blocks for 3D.

**5.7.4.3 High-order stencils**

Table 5-9 shows the performance and power efficiency of high-order stencil computation on all our evaluated hardware. The "Roofline Ratio" columns shows how much of the *roofline* [80] performance has been achieved on each hardware; this roofline performance is the same as the one used in the previous section and refers to the maximum-achievable performance on each hardware by full utilization of its external memory bandwidth with full spatial reuse but *without temporal blocking*. The numbers reported in this column effectively show the percentage of the utilized external memory bandwidth, which will be less than 1.00 unless temporal blocking is used. Hachured rows show extrapolated results. Solid green cells show the highest performance and power efficiency for each stencil with each order if extrapolated result are *excluded*. Hachured green cells show these values if extrapolated result are *included*.

For 2D stencils, *excluding* the extrapolated results, Stratix V achieves higher performance and power efficiency than the Xeon for first and second-order, and Arria 10 achieves the highest performance for first to third-order, while the Xeon Phi achieves highest performance for fourth by a small margin. However, Arria 10 achieves the best power efficiency in all cases by a clear margin. Despite the highly-optimized implementation in YASK, the Xeon and Xeon Phi devices can only utilize ~50% of their external memory bandwidth (roofline ratio). Furthermore, as explained in Section 5.6.1, temporal blocking proved to be ineffective on these

**Table 5-9 Performance and Power Efficiency of High-order Stencil Computation**

| Device | rad | Diffusion 2D Performance (GB/s\|GFLOP/s\|GCell/s) | Diffusion 2D Power Efficiency (GFLOP/s/Watt) | Diffusion 2D Roofline Ratio | Diffusion 3D Performance (GB/s\|GFLOP/s\|GCell/s) | Diffusion 3D Power Efficiency (GFLOP/s/Watt) | Diffusion 3D Roofline Ratio |
|---|---|---|---|---|---|---|---|
| Stratix V GX A7 | 1 | 100.505\| 113.068\| 12.563 | 4.054 | 3.926 | 62.105\| 100.921\| 7.763 | 3.435 | 2.426 |
| Stratix V GX A7 | 2 | 50.534\| 107.385\| 6.317 | 4.053 | 1.974 | 27.171\| 84.909\| 3.396 | 2.706 | 1.061 |
| Stratix V GX A7 | 3 | 33.879\| 105.872\| 4.235 | 4.083 | 1.323 | N/A | N/A | N/A |
| Stratix V GX A7 | 4 | 29.290\| 120.821\| 3.661 | 4.033 | 1.144 | N/A | N/A | N/A |
| Arria 10 GX 1150 | 1 | 662.655\| 745.487\| 82.832 | 11.379 | 19.43 | 232.378\| 377.614\| 29.047 | 5.863 | 6.81 |
| Arria 10 GX 1150 | 2 | 359.817\| 764.611\| 44.977 | 11.274 | 10.55 | 97.930\| 306.031\| 12.241 | 5.250 | 2.87 |
| Arria 10 GX 1150 | 3 | 225.226\| 703.831\| 28.153 | 10.931 | 6.60 | 63.963\| 295.829\| 7.995 | 4.917 | 1.88 |
| Arria 10 GX 1150 | 4 | 174.399\| 719.396\| 21.800 | 10.741 | 5.11 | 44.615\| 273.267\| 5.577 | 4.528 | 1.31 |
| Stratix 10 MX 2100 | 1 | 2349.504\|2643.192\|293.688 | 21.146 | 4.59 | 1066.630\|1733.274\|133.329 | 13.866 | 2.08 |
| Stratix 10 MX 2100 | 2 | 1321.596\|2808.390\|165.199 | 22.467 | 2.58 | 562.053\|1756.415\|070.257 | 14.051 | 1.10 |
| Stratix 10 MX 2100 | 3 | 929.898\|2905.931\|116.237 | 23.247 | 1.82 | 370.432\|1713.247\| 46.304 | 13.706 | 0.72 |
| Stratix 10 MX 2100 | 4 | 709.746\|2927.703\| 88.718 | 23.422 | 1.39 | 295.679\|1811.032\| 36.960 | 14.488 | 0.58 |
| Stratix 10 GX 2800 | 1 | 3355.470\|3774.903\|419.434 | 25.166 | 43.69 | 960.545\|1560.886\|120.068 | 10.406 | 12.51 |
| Stratix 10 GX 2800 | 2 | 1855.527\|3942.994\|231.941 | 26.287 | 24.16 | 466.036\|1456.362\| 58.254 | 9.709 | 6.07 |
| Stratix 10 GX 2800 | 3 | 1243.394\|3885.607\|155.424 | 25.904 | 16.19 | 308.438\|1426.525\| 38.555 | 9.510 | 4.02 |
| Stratix 10 GX 2800 | 4 | 1021.622\|4214.190\|127.703 | 28.095 | 13.30 | 251.012\|1537.451\| 31.377 | 10.250 | 3.27 |
| Xeon E5-2650 v4 | 1 | 40.272\| 45.306\| 5.034 | 0.521 | 0.52 | 37.712\| 61.282\| 4.714 | 0.686 | 0.49 |
| Xeon E5-2650 v4 | 2 | 40.120\| 85.255\| 5.015 | 0.942 | 0.52 | 36.872\| 115.225\| 4.609 | 1.235 | 0.48 |
| Xeon E5-2650 v4 | 3 | 39.840\| 124.500\| 4.980 | 1.331 | 0.52 | 32.864\| 151.996\| 4.108 | 1.617 | 0.43 |
| Xeon E5-2650 v4 | 4 | 40.056\| 165.231\| 5.007 | 1.737 | 0.52 | 33.592\| 205.751\| 4.199 | 2.069 | 0.44 |
| Xeon Phi 7210F | 1 | 198.048\| 222.804\| 24.756 | 1.000 | 0.50 | 177.840\| 288.990\| 22.230 | 1.279 | 0.44 |
| Xeon Phi 7210F | 2 | 187.640\| 398.735\| 23.455 | 1.774 | 0.47 | 175.776\| 549.300\| 21.972 | 2.428 | 0.44 |
| Xeon Phi 7210F | 3 | 189.520\| 592.250\| 23.690 | 2.629 | 0.47 | 170.496\| 788.544\| 21.312 | 3.480 | 0.43 |
| Xeon Phi 7210F | 4 | 184.048\| 759.198\| 23.006 | 3.369 | 0.46 | 174.576\|1069.278\| 21.822 | 4.714 | 0.44 |
| GTX 580 | 1 | N/A | N/A | N/A | 138.352\| 224.822\| 17.294 | 1.229 | 0.72 |
| GTX 580 | 2 | N/A | N/A | N/A | 114.792\| 358.725\| 14.349 | 1.960 | 0.60 |
| GTX 580 | 3 | N/A | N/A | N/A | 87.552\| 404.928\| 10.944 | 2.213 | 0.46 |
| GTX 580 | 4 | N/A | N/A | N/A | 74.032\| 453.446\| 9.254 | 2.478 | 0.38 |
| GTX 980 Ti | 1 | N/A | N/A | N/A | 242.044\| 393.322\| 30.256 | 1.907 | 0.72 |
| GTX 980 Ti | 2 | N/A | N/A | N/A | 200.826\| 627.582\| 25.103 | 3.043 | 0.60 |
| GTX 980 Ti | 3 | N/A | N/A | N/A | 153.170\| 708.414\| 19.146 | 3.435 | 0.46 |
| GTX 980 Ti | 4 | N/A | N/A | N/A | 129.518\| 793.295\| 16.190 | 3.846 | 0.38 |
| Tesla P100 | 1 | N/A | N/A | N/A | 518.389\| 842.381\| 64.799 | 4.493 | 0.72 |
| Tesla P100 | 2 | N/A | N/A | N/A | 430.112\|1344.100\| 53.764 | 7.169 | 0.60 |
| Tesla P100 | 3 | N/A | N/A | N/A | 328.047\|1517.217\| 41.006 | 8.092 | 0.46 |
| Tesla P100 | 4 | N/A | N/A | N/A | 277.389\|1699.008\| 34.674 | 9.061 | 0.38 |
| Tesla V100 | 1 | N/A | N/A | N/A | 647.177\|1051.662\| 80.897 | 4.674 | 0.72 |
| Tesla V100 | 2 | N/A | N/A | N/A | 536.969\|1678.028\| 67.121 | 7.458 | 0.60 |
| Tesla V100 | 3 | N/A | N/A | N/A | 409.547\|1894.154\| 51.193 | 8.418 | 0.46 |
| Tesla V100 | 4 | N/A | N/A | N/A | 346.304\|2121.109\| 43.288 | 9.427 | 0.38 |

devices. On the other hand, the FPGAs can achieve multiple times higher computation throughput than their external memory bandwidth due to the effectiveness of temporal blocking on this platform. For the fourth-order stencil, even with temporal blocking, the achieved computation throughput on Arria 10 is lower than the utilized memory bandwidth on the Xeon

Phi device and hence, the Xeon Phi achieves better performance despite lack of temporal blocking. We expect the Xeon Phi to be faster than the Arria 10 FPGA also for stencil orders above four. *Including* the extrapolated results, the Stratix 10 GX FPGA is expected to achieve an unprecedented level of performance in 2D stencil computation, which will very likely not only be higher than the state-of-the-art Tesla V100 GPU, but also next generation GPUs. Due to unavailability of a highly-optimized GPU implementation, an apples-to-apples comparison with GPUs for high-order 2D stencil computation is not possible at this time.

For 3D stencils, we include the results from [71] and extrapolated results based on that for newer GPUs. *Excluding* the extrapolated results, the aged Stratix V device can only outperform the Xeon processor in first-order, but loses in second-order and above. The Arria 10 FPGA achieves the highest performance for first-order, while the Xeon Phi and the GTX 580 GPU achieve higher performance for higher orders, with the Xeon Phi achieving the highest; however, Arria 10 still achieves the best power efficiency in all orders except four. *Including* the extrapolated results, we expect the Stratix 10 MX 2100 device to be able to outperform the state-of-the-art Tesla V100 GPU for first and second-order 3D stencil computation, and achieve better power efficiency in all cases. In the previous section, we showed that using the implementation from [49], which also employs temporal blocking, this modern GPU can achieve twice the performance we estimated here for the first-order 3D stencil. However, this implementation only supports first-order stencils, and the effectiveness of temporal blocking for high-order stencils on GPUs is unknown and is expected to be even less than the limited scaling with first-order stencils. On the Stratix 10 MX device, since we can rely more on vectorization rather than temporal parallelism, it is possible to increase the spatial block size to values that enable us to reduce the amount of redundant memory accesses for 3D stencil computation to less than 5%. Assuming that the memory controller efficiency is improved on this device, and operating frequency goes above our conservative estimation, matching the performance of the implementation from [49] on the Tesla V100 GPU could become possible.

Figures 5-9 and 5-10 show performance for high-order 2D and 3D stencil computation on all devices using two different metrics in a more comparable manner. Extrapolated results are hachured in these figures. These charts allow us to see the trend of performance on difference devices with respect to stencil order. On the FPGA, number of cells updated per second (GCell/s) decreases proportional to stencils order, which means the compute performance (GFLOP/s) stays relatively close. On the Xeon and Xeon Phi processors, number of cells updated per second remains similar, which means the compute performance (GFLOP/s) increases proportional to stencil order. On the GPUs, GCell/s decreases with a rate lower than the increase in radius and hence, GFLOP/s increases sub-linearly with stencil order. These performance trends show that the Xeon and Xeon Phi processors, despite being memory-bound in all cases, can utilize a fixed amount of their memory bandwidth regardless of stencil radius. Computation is also memory-bound on the GPUs; however, memory bandwidth efficiency decreases as the stencil order increases. On FPGAs, the trend is different: **we can claim the performance we have achieved resembles a *compute-bound* scenario since, regardless of stencil order, the compute performance (GFLOP/s) is nearly constant.**

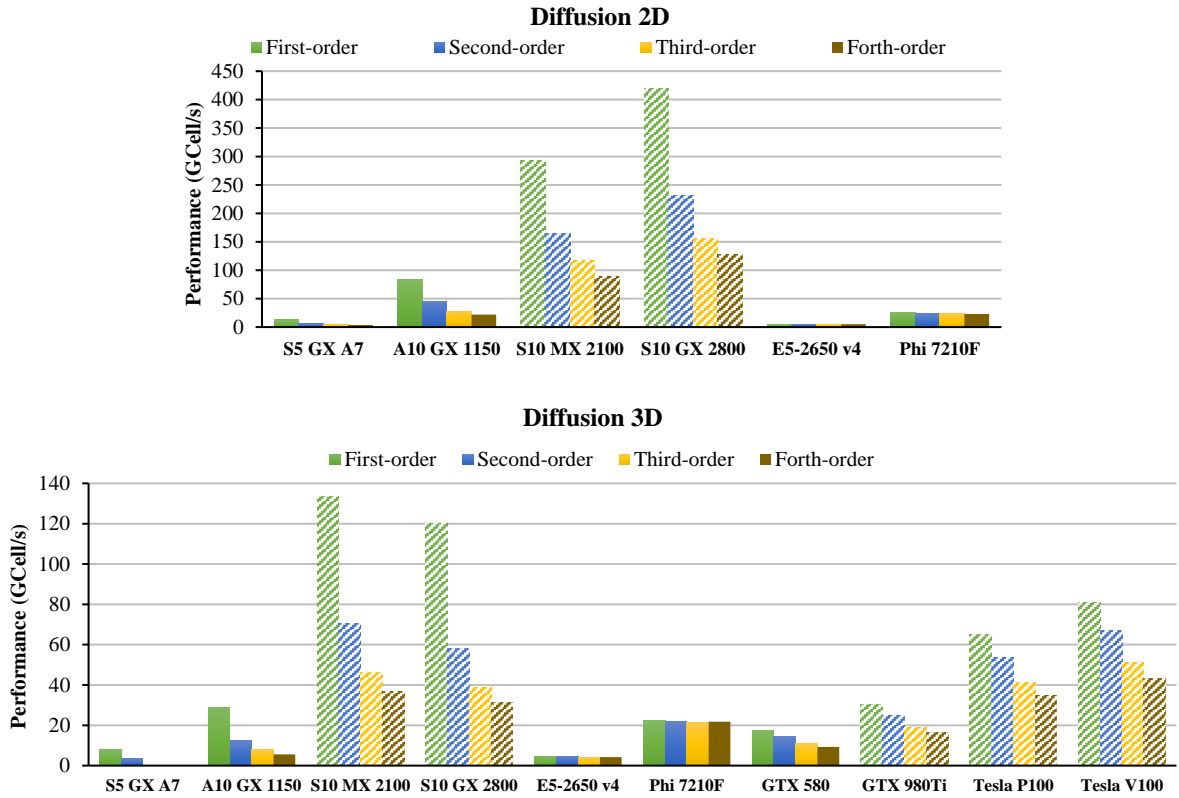

Figure 5-9 Performance of High-order Diffusion 2D and 3D in GCell/s

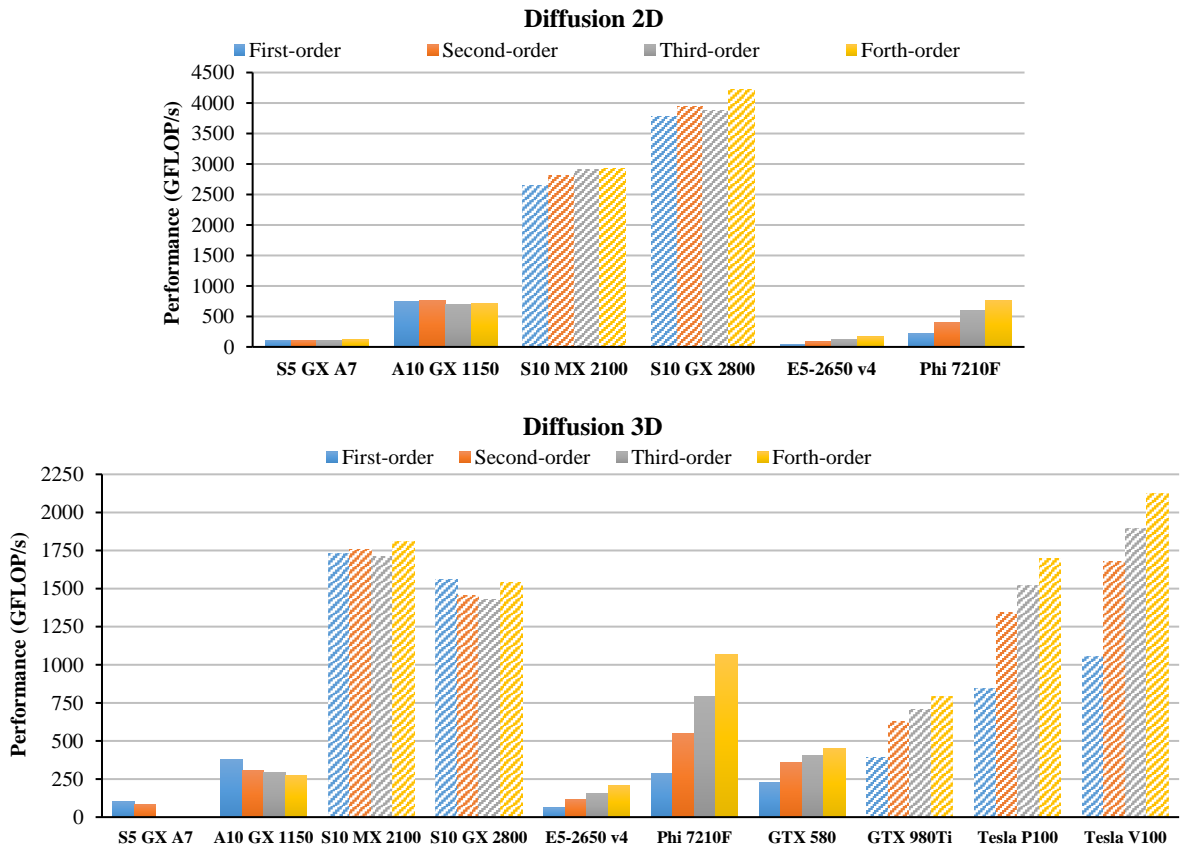

Figure 5-10 Performance of High-order Diffusion 2D and 3D in GFLOP/s

Even though our achieved compute performance (GFLOP/s) is far from the peak reported in Table 5-4 for the Arria 10 and Stratix 10 devices, we need to emphasize that these peak values can only be achieved with full utilization of all DSPs on the highest speed-grade of these devices with FMA operations running at the peak operating frequency of the DSPs (480 MHz on Arria 10 and 750 MHz on Stratix V). Such operating frequencies would be near impossible to achieve in real-world designs and hence, saturating the compute potential of these devices is generally not possible. Furthermore, parameter tuning restrictions and stencil characteristics prevent us from being able to completely utilize all the DSPs with FMA operations. As an example, for the first-order Diffusion 3D stencil on the Arria 10 device, the peak performance is ~867 GFLOP/s at the achieved $f_{max}$. Furthermore, only 1344 DSPs can be utilized for computation in this stencil due to parameter tuning restrictions, and in one out of each 7 DSPs, the adder is not used which translates to a DSP occupancy rate of ~93%. This effectively reduces the achievable peak performance to ~713 GFLOP/s. Considering the 57.2% memory access efficiency we calculated in Section 5.7.2 by eliminating the bandwidth cap, over 40% of the peak performance is also lost due to memory controller inefficiency. Adding the 7.1% redundant memory accesses due to overlapped blocking (Eq. (5-17)), we get to the measured ~378 GFLOP/s.

### 5.7.5 Comparison with Other FPGA Work

In [53], the authors only report normalized performance results compared to a previous baseline implementation, and avoid reporting any time or GLFOP/s numbers. Since the baseline implementation actually evaluates other stencils, it is not possible to reconstruct their performance results. Hence, we cannot compare our results with their implementation. Nevertheless, we do not expect such thread-based implementation with 4D blocking to able to achieve the same level of performance as our deep-pipelined implementation. [51] is another thread-based implementation that reports 8 GFLOP/s for Jacobi 2D on a Kintex-7 XC7Z045 FPGA, while we achieve over 110 GFLOP/s on Stratix V (and much more on Arria 10) for Diffusion 2D which has the exact same stencil characteristics. We achieve this large performance advantage despite the fact that their FPGA has more DSPs and roughly half of the logic and Block RAM count of our Stratix V GX A7 FPGA.

Among deep-pipeline implementations with temporal blocking, compared to [58], we achieve 24% lower performance for 2D 5-point and 9% lower performance for 3D 7-point stencil computation on the same Stratix V device, but with input sizes that are not supported by their implementation due to lack of spatial blocking. To support the 4 times larger input width that we use for Diffusion 2D, and the 36 times larger plane size we use for Diffusion 3D, the size of the shift registers in their design need to be increased by these ratios, forcing the degree of temporal parallelism to be reduced by the same factors. In that case, our implementation will have a clear performance advantage. Compared to [56], we achieve 4 times higher performance in Hotspot 2D which has similar characteristics to their FDTD 2D stencil (same $num_{acc}$ and one higher FLOP per cell update) on the same Arria 10 device. This is despite the fact that their implementation does not use spatial blocking either and restricts input width to 4096 cells. Compared to [55], we achieve 4.8x and 38x higher performance for Diffusion 2D and 3D on Stratix V A7, respectively, compared to their results for Jacobi 2D and 3D on a Virtex-7 XC7VX485T FPGA, despite lack of spatial blocking in their implementation. In [59], the implementation from [55] is further optimized by using HDL instead of HLS and using vectorization for better utilization of memory bandwidth, allowing them to achieve multiple times higher performance on the same device. However, the main shortcoming of the original work, i.e. lack of spatial blocking, is not addressed. Compared to the new results in [59], we achieve nearly 2 times better compute performance (GFLOP/s) for first-order star-shaped 2D, and nearly 3 times higher performance for first-order star-shaped 3D stencils (Heat 3D is a cubic stencil and is not comparable with our work) on Arria 10. However, their Jacobi stencils use shared coefficients and hence, have a lower FLOP per cell update compared to Diffusion 2D and 3D, which means comparing GFLOP/s could lead to incorrect conclusions. To keep the comparison fair, using 6 FLOP per cell update for Jacobi 2D and 8 FLOP per cell update for Jacobi 3D extracted from their baseline implementations, we estimate the number of cells updated per second in their implementation for these two stencils to be 68.3 and 14.9 GCell/s respectively. These numbers are still well below our results for Diffusion 2D (82.8 GCell/s) and 3D (29 GCell/s) on Arria 10. Furthermore, due to lack of spatial blocking, row size for 2D and plane size for 3D stencils are limited to 2048 and 128x128 cells in their implementation, respectively. Hence, to support the much bigger input sizes we are evaluating, they need to increase their on-chip buffer size for the 2D and 3D cases by at least 8x and 36x, respectively, which requires reducing the degree of temporal parallelism by the same amount. This lowers their performance to values that are even below what we achieved on the much smaller Stratix V device. In [57], the authors implement a first-order 2D square stencil with complex boundary conditions that is not comparable with any of our stencils; however, this implementation does not use spatial blocking either and limits input width to only 720 cells. Compared to the recent implementation from [67], we achieve nearly 3.3 times higher compute performance (GFLOP/s) and over 50% higher power efficiency with Diffusion 2D on Arria 10 compared to their implementation of Jacobi 2D on a Kintex UltraScale KU115. Furthermore, our design runs at over twice the operating frequency of their implementation. This is despite the fact their FPGA has 25% more logic elements, 40% more Block RAMs, and over 3.5 times more DSPs than ours. Taking the different FLOP per cell update of the evaluated stencils into account, if we use the GCell/s metric for comparison, we achieve 82.8 GCell/s for Diffusion 2D which is still 80% higher than the 45.6 GCell/s they achieve for Jacobi 2D (5 FLOP per cell update). Needless to say, since their implementation of Jacobi 2D is bottlenecked by FPGA area rather than external memory bandwidth, if they implement our more compute-intensive Diffusion 2D stencil on their device, they will likely achieve lower GCell/s than Jacobi 2D, resulting in even higher performance advantage for our implementation.

With respect to implementations of high-order stencil computation on FPGAs, compared to [72], we cannot directly compare performance in terms of GFLOP/s since coefficients in their stencils are shared and hence, the FLOP per cell update of their stencil is lower than ours. Comparing the number of cells updated per second, despite the fact that our stencil is more compute-intensive since we do not share the coefficients, we achieve close to twice their reported performance for fourth-order 3D stencil computation (2.783 vs. 5.577 GCell/s). The results they report are with the assumption that they have 22.24 GB/s of streaming bandwidth, while the system they use only provides 6.4 GB/s. This assumption is not reasonable since streaming bandwidth, whether it is from FPGA external memory or the link between host and the FPGA, will remain the limiting factor in performance of stencil computation for the foreseeable future, and that is why utilizing temporal blocking is crucial for stencil computation. In their case, since temporal blocking is not employed, the roofline of the performance they can achieve in practice is only 0.8 GCell/s (6.4 GB/s divided by 8 bytes per cell update). Compared to [73], since they also share coefficients, we again use the GCell/s metric for performance comparison. They report 1.54 GCell/s for a third-order 3D star-shaped stencil, while we achieve 7.995 GCell/s, which is over 5 times higher. They also estimate that a future FPGA device that is four times larger than a Virtex-6 SX475T FPGA (roughly the size of the modern Virtex Ultrascale+ VU11P device) can achieve close to 5 GCell/s, while we already achieve higher performance. It is worth noting that in both of these implementations, the more compute-intensive stencils we evaluated will likely achieve lower GCell/s than what is reported in these publications due to FPGA area bottleneck, further increasing the performance gap in our favor. **To the best of our knowledge, we achieve the highest performance for single-FPGA computation of 2D and 3D star-shaped stencils up to a radius of four, without restricting input size.**

## 5.8 Publication Errata

Compared to our publication in FPGA'18 [42], the following issues were corrected or improved in this chapter:

- All results presented here are from the final unified kernels with parameterized radius and hence, performance numbers reported here might slightly differ (both slower and faster) from that publication.
- Non-square blocks were not used for that publication. Using such blocks resulted in faster final candidates in the case of the 3D stencils, with higher best performance for Hotspot 3D on Stratix V.
- Performance projection on Stratix 10 for 2D stencils was done with an overly conservative correction factor of 80% in the publication. This was increased to 85% in this chapter since memory controller efficiency never went below 85% for any of our 2D stencils, even for the high-order cases, as long as the requirements of Eq. (5-6) were met.
- Performance projection for Stratix 10 MX 2100 was done with initial specs released by Intel. The final specs increased the size and resources of this FPGA and the new specs were used for performance projection in this document. Furthermore, the performance projection in the publication for this FPGA was done with the assumption of vector sizes that are a power of two, significantly limiting our parameter tuning freedom. This requirement was relaxed here since we expect this limitation not to exist for HBM.

Compared to the publication in IPDPS RAW'18 [43], the following corrections or improvements were made:

- All results presented here are from the final unified kernels with parameterized radius and hence, performance numbers reported here might slightly differ (both slower and faster) from that publication.
- Stratix V results were added.
- Performance projection for Stratix 10 MX 2100 and GX 2800 and Tesla V100 were added.

## 5.9 Conclusion

In this chapter, we evaluated the potential of FPGAs for 2D and 3D stencil computation and introduced our parameterized stencil accelerator, which can be used to target 2D and 3D star-shaped stencils of arbitrary radius. We showed that even though previous work avoided spatial blocking on FPGAs and only relied on temporal blocking to maximize performance, at the cost of restricting the size of input dimensions, our design with combined spatial and temporal blocking could still achieve high performance without creating such unreasonable restrictions. Moreover, we showed that our design, which takes advantage of both spatial parallelism (vectorization) and temporal parallelism, allows us to scale the computation such that the FPGA area utilization, and consequently, performance, is maximized. This allowed us to reach a compute-bound-like performance on FPGAs regardless of stencil order, while other hardware still struggle to break away from the limit imposed by their external memory bandwidth, even with temporal blocking.

To allow quick and efficient parameter tuning, we devised a performance model and showed that it can predict performance for 2D stencils with a fixed accuracy, regardless of stencil shape or radius. However, our model accuracy switched between a few different values for 3D stencils, due to the erratic behavior of the FPGA memory controller when it comes to large vector sizes that are crucial for acceleration of 3D stencils. Hence, we concluded that our model accuracy actually shows the memory controller efficiency of the device. Furthermore, we used our model alongside with a resource utilization estimation based on our measured results on Arria 10, to project performance for the upcoming Stratix 10 devices.

We achieved over 100 GFLOP/s and 700 GFLOP/s of compute performance for 2D stencil computation, on a Stratix V GX A7 and Arria 10 GX 1150 device, respectively, up to a stencil radius of four. For 3D stencils, we could achieve over 80 GFLOP/s on Stratix V, and 270 GFLOP/s on the Arria 10 device. These results are competitive or even better than same generation Xeon, Xeon Phi and GPU devices, despite multiple times lower external memory bandwidth of FPGAs – a feat that was made possible by our highly-optimized design and up to 20x performance improvement with temporal blocking compared to a design without temporal blocking. Furthermore, we achieved the highest performance for single-FPGA 2D and 3D stencil computation up to a radius of four, without restricting input size. Our performance projection shows that the upcoming Stratix 10 MX 2100 device can achieve over 1.7 GFLOP/s for 3D stencil computation thanks to its high-bandwidth memory – a level of performance that is very competitive with its same-generation GPUs. The larger but more bandwidth-constrained Stratix 10 GX 2800 FPGA is also expected to reach up to 4.2 TFLOP/s of compute performance for 2D stencils computation which will very likely outperform its same-generation and next-generation GPUs. Our evaluation showed that for 2D stencil computation, it is better if FPGA resources are spent on temporal parallelism instead of vectorization, since performance scales near-linearly with the former due to large block size, while the latter provide sub-linear scaling due to memory controller inefficiency. On the other hand, temporal blocking will have worse scaling than vectorization for 3D stencils due to small block size and large amount of redundancy and hence, vectorization is preferred. A direct result of this observation is that a large but bandwidth-constrained FPGA like the Stratix 10 GX 2800 device can still achieve great performance for 2D stencil computation, while a smaller but higher-bandwidth device like Stratix 10 MX 2100 can achieve better performance for 3D stencils.

Our study showed that FPGAs have three major advantages over GPUs for stencils computation: First, support for shift registers in FPGAs allows reducing required on-chip memory size for a fixed spatial block size compared to other hardware. Second, in a deep-pipelined FPGA design like ours where no threading is involved, thread divergence is eliminated and complex optimizations like Warp Scheduling are not required. Finally, the flexibility offered by FPGAs allows us to efficiency distribute the on-chip memory between different parallel *temporal* blocks, while the on-chip memory on GPUs is physically spread over different SMs, forcing the designer to instead split the on-chip memory between parallel *spatial* blocks. The result of this is that on FPGAs, the spatial block size can be more than two order of magnitude larger compared to GPUs, allowing much better scaling with temporal blocking up to hundreds of parallel blocks for 2D stencils, and tens of such blocks for 3D.

There are still multiple hurdles in the way of maximizing performance of stencil computation on FPGAs. Primarily, inefficiencies in the memory controller on the current devices cost us 15% of our performance in 2D, and over 40% in 3D stencil computation. This is despite our manual device buffer padding that reduces the negative effect of unaligned accesses for most parameter configurations. Second, reaching the peak $f_{max}$ of current devices is near-impossible in large designs like ours, even with our careful critical path optimizations; this further costs us a big portion of the peak compute performance. Third, DSP utilization and occupancy is also a source of major concern. The DSPs in current devices cannot perform an addition and a multiplication on two sets of different numbers and full DSP occupancy can only be obtained with FMA instructions. This results in sub-optimal DSP occupancy for low-order stencils since one out of each few DSPs will have its adder left unused. Furthermore, for high-order stencils where the number of DSPs required per PE reaches 100 or more, many DSPs will be left out since they cannot be used to implement one extra PE. Finally, the extremely low external memory bandwidth of current FPGAs forces us to heavily rely on temporal blocking to achieve high performance, while more temporal parallelism comes at the cost of more redundancy and more wasted performance. Even though the amount of redundancy is very small for 2D stencils due to large block size ($< 2\%$), it reaches up to 50% for our reported configurations on 3D stencils, and for a fixed block size, temporal blocking will eventually stop scaling for such stencils after a certain degree of temporal parallelism.

The lower-than-peak $f_{max}$ on current and possibly upcoming FPGA devices can be alleviated to some extent by double-pumping DSPs and Block RAMs used for implementing shift registers; however, this will require compiler support which might not necessarily be provided by Intel. Another possibility is using the remaining logic resources (over 40% on Arria 10) to implement mathematical functions. Even though this possibility exists in an HDL design, the OpenCL compiler automatically maps all mathematical functions that can be implemented using DSPs, to DSPs, with the exception of basic integer operations when DSP overutilization is expected. On the Stratix V device, if more DSPs are used than available on the device, the mapper will then remap the functions that cannot be implemented due to lack of DSPs to logic, but at the cost of severe degradation of operating frequency. On the Arria 10 device, however, such remapping is not performed and trying to use even one DSP more than what exists on the device will result in fitting failure. If the compiler provides the possibility of manually forcing mathematical functions to be implemented using soft logic, the programmer might be able to still fit the design in case of small DSP overutilization, or even use slightly higher degree of temporal parallelism in some cases to further improve performance.

Multiple directions can be followed to further extend our implementation:

- Careful analysis of the HDL code generated by the OpenCL compiler can allow a more in-depth understanding of the behavior of the memory controller and memory interface, opening up new possibilities for mitigating the low memory controller efficiency from the kernel side.
- Extending the kernel parameters to support square/cubic or even arbitrary-shaped stencils can help improve the generality of the design.
- Creating an automated framework to automatically transform existing stencil code in C/C++ format or a Domain-Specific Language (DSL) into the highly-optimized FPGA implementation presented here and use our performance model to automatically choose the best configuration, could prove very valuable for the community.
- Out-of-core stencil computation using Intel's recent "host pipe" extension [81] could allow processing stencils that are much larger than the FPGA device memory. Since these pipes/channels use the same interface as on-chip channels, adding support for them could be as simple as moving the *read* and *write* kernels in our design from the kernel to the host, and replacing the channel definitions. Unfortunately, at the time of writing this document, this extension is only supported by the BSP of Intel's reference Arria 10 board. Even though the throughput of host channels is limited by the PCI-E bandwidth, which is lower than the FPGA external memory bandwidth, we do not expect much of a performance degradation for 2D stencils since we rely on small vectors for these stencils and hardly utilize the external memory bandwidth as it is. Performance of 3D stencil computation, however, is expected to degrade with the ratio of the PCI-E to FPGA external memory bandwidth due to reliance on large vector sizes.
- Spatial distribution of the computation over multiple FPGAs is also an interesting path forward. Compared to the few existing cases of distributing stencil computation over multiple FPGAs [82, 83] where the computation is distributed *temporally* since the original implementations do not support spatial blocking, using spatial distribution has multiple advantages. First, for implementations that do not support spatial blocking, temporal distribution will not allow supporting larger input sizes and input size will still be limited by the on-chip memory size per FPGA. In other words, such implementations only allow accelerating very small stencils for a very large number of iterations, which has limited practical value since it is in fact larger stencils that require a higher number of iterations to achieve the higher level of accuracy required by the higher input resolution. On the other hand, spatial distribution will allow efficient acceleration of stencils as large as the cumulative size of the external memory of the available FPGAs. Second, by taking advantage of overlapped blocking with spatial distribution, inter-FPGA communication can be completely avoided. This is in complete contrast to temporal distribution where the entire input grid needs to be sent from every chained FPGA to the next due to the dependency between consecutive time-steps. Furthermore, with spatial distribution, it is possible to use the FPGA on-board network ports to create an inter-FPGA network for halo communication and further improve performance by reducing the amount of redundant computation and memory accesses associated with overlapped blocking. However, in case of Intel FPGA SDK for OpenCL, creating such network requires a rather expensive license for Intel's low-latency MAC IP Core and the board manufacturer to support the network ports in their OpenCL BSP. Finally, with temporal distribution, since data is *streamed* through the inter-FPGA link which is always slower than the FPGA external memory and only the external memory bandwidth of the first FPGA in the chain is used, maximum achievable performance will be limited to the degree of temporal parallelism multiplied by the bandwidth of the inter-FPGA link. In contrast, with spatial distribution, since the extra external memory bandwidth provided by the other FPGAs in the chain is also utilized for streaming, the maximum achievable performance will be equal to the degree of temporal parallelism multiplied by the bandwidth of the FPGA external memory. For a fixed number of FPGAs, this value will be higher than the peak achievable performance using temporal distribution by a factor equal to the ratio of the FPGA external memory bandwidth to the inter-FPGA link bandwidth.

# 6  Summary and Insights

## 6.1  Summary

In this thesis, our goal was to evaluate the usability and performance of FPGAs for accelerating typical HPC workloads using HLS, and determine which of these workloads, if any, match best with the unique architecture of FPGAs. First, we chose the Rodinia benchmarks suite as a representative of typical HPC applications and evaluated multiple different benchmarks from different domains with different levels of optimization on FPGAs. Our results showed that despite the functional portability of OpenCL on FPGAs, which allowed us to reuse existing OpenCL code that is written for GPUs, such code does not perform optimally on FPGAs. Furthermore, basic optimizations recommended by Intel proved to be insufficient to maximize the potential of FPGAs. However, equipped with multiple advanced FPGA-specific optimizations, we managed to improve the performance of the baseline implementations by up to two orders of magnitude, achieving competitive performance to that of other hardware. Even though our results showed that FPGAs could achieve better performance and power efficiency compared to their same-generation CPUs, their performance fell short of their same-generation GPUs, limiting improvement to only power efficiency in most cases. We concluded that this is largely due to the large gap in peak compute performance and external memory bandwidth of FPGAs compared to their same-generation GPUs. For compute-intensive cases, this gap proved to be hard to reduce; however, our experience with the stencil-based benchmarks in our evaluation showed that for memory-bound stencils, due to the better scaling of temporal blocking on FPGAs, it could be possible to reduce or even eliminate the gap of external memory bandwidth.

Based on the results of our initial evaluation, we concluded that stencil computation is one of the computation patterns that could be efficiently accelerated on FPGAs. Hence, we then focused our efforts on creating a general design for accelerating stencil computation on FPGAs. In the design of our stencil accelerator, we employed spatial and temporal blocking simultaneously so that unlike previous work which limited input size by avoiding spatial blocking, we could achieve high performance without limiting input size. Furthermore, we increased the generality of our design by adding support for high-order stencils which are regularly used in scientific simulations. Equipped with a performance model that allowed us to minimize parameter search space, we could limit our performance tuning per stencil to only a few placement and routing operations. The high level of optimization and flexibility of our design with respect to performance tuning allowed us to maximize FPGA area usage at high operating frequencies, achieving a compute-bound-like level of performance regardless of stencil order. Our results proved that our initial guess was correct: due to the architectural advantages of FPGAs over other devices for stencil computation which allow very efficient mapping of this type of computation to these devices, we can achieve higher performance compared to GPU and Xeon Phi devices which have multiple times higher external memory bandwidth. This was despite the low efficiency of the memory controller of our FPGAs, which was responsible for up to 43% loss of performance compared to the values predicted by our model. For 2D stencil computation, we expect FPGAs to outperform their same-generation GPUs for many years to come. However, even though FPGAs can also achieve good performance scalability with temporal blocking for 3D stencils, high external memory bandwidth is crucial for such stencils or else, temporal blocking will eventually stop scaling on a large but bandwidth-starved FPGAs.

## 6.2 Insights

Even though FPGAs are very old devices, they are at the beginning of their path to large-scale adoption in HPC. The role of the FPGA manufacturers is very crucial at this time since if FPGAs are to be able to compete with existing HPC accelerators, especially GPUs, significant improvements in both hardware and software capabilities are required. Among hardware features, the following improvements could prove valuable in HPC:

- The main source of performance bottleneck in current-generation FPGAs is external memory bandwidth. Even though the upcoming Stratix 10 MX series is the first step in addressing this issue, the higher memory bandwidth comes at the cost of less FPGA area and hence, lower peak compute performance compared to the Stratix 10 GX series. On the other hand, the larger GX series can only be coupled with a few banks of DDR4 memory, which will not be able to provide sufficient bandwidth to allow full utilization of the compute capabilities of these FPGAs for the majority of HPC applications. What the manufacturers should focus on is keeping a *balance* between the external memory bandwidth and the compute performance of the FPGAs so that a large set of applications can be accelerated by these devices using a majority of both the external memory bandwidth and the compute resources, rather than being severely bottlenecked by one of them while the other is heavily underutilized.
- Apart from improving the external memory itself, the memory controllers also need major improvements to be able to keep up with the highly-efficient controllers on modern CPU and GPUs. The fact that we could improve the performance of our stencil kernels by up to 30% just by padding the device buffers by a few bytes (Section 5.3.3) clearly shows that the memory controllers on current-generation Intel FPGAs are not even capable of performing basic access realignment. We also showed that these controllers significantly lose their efficiency with large vector sizes. At the end of the day, an inefficient memory controller coupled with high-bandwidth memory could potentially provide even less effective bandwidth than an efficient memory controller controlling a low-bandwidth memory.
- Lower-than-peak $f_{max}$ is a major source of loss of peak compute performance in current-generation FPGAs. With future FPGAs like Stratix 10 being able to theoretically operate at 800 MHz or higher, this loss of performance can become even more severe. We expect that just as how Block RAM double-pumping was shown to be effective on current-generation FPGAs, it could actually become necessary for future FPGAs. In fact, even triple-pumping could prove to be effective on such FPGAs. Apart from that, DSPs could also be double or triple-pumped in next-generation FPGAs, which could allow getting closer to the peak compute performance of these devices without needing the kernel to run near the peak $f_{max}$ of the device DSPs. Unfortunately, however, based on the latest version of the "Intel FPGA SDK for OpenCL Pro Edition: Best Practices Guide" [18], even Block RAM double-pumping which is supported in current-generation FPGAs might not be supported on Stratix 10, let alone Block RAM triple-pumping or DSP multi-pumping. This new limitation seems to stem from the fact that resource multi-pumping will limit the operating frequency of the kernel to values well below the *unrealistic* 800-900 MHz values used as a selling point for Stratix 10, while even without resource multi-pumping, the kernel frequency will very likely still be limited to values much lower than 800 MHz for most designs due to chain of variable updates in collapsed loops (Section 3.2.4.4).

- The addition of support for single-precision floating-point operations in the DSPs of the Arria 10 FPGA was a step in the correct direction to allow easier adoption of FPGAs in HPC. However, a big portion of HPC applications rely on double-precision (or even higher) computation which cannot be efficiently realized on current FPGAs. Configurable DSPs with simultaneous support for double, single and half-precision FPGAs, similar to how ALUs in recent NVIDIA Tesla GPUs (P100 and V100) simultaneously support single and half-precision, could turn FPGAs into full-fledged HPC devices that can efficiently implement any type of workload with any precision.
- Current-generation Intel FPGAs do not allow the adder and the multiplier in the DSPs to be used for two different sets of numbers; this leads to low DSP occupancy and large loss of peak performance unless the target application can be efficiently mapped to FMA operations. It is expected that this limitation could be eliminated by small FPGA area overhead in form of extra DSP inputs/outputs and global routing wires.
- Improvements in Block RAMs, especially in form of extra ports that can allow lowering replication factor to allow parallel access to on-chip buffers, could allow noticeable improvements in every application. For a fixed total Block RAM size, fewer but larger Block RAMs could also be advantageous compared to more but smaller ones in cases where large but infrequently-accessed on-chip buffers are required, specially shift registers and FIFOs, since less global routing will be required to chain the blocks and routing congestion will be lowered.
- Based on our experience, Partial Reconfiguration on Arria 10 creates multiple additional issues not just with respect to performance and placement and routing quality, but also in terms of reliability since run-time partial reconfiguration through PCI-E is an unreliable operation with high chance of failure (application or OS crash). The latter issue can severely hinder adoption of FPGAs in cloud systems since such systems require high reliability. Maybe the time has come for FPGAs to adopt static non-reconfigurable PCI-E controllers like GPUs, eliminating the need for Partial Reconfiguration to use FPGAs as a PCI-E-attached accelerator. Furthermore, this will also eliminate the need for the complex initial set-up of the PCI-E core on the FPGA to be used with OpenCL, turning these devices into a full-fledged accelerator that can be installed in a host machine and used right away.

With respect to software, the following improvements could ease the adoption of FPGAs in HPC:

- Placement and routing time on FPGAs is a major limiting factor in performance evaluation of these devices, and as the devices get larger, this problem becomes even more pronounced. What is specifically lacking in current HLS tools is a fast *clock-accurate* simulator that would allow users to evaluate the performance of their designs *without* needing to actually place and route them.
- Improvements in the HLS compilers are slow, and sometimes large performance or area utilization regressions are observed with new versions. This becomes even more problematic when the long update cycle of such tools is taken into account, with only one or two major updates per year. Performance consistency should be improved so that new versions perform at least as fast as older versions.
- For the particular case of Intel FPGA SDK for OpenCL, the BSPs are a major source of concern. Since BSPs are provided by board manufactures who generally do not have enough incentive to regularly update their BSPs, specially due to difficulty of timing closure, features introduced in new versions of the compiler cannot be used for months until the manufactures release a compatible BSP. Furthermore, some manufacturers never support certain components of their boards in their OpenCL BSP (network ports, QDR memory, multiple DDR banks, etc.), which prevents OpenCL users from accessing these components even though they have paid full price for the boards. Minimizing the components that need to be supported by the BSP for correct operation and instead, allowing the user to optionally add other components at compile-time alongside with the OpenCL kernel could largely alleviate this issue. After all, all such components use IP Cores provided by Intel and the only board-specific information are the board-related timing values that can be provided by the manufacturer in the BSP. In such case, the compiler can just read the timing values from the BSP and instantiate the necessary IP Cores alongside with the kernel.
- The high price of the FPGA tools and lack of libraries and open-source projects significantly hinder the ability of a large part of the community in adopting FPGAs. A person trying to start coding on GPUs only needs to install a CUDA or OpenCL compiler on his machine, which are provided for free, and he can immediately start coding. Furthermore, there are a plethora of highly-optimized libraries and numerous existing open-source projects that he can integrate into his project to maximize his work efficiency. Until FPGA manufacturers can provide a similar ecosystem, FPGAs will not get close to the adoption rate of GPUs in the HPC community or among hobbyists.

# References

[1]  G. E. Moore, "Cramming More Components Onto Integrated Circuits," *Proceedings of the IEEE,* vol. 86, no. 1, pp. 82-85, Jan. 1998.

[2]  R. H. Dennard, F. H. Gaensslen, V. L. Rideout, E. Bassous and A. R. LeBlanc, "Design of Ion-Implanted MOSFET's with Very Small Physical Dimensions," *IEEE Journal of Solid-State Circuits,* vol. 9, no. 5, pp. 256-268, Oct. 1974.

[3]  H. Esmaeilzadeh, E. Blem, R. S. Amant, K. Sankaralingam and D. Burger, "Dark Silicon and the End of Multicore Scaling," in *38th Annual International Symposium on Computer Architecture (ISCA)*, San Jose, CA, 2011.

[4]  Z. Zhang, Y. Fan, W. Jiang, G. Han, C. Yang and J. Cong, "AutoPilot: A Platform-Based ESL Synthesis System," in *High-Level Synthesis: From Algorithm to Digital Circuit*, P. Coussy and A. Morawiec, Eds., Dordrecht, Springer Netherlands, 2008, pp. 99-112.

[5]  T. Feist, "White Paper: Vivado Design Suite," 22 June 2012. [Online]. Available: https://www.xilinx.com/support/documentation/white_papers/wp416-Vivado-Design-Suite.pdf.

[6]  T. S. Czajkowski, U. Aydonat, D. Denisenko, J. Freeman, M. Kinsner, D. Neto, J. Wong, P. Yiannacouras and D. P. Singh, "From Opencl to High-Performance Hardware on FPGAS," in *22nd International Conference on Field Programmable Logic and Applications (FPL)*, Oslo, 2012.

[7]  Xilinx, Inc., "UG1023 (v2017.4): SDAccel Environment User Guide," 30 Mar. 2018. [Online]. Available: https://www.xilinx.com/support/documentation/sw_manuals/xilinx2017_4/ug1023-sdaccel-user-guide.pdf.

[8]  A. Putnam, A. M. Caulfield, E. S. Chung, D. Chiou, K. Constantinides, J. Demme, H. Esmaeilzadeh, J. Fowers, G. P. Gopal, J. Gray, M. Haselman, S. Hauck, S. Heil, A. Hormati, J.-Y. Kim, S. Lanka, J. Larus, E. Peterson, S. Pope, A. Smith, J. Thong, P. Y. Xiao and D. Burger, "A Reconfigurable Fabric for Accelerating Large-scale Datacenter Services," in *Proceeding of the 41st Annual International Symposium on Computer Architecuture (ISCA)*, Minneapolis, MN, 2014.

[9]  Intel Corporation, "Intel Arria 10 Device Datasheet," 15 June 2018. [Online]. Available: https://www.altera.com/en_US/pdfs/literature/hb/arria-10/a10_datasheet.pdf.

[10]  Amazon Web Services, Inc., "Amazon Elastic Compute Cloud: User Guide for Linux Instances," July 2018. [Online]. Available: https://docs.aws.amazon.com/AWSEC2/latest/UserGuide/ec2-ug.pdf#fpga-getting-started.

[11] Xilinx, Inc., "UG1240 (v2017.2): Getting Started with the SDAccel Environment on Nimbix Cloud," 16 Aug. 2017. [Online]. Available: https://www.xilinx.com/support/documentation/sw_manuals/xilinx2017_2/ug1240-sdaccel-nimbix-getting-started.pdf.

[12] Nallatech, Inc., "Nallatech 385A FPGA Acceleration Card," 2 Oct. 2017. [Online]. Available: http://www.nallatech.com/wp-content/uploads/Nallatech-385A-Product-Brief-v2.3.pdf.

[13] S. Che, M. Boyer, J. Meng, D. Tarjan, J. W. Sheaffer, S. H. Lee and K. Skadron, "Rodinia: A Benchmark Suite for Heterogeneous Computing," in *IEEE International Symposium on Workload Characterization (IISWC)*, Austin, TX, 2009.

[14] Intel Corporation, "Intel Arria 10 Device Overview," 4 Apr. 2018. [Online]. Available: https://www.altera.com/en_US/pdfs/literature/hb/arria-10/a10_overview.pdf.

[15] Intel Corporation, "Intel Arria 10 Core Fabric and General Purpose I/Os Handbook," 7 May 2018. [Online]. Available: https://www.altera.com/en_US/pdfs/literature/hb/arria-10/a10_handbook.pdf.

[16] Khronos OpenCL Working Group, "The OpenCL Specification: Version 1.0," 10 June 2009. [Online]. Available: https://www.khronos.org/registry/OpenCL/specs/opencl-1.0.pdf.

[17] H. R. Zohouri, N. Maruyama, A. Smith, M. Matsuda and S. Matsuoka, "Evaluating and Optimizing OpenCL Kernels for High Performance Computing with FPGAs," in *Proceedings of the International Conference for High Performance Computing, Networking, Storage and Analysis (SC)*, Salt Lake City, UT, 2016.

[18] Intel Corporation, "Intel FPGA SDK for OpenCL: Best Practices Guide," 4 May 2018. [Online]. Available: https://www.altera.com/en_US/pdfs/literature/hb/opencl-sdk/aocl-best-practices-guide.pdf.

[19] Intel Corporation, "Intel FPGA SDK for OpenCL: Programming Guide," 14 June 2018. [Online]. Available: https://www.altera.com/en_US/pdfs/literature/hb/opencl-sdk/aocl_programming_guide.pdf.

[20] Intel Corporation, "Configurationvia Protocol (CvP) Implementation in V-series FPGA Devices User Guide," 31 Oct. 2016. [Online]. Available: https://www.altera.com/en_US/pdfs/literature/ug/ug_cvp.pdf.

[21] Intel Corporation, "Arria 10 CvP Initialization and Partial Reconfiguration over PCI Express User Guide," 31 Oct. 2016. [Online]. Available: https://www.altera.com/en_US/pdfs/literature/ug/ug_a10_cvp_prop.pdf.

[22] H. R. Zohouri, N. Maruyama, A. Smith, M. Matsuda and S. Matsuoka, "Towards Understanding the Performance of FPGAs using OpenCL Benchmarks," in *10th HiPEAC Workshop on Reconfigurable Computing*, Prague, 2016.

[23] A. Danalis, G. Marin, C. McCurdy, J. S. Meredith, P. C. Roth, K. Spafford, V. Tipparaju and J. S. Vetter, "The Scalable Heterogeneous Computing (SHOC) Benchmark Suite," in *Proceedings of the 3rd Workshop on General-Purpose Computation on Graphics Processing Units*, Pittsburgh, PA, 2010.

[24] K. Krommydas, W. Feng, C. D. Antonopoulos and N. Bellas, "OpenDwarfs: Characterization of Dwarf-Based Benchmarks on Fixed and Reconfigurable Architectures," *Journal of Signal Processing Systems,* vol. 85, no. 3, pp. 373-392, Dec. 2016.

[25] J. A. Stratton, C. Rodrigues, I.-J. Sung, N. Obeid, L.-W. Chang, N. Anssari, G. D. Liu and W.-m. W. Hwu, "Parboil: A Revised Benchmark Suite for Scientific and Commercial Throughput Computing," IMPACT Technical Report, 2012.

[26] L.-N. Pouchet, "Polybench: The Polyhedral Benchmark Suite," 5 May 2015. [Online]. Available: http://web.cse.ohio-state.edu/~pouchet.2/software/polybench/.

[27] K. Asanović, R. Bodik, B. C. Catanzaro, J. J. Gebis, P. Husbands, K. Keutzer, D. A. Patterson, W. L. Plishker, J. Shalf, S. W. Williams and K. A. Yelick, "The Landscape of Parallel Computing Research: A View from Berkeley," 2006.

[28] Kingston Technology, "Kingston KVR16S11S6/2 Memory Module Specification," 11 Dec. 2013. [Online]. Available: https://www.kingston.com/dataSheets/KVR16S11S6_2.pdf.

[29] Nvidia Corp, "NVML," Oct. 2017. [Online]. Available: https://docs.nvidia.com/pdf/NVML_API_Reference_Guide.pdf.

[30] "Linux Programmer's Manual: MSR," 31 Mar. 2009. [Online]. Available: http://man7.org/linux/man-pages/man4/msr.4.html.

[31] M. Owaida, N. Bellas, K. Daloukas and C. D. Antonopoulos, "Synthesis of Platform Architectures from OpenCL Programs," in *IEEE 19th Annual International Symposium on Field-Programmable Custom Computing Machines (FCCM)*, Salt Lake City, UT, 2011.

[32] K. Krommydas, R. Sasanka and W. Feng, "Bridging the FPGA Programmability-Portability Gap via Automatic OpenCL Code Generation and Tuning," in *IEEE 27th International Conference on Application-specific Systems, Architectures and Processors (ASAP)*, London, 2016.

[33] S. Che, J. Li, J. W. Sheaffer, K. Skadron and J. Lach, "Accelerating Compute-Intensive Applications with GPUs and FPGAs," in *Symposium on Application Specific Processors*, Anaheim, CA, 2008.

[34] S. Lee, J. Kim and J. S. Vetter, "OpenACC to FPGA: A Framework for Directive-Based High-Performance Reconfigurable Computing," in *IEEE International Parallel and Distributed Processing Symposium (IPDPS)*, Chicago, IL, 2016.

[35] J. Lambert, S. Lee, J. Kim, J. S. Vetter and A. D. Malony, "Directive-Based, High-Level Programming and Optimizations for High-Performance Computing with FPGAs," in *Proceedings of 2018 International Conference on Supercomputing (ICS)*, Beijing, 2018.

[36] F. B. Muslim, L. Ma, M. Roozmeh and L. Lavagno, "Efficient FPGA Implementation of OpenCL High-Performance Computing Applications via High-Level Synthesis," *IEEE Access,* vol. 5, pp. 2747-2762, 2017.

[37] Q. Gautier, A. Althoff, P. Meng and R. Kastner, "Spector: An OpenCL FPGA Benchmark Suite," in *International Conference on Field-Programmable Technology (FPT)*, Xi'an, 2016.

[38] T. Lloyd, A. Chikin, E. Ochoa, K. Ali and J. N. Amaral, "A Case for Better Integration of Host and Target Compilation When Using OpenCL for FPGAs," in *Fourth International Workshop on FPGAs for Software Programmers (FSP)*, Ghent, 2017.

[39] J. Cong, Z. Fang, M. Lo, H. Wang, J. Xu and S. Zhang, "Understanding Performance Differences of FPGAs and GPUs," in *IEEE 26th Annual International Symposium on Field-Programmable Custom Computing Machines (FCCM)*, Boulder, CO, 2018.

[40] "Understanding Performance Differences of FPGAs and GPUs," [Online]. Available: http://hwang.me/HanruiWang_FPGA_paper.pdf.

[41] J. Fine Licht, S. Meierhans and T. Hoefler, "Transformations of High-Level Synthesis Codes for High-Performance Computing," Aug. 2018. [Online]. Available: https://arxiv.org/pdf/1805.08288v3.pdf.

[42] H. R. Zohouri, A. Podobas and S. Matsuoka, "Combined Spatial and Temporal Blocking for High-Performance Stencil Computation on FPGAs Using OpenCL," in *Proceedings of the 2018 ACM/SIGDA International Symposium on Field-Programmable Gate Arrays (FPGA)*, Monterey, CA, 2018.

[43] H. R. Zohouri, A. Podobas and S. Matsuoka, "High-Performance High-Order Stencil Computation on FPGAs Using OpenCL," in *IEEE International Parallel and Distributed Processing Symposium Workshops (IPDPSW)*, Vancouver, BC, 2018.

[44] A. Nguyen, N. Satish, J. Chhugani, C. Kim and P. Dubey, "3.5D Blocking Optimization for Stencil Computations on Modern CPUs and GPUs," in *ACM/IEEE International*

*Conference for High Performance Computing, Networking, Storage and Analysis (SC)*, New Orleans, LA, 2010.

[45] V. Bandishti, I. Pananilath and U. Bondhugula, "Tiling Stencil Computations to Maximize Parallelism," in *Proceedings of the International Conference on High Performance Computing, Networking, Storage and Analysis*, Los Alamitos, CA, 2012.

[46] T. Grosser, A. Cohen, J. Holewinski, P. Sadayappan and S. Verdoolaege, "Hybrid Hexagonal/Classical Tiling for GPUs," in *Proceedings of Annual IEEE/ACM International Symposium on Code Generation and Optimization (CGO)*, Orlando, FL, 2014.

[47] C. Yount, "Vector Folding: Improving Stencil Performance via Multi-dimensional SIMD-vector Representation," in *IEEE 17th International Conference on High Performance Computing and Communications, 2015 IEEE 7th International Symposium on Cyberspace Safety and Security, and 2015 IEEE 12th International Conference on Embedded Software and Systems*, New York, NY, 2015.

[48] C. Yount, J. Tobin, A. Breuer and A. Duran, "YASK–Yet Another Stencil Kernel: A Framework for HPC Stencil Code-Generation and Tuning," in *Sixth International Workshop on Domain-Specific Languages and High-Level Frameworks for High Performance Computing (WOLFHPC)*, Salt Lake City, UT, 2016.

[49] N. Maruyama and T. Aoki, "Optimizing Stencil Computations for NVIDIA Kepler GPUs," in *Proceedings of the 1st International Workshop on High-Performance Stencil Computations (HiStencils)*, Vienna, 2014.

[50] N. Prajapati, W. Ranasinghe, S. Rajopadhye, R. Andonov, H. Djidjev and T. Grosser, "Simple, Accurate, Analytical Time Modeling and Optimal Tile Size Selection for GPGPU Stencils," in *Proceedings of the 22nd ACM SIGPLAN Symposium on Principles and Practice of Parallel Programming (PPoPP)*, Austin, TX, 2017.

[51] G. Deest, T. Yuki, S. Rajopadhye and S. Derrien, "One Size Does Not Fit All: Implementation Trade-Offs for Iterative Stencil Computations on FPGAs," in *27th International Conference on Field Programmable Logic and Applications (FPL)*, Ghent, 2017.

[52] P. Rawat, M. Kong, T. Henretty, J. Holewinski, K. Stock, L.-N. Pouchet, J. Ramanujam, A. Rountev and P. Sadayappan, "SDSLc: A Multi-target Domain-specific Compiler for Stencil Computations," in *Proceedings of the 5th International Workshop on Domain-Specific Languages and High-Level Frameworks for High Performance Computing (WOLFHPC)*, Austin, TX, 2015.

[53] S. Wang and Y. Liang, "A Comprehensive Framework for Synthesizing Stencil Algorithms on FPGAs using OpenCL Model," in *54th ACM/EDAC/IEEE Design Automation Conference (DAC)*, Austin, TX, 2017.

[54] A. A. Nacci, V. Rana, F. Bruschi, D. Sciuto, I. Beretta and D. Atienza, "A High-level Synthesis Flow for the Implementation of Iterative Stencil Loop Algorithms on FPGA Devices," in *Proceedings of the 50th Annual Design Automation Conference (DAC)*, Austin, TX, 2013.

[55] R. Cattaneo, G. Natale, C. Sicignano, D. Sciuto and M. D. Santambrogio, "On How to Accelerate Iterative Stencil Loops: A Scalable Streaming-Based Approach," *ACM Trans. Archit. Code Optim.,* vol. 12, no. 4, pp. 1-26, Jan. 2016.

[56] T. Kenter, J. Förstner and C. Plessl, "Flexible FPGA design for FDTD using OpenCL," in *27th International Conference on Field Programmable Logic and Applications (FPL)*, Ghent, 2017.

[57] K. Sano and S. Yamamoto, "FPGA-Based Scalable and Power-Efficient Fluid Simulation using Floating-Point DSP Blocks," *IEEE Transactions on Parallel and Distributed Systems,* vol. 28, no. 10, pp. 2823-2837, Oct. 2017.

[58] H. M. Waidyasooriya, Y. Takei, S. Tatsumi and M. Hariyama, "OpenCL-Based FPGA-Platform for Stencil Computation and Its Optimization Methodology," *IEEE Transactions on Parallel and Distributed Systems,* vol. 28, no. 5, pp. 1390-1402, May 2017.

[59] E. Reggiani, G. Natale, C. Moroni and M. D. Santambrogio, "An FPGA-based Acceleration Methodology and Performance Model for Iterative Stencils," in *IEEE International Parallel and Distributed Processing Symposium Workshops (IPDPSW)*, Vancouver, BC, 2018.

[60] Y. Cui, E. Poyraz, K. B. Olsen, J. Zhou, K. Withers, S. Callaghan, J. Larkin, C. Guest, D. Choi, A. Chourasia, Z. Shi, S. M. Day, P. J. Maechling and T. H. Jordan, "Physics-based Seismic Hazard Analysis on Petascale Heterogeneous Supercomputers," in *Proceedings of the International Conference on High Performance Computing, Networking, Storage and Analysis*, Denver, CO, 2013.

[61] T. Shimokawabe, T. Aoki and N. Onodera, "High-productivity Framework on GPU-rich Supercomputers for Operational Weather Prediction Code ASUCA," in *Proceedings of the International Conference for High Performance Computing, Networking, Storage and Analysis (SC)*, New Orleans, LA, 2014.

[62] W. Xue, C. Yang, H. Fu, X. Wang, Y. Xu, J. Liao, L. Gan, Y. Lu, R. Ranjan and L. Wang, "Ultra-Scalable CPU-MIC Acceleration of Mesoscale Atmospheric Modeling on Tianhe-2," *IEEE Transactions on Computers,* vol. 64, no. 8, pp. 2382-2393, Aug. 2015.

[63] F. Richter, M. Schmidt and D. Fey, "A Configurable VHDL Template for Parallelization of 3D Stencil Codes on FPGAs," in *Proceedings of the International Conference on Engineering of Reconfigurable Systems and Algorithms (ERSA)*, Las Vegas, NV, 2012.

[64] X. Niu, J. G. F. Coutinho, Y. Wang and W. Luk, "Dynamic Stencil: Effective Exploitation of Run-time Resources in Reconfigurable Clusters," in *International Conference on Field-Programmable Technology (FPT)*, Kyoto, 2013.

[65] O. Lindtjorn, R. Clapp, O. Pell, H. Fu, F. M. and O. Mencer, "Beyond Traditional Microprocessors for Geoscience High-Performance Computing Applications," *IEEE Micro,* vol. 31, no. 2, pp. 41-49, March-April 2011.

[66] O. Pell, J. Bower, R. Dimond, O. Mencer and M. J. Flynn, "Finite-Difference Wave Propagation Modeling on Special-Purpose Dataflow Machines," *IEEE Transactions on Parallel and Distributed Systems,* vol. 24, no. 5, pp. 906-915, May 2013.

[67] J. Fine Licht, M. Blott and T. Hoefler, "Designing Scalable FPGA Architectures Using High-level Synthesis," in *Proceedings of the 23rd ACM SIGPLAN Symposium on Principles and Practice of Parallel Programming*, New York, NY, USA, 2018.

[68] H. Fu, C. He, B. Chen, Z. Yin, Z. Zhang, W. Zhang, T. Zhang, W. Xue, W. Liu, W. Yin, G. Yang and X. Chen, "18.9-Pflops Nonlinear Earthquake Simulation on Sunway TaihuLight: Enabling Depiction of 18-Hz and 8-meter Scenarios," in *Proceedings of the International Conference for High Performance Computing, Networking, Storage and Analysis (SC)*, Denver, CO, 2017.

[69] T. Muranushi, H. Hotta, J. Makino, S. Nishizawa, H. Tomita, K. Nitadori, M. Iwasawa, N. Hosono, Y. Maruyama, H. Inoue, H. Yashiro and Y. Nakamura, "Simulations of Below-ground Dynamics of Fungi: 1.184 Pflops Attained by Automated Generation and Autotuning of Temporal Blocking Codes," in *Proceedings of the International Conference for High Performance Computing, Networking, Storage and Analysis (SC)*, Salt Lake City, UT, 2016.

[70] C. Yang, W. Xue, H. Fu, H. You, X. Wang, Y. Ao, F. Liu, L. Gan, P. Xu, L. Wang, G. Yang and W. Zheng, "10M-Core Scalable Fully-Implicit Solver for Nonhydrostatic Atmospheric Dynamics," in *International Conference for High Performance Computing, Networking, Storage and Analysis (SC)*, Salt Lake City, UT, 2016.

[71] W. T. Tang, W. J. Tan, R. Krishnamoorthy, Y. W. Wong, S. H. Kuo, R. S. M. Goh, S. J. Turner and W. F. Wong, "Optimizing and Auto-Tuning Iterative Stencil Loops for GPUs with the In-Plane Method," in *IEEE 27th International Symposium on Parallel and Distributed Processing (IPDPS)*, Boston, MA, 2013.

[72] M. Shafiq, M. Pericàs, R. Cruz, M. Araya-Polo, N. Navarro and E. Ayguadé, "Exploiting Memory Customization in FPGA for 3D Stencil Computations," in *International Conference on Field-Programmable Technology (FPT)*, Sydney, NSW, 2009.

[73] H. Fu and R. G. Clapp, "Eliminating the Memory Bottleneck: An FPGA-based Solution for 3D Reverse Time Migration," in *Proceedings of the 19th ACM/SIGDA International Symposium on Field Programmable Gate Arrays (FPGA)*, Monterey, CA, 2011.

[74] Bittware, Inc., "S10VM4," 28 Nov. 2017. [Online]. Available: http://www.bittware.com/wp-content/uploads/datasheets/ds-s10vm4.pdf.

[75] Nallatech, "Nallatech 520C Compute Acceleration Card," 2 Oct. 2017. [Online]. Available: http://www.nallatech.com/wp-content/uploads/Nallatech-520C-Product-Brief-V9.pdf.

[76] E. Nurvitadhi, G. Venkatesh, J. Sim, D. Marr, R. Huang, J. Ong Gee Hock, Y. T. Liew, K. Srivatsan, D. Moss, S. Subhaschandra and G. Boudoukh, "Can FPGAs Beat GPUs in Accelerating Next-Generation Deep Neural Networks?," in *Proceedings of the 2017 ACM/SIGDA International Symposium on Field-Programmable Gate Arrays (FPGA)*, Monterey, CA, 2017.

[77] C. Yount and A. Duran, "Effective Use of Large High-Bandwidth Memory Caches in HPC Stencil Computation via Temporal Wave-Front Tiling," in *7th International Workshop on Performance Modeling, Benchmarking and Simulation of High Performance Computer Systems (PMBS)*, Salt Lake City, UT, 2016.

[78] Intel Corporation, "Intel Stratix 10 MX Devices Solve the Memory Bandwidth Challenge," 2017. [Online]. Available: https://www.altera.com/content/dam/altera-www/global/en_US/pdfs/literature/wp/wp-01264-stratix10mx-devices-solve-memory-bandwidth-challenge.pdf.

[79] M. Hutton, "Stratix 10: 14nm FPGA delivering 1GHz," in *IEEE Hot Chips 27 Symposium (HCS)*, Cupertino, CA, 2015.

[80] S. Williams, A. Waterman and D. Patterson, "Roofline: An Insightful Visual Performance Model for Multicore Architectures," *Commun. ACM,* vol. 52, no. 4, pp. 65-76, Apr. 2009.

[81] K. Kang and P. Yiannacouras, "Host Pipes: Direct Streaming Interface Between OpenCL Host and Kernel," in *Proceedings of the 5th International Workshop on OpenCL (IWOCL)*, Toronto, 2017.

[82] A. Mondigo, T. Ueno, D. Tanaka, K. Sano and S. Yamamoto, "Design and Scalability Analysis of Bandwidth-Compressed Stream Computing with Multiple FPGAs," in *12th International Symposium on Reconfigurable Communication-centric Systems-on-Chip (ReCoSoC)*, Madrid, 2017.

[83] G. Natale, G. Stramondo, P. Bressana, R. Cattaneo, D. Sciuto and M. D. Santambrogio, "A Polyhedral Model-based Framework for Dataflow Implementation on FPGA Devices of Iterative Stencil Loops," in *Proceedings of the 35th International Conference on Computer-Aided Design (ICCAD)*, Austin, TX, 2016.

# Publications

## Refereed Conferences

**Hamid Reza Zohouri**, Naoya Maruyama, Aaron Smith, Motohiko Matsuda, and Satoshi Matsuoka, "Evaluating and Optimizing OpenCL Kernels for High Performance Computing with FPGAs," in *Proceedings of the International Conference for High Performance Computing, Networking, Storage and Analysis (SC)*, Salt Lake City, UT, Nov. 2016, p. 1-12.

Artur Podobas, **Hamid Reza Zohouri**, Naoya Maruyama, and Satoshi Matsuoka, "Evaluating High-Level Design Strategies on FPGAs for High-Performance Computing," in *27th International Conference on Field Programmable Logic and Applications (FPL)*, Ghent, Sept. 2017, pp. 1-4.

**Hamid Reza Zohouri**, Artur Podobas, and Satoshi Matsuoka, "Combined Spatial and Temporal Blocking for High-Performance Stencil Computation on FPGAs Using OpenCL," in *Proceedings of the 2018 ACM/SIGDA International Symposium on Field-Programmable Gate Arrays (FPGA)*, Monterey, CA, Feb. 2018, pp. 153-162.

## Refereed Workshops

**Hamid Reza Zohouri**, Naoya Maruyama, A. Smith, M. Matsuda, and Satoshi Matsuoka, "Towards Understanding the Performance of FPGAs using OpenCL Benchmarks," in *10th HiPEAC Workshop on Reconfigurable Computing (WRC)*, Prague, Feb. 2016.

**Hamid Reza Zohouri**, Artur Podobas, and Satoshi Matsuoka, "High-Performance High-Order Stencil Computation on FPGAs Using OpenCL," in *IEEE International Parallel and Distributed Processing Symposium Workshops (IPDPSW)*, Vancouver, BC, May 2018, pp. 123-130.

## Refereed Poster

**Hamid Reza Zohouri**, Artur Podobas, Naoya Maruyama, and Satoshi Matsuoka, "OpenCL-Based High-Performance 3D Stencil Computation on FPGAs," *International Conference for High Performance Computing, Networking, Storage and Analysis (SC)*, Denver, CO, Nov. 2017.

## Non-Refereed Workshop

**Hamid Reza Zohouri,** Naoya Maruyama, Aaron Smith, Motohiko Matsuda and Satoshi Matsuoka, "Optimizing the Rodinia Benchmark for FPGAs (Unrefereed Workshop Manuscript)," *IPSJ SIG Technical Reports*, vol. 2015-HPC-125, no. 16, Dec. 2015.